\documentclass{emulateapj}
\usepackage{graphicx}
\usepackage{natbib}

\shorttitle{THE OPTX PROJECT I}
\shortauthors{Trouille et al.}

\begin{document}

\title{THE OPTX PROJECT I: THE FLUX AND REDSHIFT CATALOGS FOR THE\\
  CLANS, CLASXS, AND CDF-N FIELDS\altaffilmark{1}}

\author{L. Trouille\altaffilmark{2}, A. J. Barger\altaffilmark{2,3,4},
  L. L. Cowie\altaffilmark{4}, Y. Yang\altaffilmark{5}, and
 R. F. Mushotzky\altaffilmark{6}}

\altaffiltext{1}{Some of the data presented herein were obtained at the W. M. Keck Observatory, which is operated as a scientific partnership among the California Institute of Technology, the University of California, and the National Aeronautics and Space Administration. The observatory was made possible by the generous financial support of the W. M. Keck Foundation.}
\altaffiltext{2}{Department of Astronomy, University of Wisconsin-Madison, 475 N. Charter Street, Madison, WI 53706}
\altaffiltext{3}{Department of Physics and Astronomy, University of Hawaii, 2505 Correa Road, Honolulu, HI 96822}
\altaffiltext{4}{Institute for Astronomy, University of Hawaii, 2680 Woodlawn Drive, Honolulu, HI 96822}
\altaffiltext{5}{Department of Astronomy, University of Illinois, 1002 W. Green St., Urbana, IL 61801}
\altaffiltext{6}{NASA Goddard Space Flight Center, Code 662, Greenbelt, MD 20771}

\begin{abstract}

We present the redshift catalogs for the X-ray sources detected in the
\emph{Chandra} Deep Field
North (CDF-N), the \emph{Chandra} Large Area Synoptic X-ray Survey
(CLASXS), and the \emph{Chandra} Lockman Area North Survey
(CLANS). The catalogs for the CDF-N and CLASXS fields include
redshifts from previous work, while the redshifts for the CLANS field
are all new. For fluxes above $10^{-14}$ ergs~cm$^{-2}$~s$^{-1}$ ($2-8~\rm
keV$) we have redshifts for $76$\% of the sources. We extend
the redshift information for the full sample
using photometric redshifts. The goal of the OPTX Project is to use
these three surveys, which are among the most spectroscopically
complete surveys to date, to analyze the effect of spectral
type on the shape and evolution of the X-ray luminosity functions and
to compare the optical spectral types with the X-ray spectral
properties.

We also present the CLANS X-ray catalog. The nine ACIS-I
fields cover a solid angle of $\sim$0.6 deg$^2$ and reach fluxes of
$7\times10^{-16}$ ergs~cm$^{-2}$~s$^{-1}$ ($0.5-2~{\rm keV}$) and
$3.5\times10^{-15}$ ergs~cm$^{-2}$~s$^{-1}$ ($2-8~{\rm keV}$). We find
a total of 761 X-ray point sources. Additionally, we present the
optical and infrared photometric catalog for the CLANS X-ray sources,
as well as updated optical and infrared photometric catalogs for the
X-ray sources in the CLASXS and CDF-N fields.  

The CLANS and CLASXS surveys bridge the gap between
the ultradeep pencil-beam surveys, such as the CDFs, and the
shallower, very large-area surveys. As a result, they probe the X-ray
sources that contribute the bulk of the $2-8~{\rm keV}$ X-ray
background and cover the flux range of the observed break in the
log$N-$log$S$ distribution. We construct differential number counts
for each individual field and for the full sample.   

\end{abstract}

\keywords{cosmology: observations  --- galaxies: active}

\section{INTRODUCTION}

We combine data from the \emph{Chandra} Deep Field
North (CDF-N), the \emph{Chandra} Large Area Synoptic X-ray Survey
(CLASXS), and the \emph{Chandra} Lockman Area North Survey (CLANS) to
provide one of the most spectroscopically complete large samples of
\emph{Chandra} X-ray sources for use in the OPTX Project. In this
article, the first in the OPTX Project series, we present the database
that we use in Yencho et al.~(2008) to examine the
X-ray luminosity functions and their dependence on spectral
classification and in L.~Trouille et al.~(2008, in preparation) to
compare the optical spectral classifications with the X-ray spectral
properties.

As a community, we have obtained a myriad of X-ray surveys, from deep
pencil-beam to shallow wide-field surveys (see Figure 1 in
Brandt \& Hasinger 2005). The ultradeep surveys (CDF-N, Brandt et al.~2001 and
Alexander et al.~2003; \emph{Chandra} Deep Field South or CDF-S,
Giacconi et al.~2002) have resolved nearly $100$\% of
the $2-8~\rm keV$ X-ray background (hereafter XRB; see Churazov et
al.~2007 for a recent measurement of the XRB using INTEGRAL and Gilli et
al.~2007 and Frontera et al.~2007 for in-depth comparisons of the
various XRB measurements to date). The shallow (ASCA, Ueda et
al.~1999; XBo\"{o}tes,
Murray et al.~2005) and intermediate-depth (SEXSI, Harrison et
al. 2003; CLASXS, Yang et al.~2004; AEGIS, Nandra et
al.~2005, Davis et al.~2007; extended-CDF-S, Lehmer et al.~2005,
Virani et al.~2006a;
XMM-COSMOS, Cappelluti et al.~2007; ChaMP, Kim et al.~2007a; and
CLANS, this article) wide-area surveys improve statistics on active
galactic nuclei (AGN) evolution and luminosity functions,
detect the rarer sources (obscured QSOs and high-redshift AGNs), and
uncover the extent of large-scale structure.     

However, it is essential that these surveys be followed up spectroscopically as
completely as possible. While deep multi-band photometry for surveys has made
determining photometric redshifts possible (e.g., Rowan-Robinson et
al.~2008 and references therein) and the reliability can be improved
by incorporating near-infrared (NIR) and mid-infrared (MIR) data (Wang et al.~2006), with optical spectra of the X-ray sources we can both make redshift
identifications and spectrally classify the sources. Throughout this
paper, ``identification'' of  redshifts is defined as the robust
determination of a redshift from the observed optical spectrum. We can
also use the optical spectra
to compare optical emission line luminosities with X-ray luminosities
(Mulchaey et al.~1994; Alonso-Herrero et al.~1997).

While other X-ray surveys have numerous redshifts, they are relatively
incomplete in heterogeneous ways (see Table
\ref{compl table}). In contrast, the present article is one in a series
focusing on three of the most uniformly observed and spectroscopically
complete surveys to date. The deep pencil-beam CDF-N survey is the
most spectroscopically
complete of all of the X-ray survey fields. Our group (Barger et
al.~2002, 2003, 2005; present work) has spectroscopically observed 459
of the 503 X-ray sources in this field and has obtained reliable
redshift identifications for 312. In comparison, of
the 349 X-ray sources in the 1 Ms CDF-S, 251 have been spectroscopically
observed and 168 have redshift identifications (Szokoly et al.~2004).     

Our two intermediate-depth wide-area surveys provide an essential
step between the ultradeep narrow \emph{Chandra} surveys and the
shallow wide-area surveys. They cover large cosmological
volumes, detect rare, high-luminosity AGNs, and robustly probe AGN
evolution between $z\sim 0$ and 1.

Yang et al.~(2004) undertook an $\sim$0.4 deg$^2$ contiguous
\emph{Chandra} survey of the Lockman Hole-Northwest field. The
\emph{Chandra} Large Area Synoptic X-ray Survey (CLASXS) was designed
to sample a large, contiguous solid angle, while remaining sensitive
enough to measure $2-3$ times fainter than the observed break in the
$2-8~\rm{keV}$ log$N-$log$S$ distribution. Our group (Steffen et
al.~2004; present work) has spectroscopically observed 468 of the 525
X-ray sources in the CLASXS field and obtained reliable redshift
identifications for 280. 

In 2004 the \emph{Chandra}/SWIRE team (PI, B. Wilkes) observed
a solid angle of $\sim$0.6 deg$^2$ in a second field in the Lockman
Area to a $2-8~\rm keV$ limiting flux of $3.5\times10^{-15}$
ergs~cm$^{-2}$~s$^{-1}$. This field is also part of the \emph{Spitzer}
Wide-Area Infrared
Extragalactic Survey (SWIRE; Lonsdale et al.~2003, 2004), which
surveyed approximately 65 deg$^2$ distributed over 7 fields in
the northern and southern sky. Polletta et al.~(2006) used
\emph{Chandra}/SWIRE to study the \emph{Spitzer} selected
sources in the field. Here we describe the \emph{Chandra} survey in
detail for the first time. For
clarity, we have renamed it the \emph{Chandra} Lockman Area North
Survey (CLANS). We have spectroscopically observed 533 of the 761
X-ray sources in the CLANS field and have obtained reliable redshift
identifications for 336. 

In this paper we present the X-ray data for the CLANS field and the
most up-to-date photometric and spectroscopic data of the optical and infrared
counterparts to the X-ray sources for the CLANS, CLASXS, and CDF-N
fields. In \S \ref{xray} we 
describe the CLANS X-ray observations and provide the X-ray
catalog. We discuss our optical and infrared imaging data in \S
\ref{photometric}, our redshift information in \S \ref{spectroscopictop},
and our optical spectral classifications in \S \ref{optclass}. In \S
\ref{catalogs} we present the CLANS, CLASXS, and CDF-N optical and infrared
photometric and spectroscopic catalogs. We calculate the rest-frame
$2-8~\rm keV$ X-ray luminosities in \S
\ref{Lx}, and in \S \ref{numcts sec} we construct differential
log$N-$log$S$ relations and compare our results with those of other surveys.   

We use J2000 coordinates and assume $\Omega_M=0.3,
\Omega_{\Lambda}=0.7$, and H$_0=70$ km s$^{-1}$ Mpc$^{-1}$. All
magnitudes are in the AB magnitude system and, unless otherwise
specified, fluxes are in ergs~cm$^{-2}$~s$^{-1}$.  

\vspace{1in}
\section{X-RAY PROPERTIES}
\label{xray} 

\subsection{CLANS X-ray Observations}

\begin{figure}[!h]
\epsscale{1}
\plotone{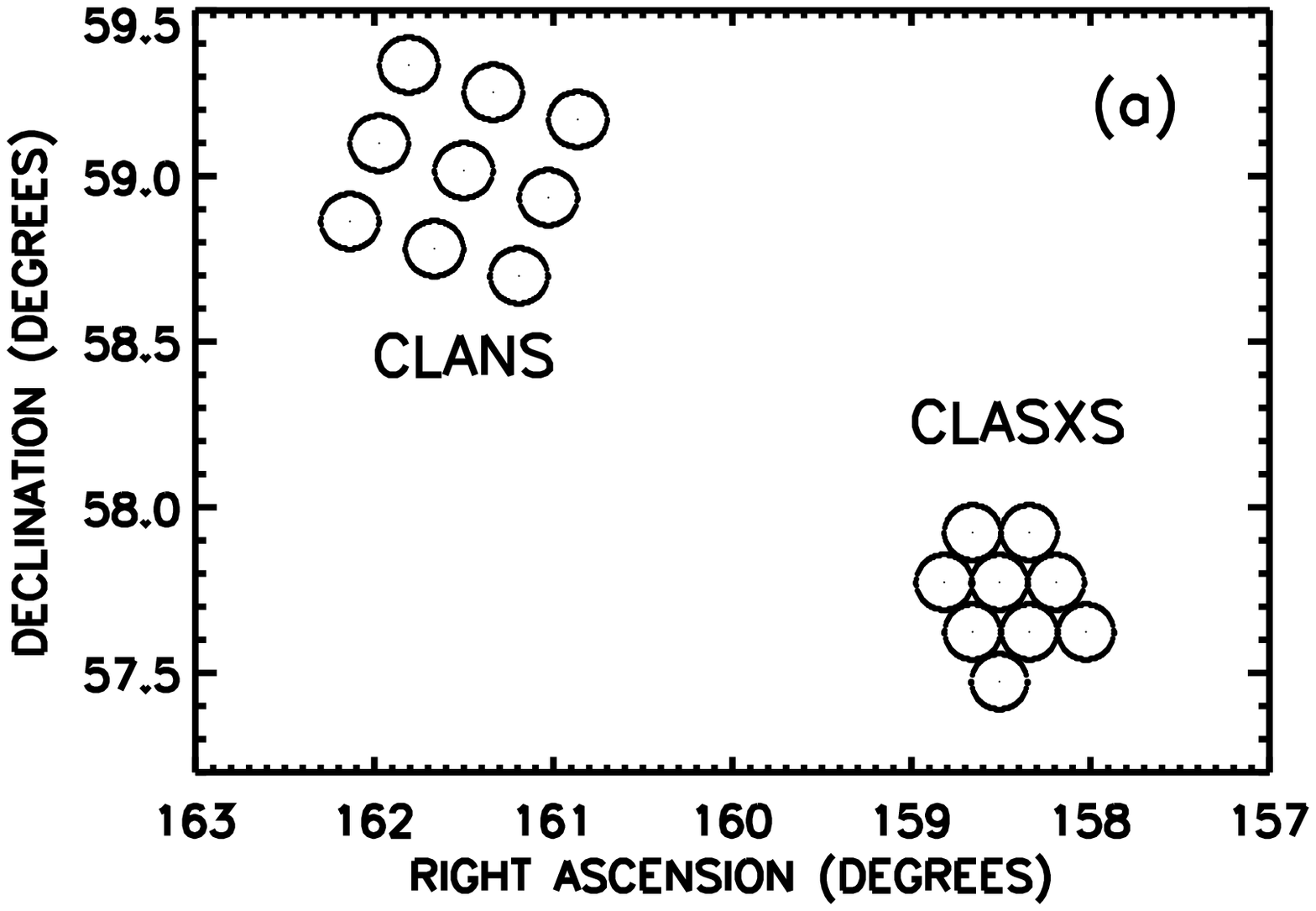}
\plotone{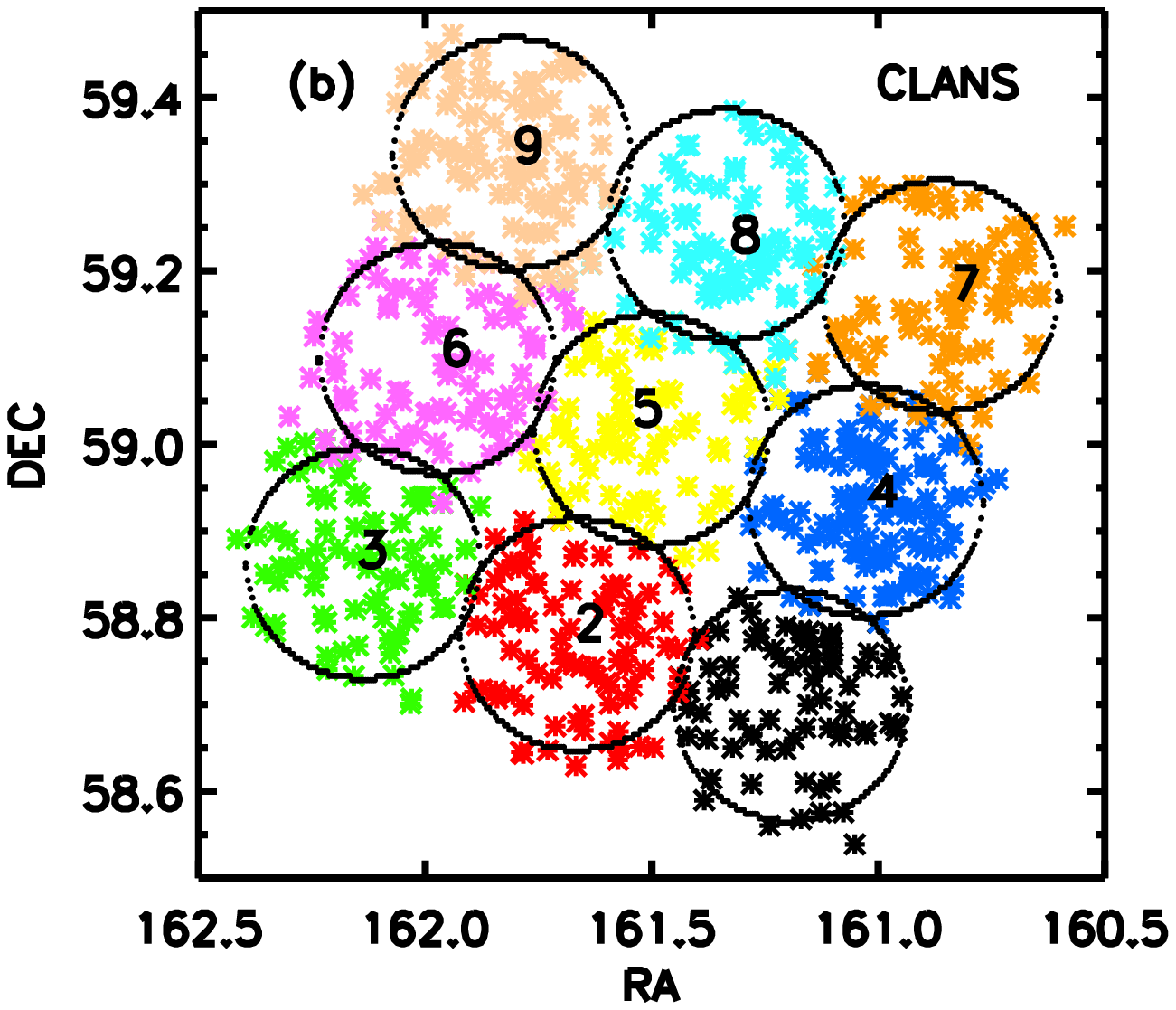}
\plotone{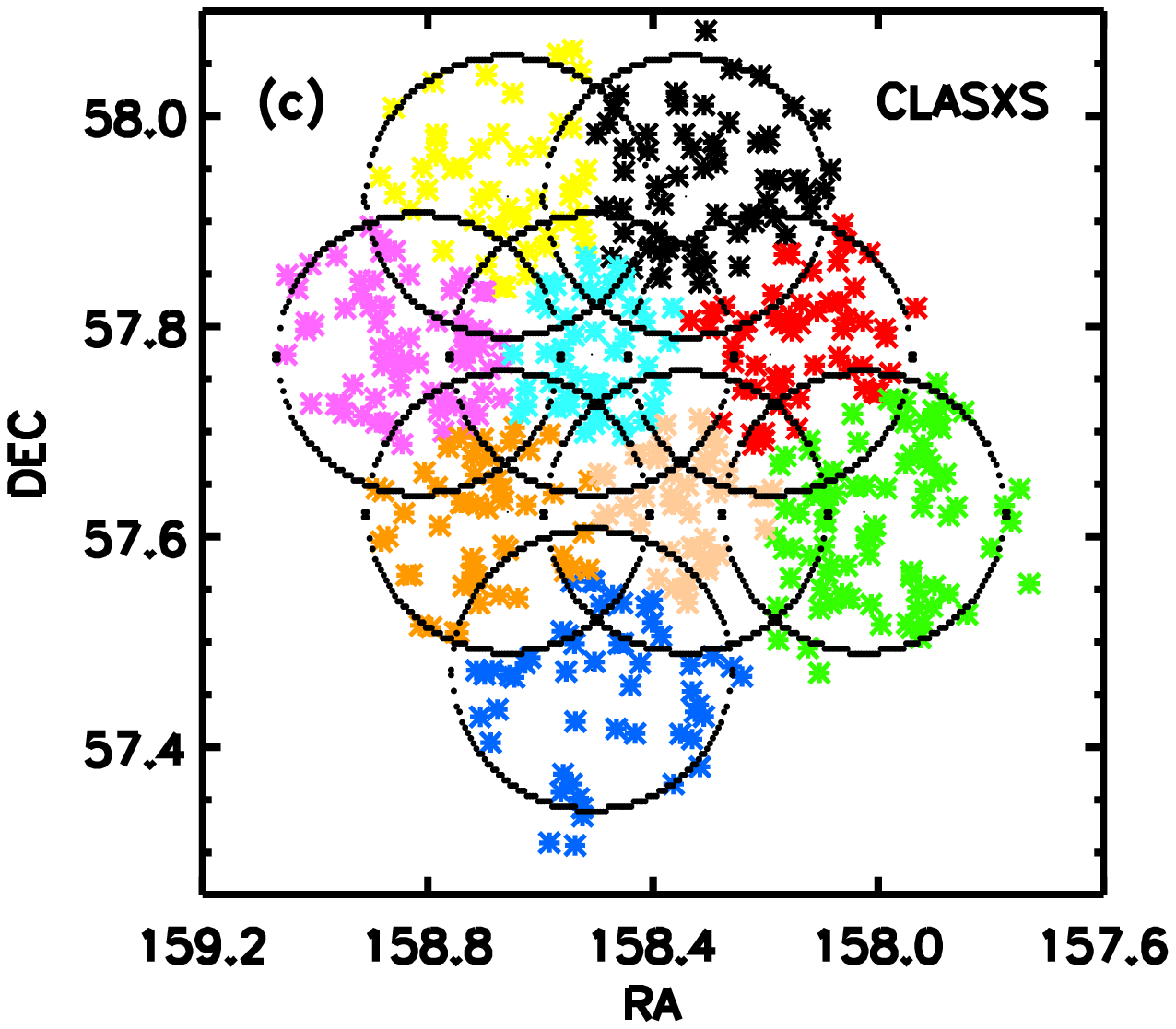}
\caption{(a) Location of the CLANS and CLASXS
  pointings in the Lockman Hole. The circles delimit $5\arcmin$ from
  the pointing centers (the area within which the sensitivity of
  source detection in the ACIS-I images is approximately
  uniform). Location of the (b) CLANS and
  (c) CLASXS X-ray sources. The circles delimit $8\arcmin$ from the
  pointing centers, which is the limiting off-axis angle used in \S
\ref{numcts sec} when calculating the log$N-$log$S$ distributions for these
fields. The numbers 1-9 in (b) correspond to the
  \emph{Chandra}/SWIRE Lockman pointings listed in Table \ref{obssum table}.}
\label{position figure}
\end{figure}
%\clearpage

To minimize the effects of Galactic attenuation, both the CLANS and
CLASXS fields reside in the Lockman Hole high latitude region of
extremely low Galactic HI column density ($5.7\times10^{19}$
cm$^{-2}$; Lockman et al.~1986). The Galactic HI column density along
the line of sight to the CDF-N is $1.6\times10^{20}$ cm$^{-2}$ (Stark
et al.~1992). 

CLANS consists of nine separate $\sim$70 ks \emph{Chandra} ACIS-I
exposures centered at $(\alpha,\delta)_{J2000}=(10^h46^m,
+59^{\circ}01 \arcmin$) (see Table \ref{obssum table}) combined to
create an $\sim$0.6 deg$^2$ image containing 761 sources. The X-ray
flux limits are $f_{2-8~{\rm keV}}\sim 3.5\times10^{-15}$
ergs~cm$^{-2}$~s$^{-1}$ and $f_{0.5-2~{\rm keV}}\sim 7\times10^{-16}$
ergs~cm$^{-2}$~s$^{-1}$.  

While the ACIS-I field of view is $17 \arcmin \times 17 \arcmin $, the
sensitivity of the source detection is approximately uniform only
within off-axis angles less than $\sim5 \arcmin$. As the off-axis
angle increases, the sensitivity drops due to vignetting
effects, quantum efficiency changes across the field, and the
broadening of the point-spread functions (see Figure \ref{sims} and \S
\ref{effarea}). Therefore, while the CLANS ACIS-I pointings exposed
an almost contiguous area, the actual coverage, as evidenced by the
gaps between the
$5 \arcmin $ circles in Figure \ref{position figure}a, is
non-uniform. The survey was optimized to obtain the largest solid
angle possible. The CLASXS survey, on the other hand, was designed to
achieve uniform field coverage (at off-axis angles $\approx5 \arcmin$,
the CLASXS pointings overlap slightly as shown in Figure \ref{position
figure}a).

In Figures \ref{position figure}b and
\ref{position figure}c
we show $8\arcmin$ circles around the CLANS and CLASXS
pointings. This is the limiting off-axis angle used in \S
\ref{numcts sec} when calculating the log$N-$log$S$ distributions for these
fields. While there is substantial overlap between the CLASXS
pointings at these off-axis angles, the CLANS pointings overlap very
little. This is important in the determination of the effective areas
for each field (see \S \ref{effarea}). 

In Table \ref{xray table} we list the characteristics of the CLANS
field in terms of the \emph{Chandra} exposure times, areas covered,
X-ray flux limits, and total number of X-ray sources. We include the CLASXS and
CDF-N characteristics for comparison. 

\subsection{CLANS X-ray Source Detection and Fluxes}
\label{hr}

Brandt et al.~(2001) and Alexander et al.~(2003) used the
\emph{wavdetect} tool included in the CIAO package (Freeman et
al.~2002) to detect the X-ray sources in the CDF-N. Yang et al.~(2004) also
used \emph{wavdetect} on the CLASXS field, in particular because the
program uses a set of scales to optimize the source detection, making
it excellent at separating nearby sources in crowded fields. In
general, \emph{wavdetect} provides better sensitivity than the
classical sliding-box methods.  

We ran \emph{wavdetect} on the full-resolution CLANS images with wavelet
scales of 1, $\sqrt2$, 2, 2$\sqrt2$, 4, 4$\sqrt2$, and 8. We used a
significance threshold of $10^{-7}$, which translates to a probability
of a false detection of 0.4 per ACIS-I field based on Monte Carlo
simulation results (Freeman et al.~2002). 

We used the aperture photometry tool for source flux extraction
described in detail in Yang et al.~(2004). The method uses circular extraction
cells. We linearly interpolated the CIAO Library PSFs to the off-axis
angles and $\sim95$\% of the enclosed energies for each source. In the
$0.5-2~\rm keV$ ($2-8~\rm keV$) band, the 95\% encircled radius is equal to
$2.5\arcsec$ at an off-axis angle of $3\arcmin$ ($2\arcmin$) and is
equal to $9.5\arcsec$ ($10\arcsec$) at an off-axis angle of
$8\arcmin$. For a
plot of the variation of the 95\% encircled radius with
off-axis angle, see Figure 3 in Yang et al.~(2004). If the
determined PSF value were greater than 2$\farcs$5, then we used it as
the radius of the circular extraction cell for that source and applied
an aperture correction to the final flux determination. If it were less
than 2$\farcs$5, then we used a fixed 2$\farcs$5 radius. As described in Yang et
al.~(2004), we estimated the background using an
eight-piece segmented annulus region four times as large as the source
cell area, with an inner radius $5\arcsec$ larger than the source cell
radius. We excluded any segments
containing more counts than the 3$\sigma$ Poisson upper limit in order
to avoid nearby sources. We then determined the background surface
brightness by dividing the counts by the area enclosed in the
remaining segments. By multiplying this background surface
brightness by the area within the circular extraction cell, we
determined the background counts. We obtained the net counts by
subtracting the background counts from the counts within the circular
extraction cell. 

We then made full-resolution spectrally weighted exposure maps. We used
these exposure maps only to correct for vignetting. We computed the
flux conversion at the aim point using spectral modeling (using
monochromatic maps does not change the results significantly). For each
source we convolved the exposure map with
the PSF generated using \emph{mkpsf} and normalized it to the exposure
time at the aim point. This gives the effective exposure time if the
source is at the aim point. In XSPEC we obtained the count rate to
flux conversion factor at the aim point by assuming each source has a
single power law spectrum with Galactic absorption (using the
CIAO-scripts \emph{mkacisrmf} and \emph{mkarf} to generate the
necessary RMF and ARF files). We calculated the power
law photon index, $\Gamma$, for each source using the hardness ratio, defined
as HR$\equiv$C$_{2-8~\rm keV}$/C$_{0.5-2~\rm keV}$, where C$_{2-8~\rm keV}$ and
C$_{0.5-2~\rm keV}$ are the count rates. For
a more detailed description and analysis of this flux conversion
method, see Yang et al.~(2004).  

%\clearpage
\begin{table*}
\begin{small}
\centering
\caption{Spectroscopic Completeness of Selected Surveys from the Literature}
\label{compl table}
\begin{tabular}{lcccc}
\tableline
\tableline
           & eCDF-S            & AEGIS           & SEXSI           & ChaMP \\
\tableline
Area (deg$^2$)                                 &   0.3  & 0.5    &  2     &  9.6  \\
X-ray Flux Limit (10$^{-16}$ ergs cm$^2$ s$^{-1}$)  &  6.7\tablenotemark{a} & 8.2\tablenotemark{b}     & 300\tablenotemark{c}     & 9.0\tablenotemark{d} \\
Optical Limit for Spectroscopic Follow-up    & $R_{AB}<25$ & $R_{AB}<24.1$ & $R_{AB}<24$ & $r_{AB}'<24$ \\
Spectroscopic Completeness\tablenotemark{e}               & 35\%\tablenotemark{f} &  \nodata\tablenotemark{g}  &  40-70\%\tablenotemark{h} & 54\%\tablenotemark{i}  \\
\tableline
\end{tabular}
\end{small}
\footnotesize
\tablecomments{$^a$ $2-8~\rm{keV}$: average on-axis flux limit (Lehmer et al.~2005).\\
$^b$ $2-7~\rm{keV}$: defined as the flux to which at least 1\% of the survey area is sensitive (Nandra et al.~2005).\\
$^c$ $2-10~\rm{keV}$: corresponds to the deepest flux reached by all 27 non-contiguous fields in SEXSI. 1 deg$^2$ of the survey probes to a deeper flux limit of $f_{2-10~\rm{keV}}=10^{-14}$ ergs cm$^2$ s$^{-1}$ (Harrison et al.~2003).\\
$^d$ $0.5-8~\rm{keV}$: corresponds to the flux limit for the deepest ChaMP exposure. ChaMP is a non-contiguous survey with exposure times ranging from 0.9 to 124~ks (Kim et al.~2007a).\\
$^e$ Fraction of X-ray sources with optical magnitudes brighter than the optical limit for spectroscopic follow-up for which spectroscopic redshifts have been determined.\\
$^f$ Virani et al.~(2006b).\\
$^g$ Spectroscopic redshifts for 84 AEGIS X-ray sources have been obtained to date from the DEEP2 Galaxy Redshift Survey. We have not found in the literature the number of AEGIS X-ray sources with $R_{AB}<24.1$ so are unable to provide a percentage for spectroscopic completeness (Davis et al.~2003, 2007; Bundy et al.~2008; Georgakakis et al.~2007).\\
$^h$ Eckart et al.~(2006).\\
$^i$ Silverman et al.~(2008).}
%\tablenotetext{a}{$2-8~\rm{keV}$: average on-axis flux limit (Lehmer et al.~2005).}
%\tablenotetext{b}{$2-7~\rm{keV}$: defined as the flux to which at least 1\% of the survey area is sensitive (Nandra et al.~2005).}
%\tablenotetext{c}{$2-10~\rm{keV}$: corresponds to the deepest flux reached by all 27 non-contiguous fields in SEXSI. 1 deg$^2$ of the survey probes to a deeper flux limit of $f_{2-10~\rm{keV}}=10^{-14}$ ergs cm$^2$ s$^{-1}$ (Harrison et al.~2003).}
%\tablenotetext{d}{$0.5-8~\rm{keV}$: corresponds to the flux limit for the deepest ChaMP exposure. ChaMP is a non-contiguous survey with exposure times ranging from 0.9 to 124~ks (Kim et al.~2007a).} 
%\tablenotetext{e}{Fraction of X-ray sources with optical magnitudes brighter than the optical limit for spectroscopic follow-up for which spectroscopic redshifts have been determined.}
%\tablenotetext{f}{Virani et al.~(2006b).}
%\tablenotetext{g}{Spectroscopic redshifts for 84 AEGIS X-ray sources have been obtained to date from the DEEP2 Galaxy Redshift Survey. We have not found in the literature the number of AEGIS X-ray sources with $R_{AB}<24.1$ so are unable to provide a percentage for spectroscopic completeness (Davis et al.~2003, 2007; Bundy et al.~2008; Georgakakis et al.~2007).}
%\tablenotetext{h}{Eckart et al.~(2006).}
%\tablenotetext{i}{Silverman et al.~(2008).}
\end{table*}

\begin{table*}
\begin{scriptsize}
\centering
\caption{CLANS Observation Summary}
\label{obssum table}
\begin{tabular}{lccccc}
\tableline\tableline
& & & & & Exposure\tablenotemark{a}\\
Target Name & Observation ID & $\alpha_{2000}$ & $\delta_{2000}$ &  Observation Start Date & (ks) \\
\tableline
SWIRE LOCKMAN 5 (center) & 5023 &10 46 00.00 & +59 01 00.00  &2004 Sept 12 21:30:56 & 67\\
SWIRE LOCKMAN 1&5024 &10 44 46.15 & +58 41 55.45  &2004 Sept 16 06:53:47 & 65\\
SWIRE LOCKMAN 2&5025 &10 46 39.44 & +58 46 51.24  &2004 Sept 17 20:30:04& 70\\
SWIRE LOCKMAN 3&5026 &10 48 32.77 & +58 51 47.33  &2004 Sept 18 16:17:12& 69\\
SWIRE LOCKMAN 4&5027 &10 44 06.67 & +58 56 05.28  &2004 Sept 20 14:40:35& 67\\
SWIRE LOCKMAN 6&5028 &10 47 53.44 & +59 05 57.00  &2004 Sept 23 03:36:12& 71\\
SWIRE LOCKMAN 7&5029 &10 43 27.23 & +59 10 15.07  &2004 Sept 24 03:43:15& 71\\
SWIRE LOCKMAN 8&5030 &10 45 20.56 & +59 15 11.16  &2004 Sept 25 19:47:08 & 66\\
SWIRE LOCKMAN 9&5031 &10 47 13.85 & +59 20 06.95  &2004 Sept 26 14:47:00& 65\\
\tableline
\end{tabular}
\end{scriptsize}
\footnotesize\\
\tablenotetext{a}{Total good time with dead-time correction}
\end{table*}

\begin{table*}
\begin{small}
\centering
\caption{X-ray Data Specifications}
\label{xray table}
\begin{tabular}{rrrr}
\tableline\tableline
Category & CLANS & CLASXS & CDF-N \\
\tableline
Exposure Time &$\sim70$~ks &$\sim40$~ks (73~ks\tablenotemark{a}) &2~Ms \\
Area (deg$^2$) &0.6  &0.45  &0.124 \\
$0.5-2~\rm keV$ Flux Limit\tablenotemark{b} &$7$ & $12$ ($7$)\tablenotemark{c} & $0.25$ \\
$2-8~\rm keV$ Flux Limit\tablenotemark{b} &$35$ &$60$ ($35$)\tablenotemark{c} & $1.5$ \\
Total \# of X-ray Sources &761 &525 &503\\
\tableline
\end{tabular}
\end{small}
\footnotesize
\tablenotetext{a}{Exposure time for the central pointing. All other pointings have exposure times of $\sim40$~ks.}
\tablenotetext{b}{Flux limit (for a $S/N=3$) at the pointing center in $10^{-16}$ ergs cm$^{-2}$ s$^{-1}$.}
\tablenotetext{c}{Flux limit (for a $S/N=3$) at the pointing center of the $\sim40~$ks ($73~$ks) exposure.}
\end{table*}

%\clearpage
%\pagebreak

\begin{table*}
\begin{small}
\caption{CLANS X-ray Catalog: Basic Source Properties}
\label{xcat table1}
\begin{tabular}{cccccccccc}
\tableline\tableline
  &  & $\Delta\alpha$ &  $\Delta\delta$  &  &  &  &  & &  \\
Num. & $\alpha_{2000}$ & $\delta_{2000}$ &(arcsec)  & (arcsec) & $f_{0.5-2~{\rm keV}}$\tablenotemark{a}  & $f_{2-8~{\rm keV}}$\tablenotemark{a}  & $f_{0.5-8~{\rm keV}}$ & HR\tablenotemark{b} & $\Gamma$\tablenotemark{c} \\
(1) & (2) & (3) & (4) & (5) & (6)  & (7) & (8) & (9) & (10)\\
\tableline
    1 &   160.58846 &    59.25157 &       0.831 &       0.457 &        3.37$_{-        0.51}^{+        0.70}$ &       11.82$_{-        2.31}^{+        2.80}$ &       14.36$_{-        1.80}^{+        2.17}$ &        0.79$_{-   0.20}^{+   0.25}$ &   1.05$_{-   0.26}^{+   0.25}$\\
    2 &   160.64172 &    59.17641 &       0.301 &       0.239 &        7.79$_{-        0.83}^{+        0.97}$ &       10.01$_{-        1.75}^{+        2.28}$ &       17.98$_{-        1.62}^{+        1.94}$ &        0.37$_{-   0.08}^{+   0.10}$ &   1.71$_{-   0.19}^{+   0.20}$\\
    3 &   160.64175 &    59.16873 &       0.710 &       0.376 &        0.63$_{-        0.22}^{+        0.38}$ &        0.75$_{-        0.52}^{+        0.79}$ &        1.45$_{-        0.50}^{+        0.60}$ &        0.34$_{-   0.27}^{+   0.42}$ &   1.76$_{-   0.69}^{+   0.78}$\\
    4 &   160.65615 &    59.11504 &       1.231 &       0.581 &        1.52$_{-        0.57}^{+        1.01}$ &        0.51$_{-        0.42}^{+        1.74}$ &        2.20$_{-        0.67}^{+        1.30}$ &        0.13$_{-   0.12}^{+   0.46}$ &   2.37$_{-   1.08}^{+   0.36}$\\
    5 &   160.66704 &    59.07057 &       0.983 &       0.559 &        1.19$_{-        0.31}^{+        0.46}$ &        6.34$_{-        1.83}^{+        2.12}$ &        7.88$_{-        1.51}^{+        1.83}$ &        1.09$_{-   0.42}^{+   0.56}$ &   0.74$_{-   0.13}^{+   0.44}$\\
    6 &   160.66924 &    59.25401 &       1.004 &       0.460 &        0.93$_{-        0.28}^{+        0.37}$ &        4.63$_{-        1.38}^{+        1.86}$ &        5.86$_{-        1.21}^{+        1.49}$ &        1.03$_{-   0.44}^{+   0.59}$ &   0.80$_{-   0.18}^{+   0.49}$\\
..&  ..   & ..    &    .. &    ..&    ..& ..&  ..   & ..    &    ..\\
\tableline
\end{tabular}
\end{small}
\footnotesize
\tablecomments{Table \ref{xcat table1} is presented in its entirety in the electronic edition of the \emph{Astrophysical Journal Supplement}. A portion is shown here for guidance regarding its form and content.\\
$^a$ In units of $10^{-15}$~ergs~s$^{-1}$. For any source detected in one band but with a very weak signal in the other, the background-subtracted flux could be negative. In this case, only the upper Poisson error is quoted and we flag it with a $<$ symbol.\\
$^b$ For sources with a very weak signal in the $0.5-2~\rm keV$ ($2-8~\rm keV$) band, we list the lower (upper) limit and flag it with a $>$ ($<$) symbol.\\
$^c$ For sources detected in one band but not in the other, $\Gamma=-99$.}
\end{table*}

\begin{table*}
\begin{small}
\caption{CLANS X-ray Catalog: Additional Source Properties}
\label{xcat table2}
\begin{tabular}{ccccccccc}
\tableline\tableline
 &  &  & & t$_{0.5-2~\rm keV}$  & t$_{2-8~\rm keV}$  & t$_{0.5-8~\rm keV}$  & R\\
Num.   &  n$_{0.5-2~\rm keV}$\tablenotemark{a} & n$_{2-8~\rm keV}$\tablenotemark{a}  &  n$_{0.5-8~\rm keV}$\tablenotemark{a}  &($10^4$~s) &($10^4$~s)& ($10^4$~s)& (arcminutes)\\
(1)&(2)&(3)&(4)&(5)&(6)&(7)&(8) \\
\tableline
   1 &  36.50$_{-   5.56}^{+   7.63}$ &  26.00$_{-   5.07}^{+   6.16}$ &  59.75$_{-   7.48}^{+   9.04}$ &   5.90 &   5.34 &   5.80 &    9.7\\
   2 &  83.75$_{-   8.90}^{+  10.45}$ &  29.75$_{-   5.20}^{+   6.78}$ & 112.50$_{-  10.12}^{+  12.16}$ &   6.47 &   6.26 &   6.40 &    6.8\\
   3 &   6.75$_{-   2.35}^{+   4.01}$ &   2.25$_{-   1.57}^{+   2.38}$ &   9.25$_{-   3.21}^{+   3.85}$ &   6.48 &   6.28 &   6.41 &    6.8\\
   4 &   5.75$_{-   2.15}^{+   3.83}$ &   0.75$_{-   0.62}^{+   2.54}$ &   7.50$_{-   2.28}^{+   4.44}$ &   2.71 &   2.62 &   2.65 &    7.2\\
   5 &  12.75$_{-   3.32}^{+   4.94}$ &  13.25$_{-   3.82}^{+   4.44}$ &  27.00$_{-   5.17}^{+   6.26}$ &   5.65 &   5.40 &   5.54 &    8.5\\
   6 &  11.00$_{-   3.28}^{+   4.41}$ &  11.00$_{-   3.28}^{+   4.41}$ &  23.00$_{-   4.77}^{+   5.86}$ &   6.27 &   6.07 &   6.20 &    7.8\\
 ..&  ..   & ..    &    .. &    ..&    ..&    ..& .. \\
\tableline
\end{tabular}
\end{small}
\footnotesize
\tablecomments{Table \ref{xcat table2} is available in its entirety in the electronic edition of the \emph{Astrophysical Journal Supplement}. A portion is shown here for guidance regarding its form and content.\\
$^a$ The eight-piece
segmented annulus region we use to estimate the
background for each source is 4 times as large as the source cell
area. Therefore, the net counts for the majority of the sources are multiples of
0.25. For a few sources we exclude background segments
containing more counts than the
3$\sigma$ Poisson upper limit in order to avoid nearby
sources. Therefore the net counts for these sources are not multiples of 0.25.}
\end{table*}

\clearpage
%see tab4.tex and tab5.tex for the full version of these tables (to be included in the electronic edition).

\subsection{CLANS X-ray Catalog}

The CLANS observations consist of a 3$\times$3 raster with an $\sim2\arcmin$
overlap between continguous pointings (Polletta et al.~2006; see our
Figure \ref{position figure}). Following the prescription in Yang et
al.~(2004) for
the CLASXS field, we merged the nine individual pointing catalogs to
create the final CLANS X-ray catalog. For sources with more than one 
detection in the nine fields, we used the detection from the
observation in which the effective area of the source was the largest. 

We present the CLANS X-ray catalog in Tables \ref{xcat
  table1} and \ref{xcat table2}. In both tables, column (1) is
the source number. The numbers correspond to ascending order in right
ascension. In Table \ref{xcat table1}, columns (2) and (3) give
the right ascension and
declination coordinates of the X-ray sources. Columns (4) and (5) list
the 95\% confidence errors from \emph{wavdetect} on these right ascension 
and declination coordinates. Columns (6), (7), and (8) provide the
X-ray fluxes in the $0.5-2~\rm keV$, $2-8~\rm keV$, and $0.5-8~\rm
keV$ bands, respectively, in units of 10$^{-15}$
ergs~cm$^{-2}$~s$^{-1}$. The errors quoted are the $1\sigma$ Poisson
errors, using
the approximations from Gehrels (1986). They do not include the
uncertainty in the flux conversion factor; however, the errors are
generally dominated by the Poisson errors. Column (9) gives the hardness
ratio, as defined in \S \ref{hr}. Column (10) lists the value of
$\Gamma$ for each source. The errors in Columns 9 and 10 are the
$1\sigma$ errors propagated from the $1\sigma$ errors on the counts
listed in Table \ref{xcat table2}. 

In Table \ref{xcat table2}, columns (2), (3), and (4) list the net
counts in the $0.5-2~\rm keV$, $2-8~\rm keV$, and $0.5-8~\rm
keV$ bands, respectively. The errors quoted are the $1\sigma$ Poisson
errors, using the approximations from Gehrels (1986). Columns (5),
(6), and (7) give the effective exposure times from the exposure maps
in each of the three energy bands. Column (8) provides the distance from the
pointing center for each source.  

\subsection{Flux Limits}

Following the prescription in Alexander et al.~(2003), we determined
the flux limits for a signal-to-noise ratio ($S/N$) of 3 for the CLANS
and CLASXS fields. To determine the sensitivity across the field it is
necessary to take into account
the broadening of the PSF with off-axis angle, as well as changes in
the effective exposure and background rates across the field. Under
the simplifying assumption of $\sqrt{N}$ uncertainties, we can
determine the sensitivity across the field following Muno et al.~(2003) as 
\begin{equation}
S=\frac{n_{\sigma}^2}{2}(1+[1+\frac{8b}{n_{\sigma}^2}]^{1/2}),
\end{equation}
where $S$ is the number of source counts and $b$ is the number of
background counts in a source cell. The $n_{\sigma}$ is the required
signal-to-noise ratio. 

After setting $n_{\sigma}=3$, the only component within Equation 1
that we need to measure is the background counts. We determined the
median number of background counts at each off-axis angle (using the
method described in \S \ref{hr}) and inputted
this into Equation 1 to determine $S$ at each off-axis
angle. Using the flux conversion method discussed in \S \ref{hr}, we
converted $S$ from counts to flux values. 

\begin{figure}
\epsscale{2.2}
\plottwo{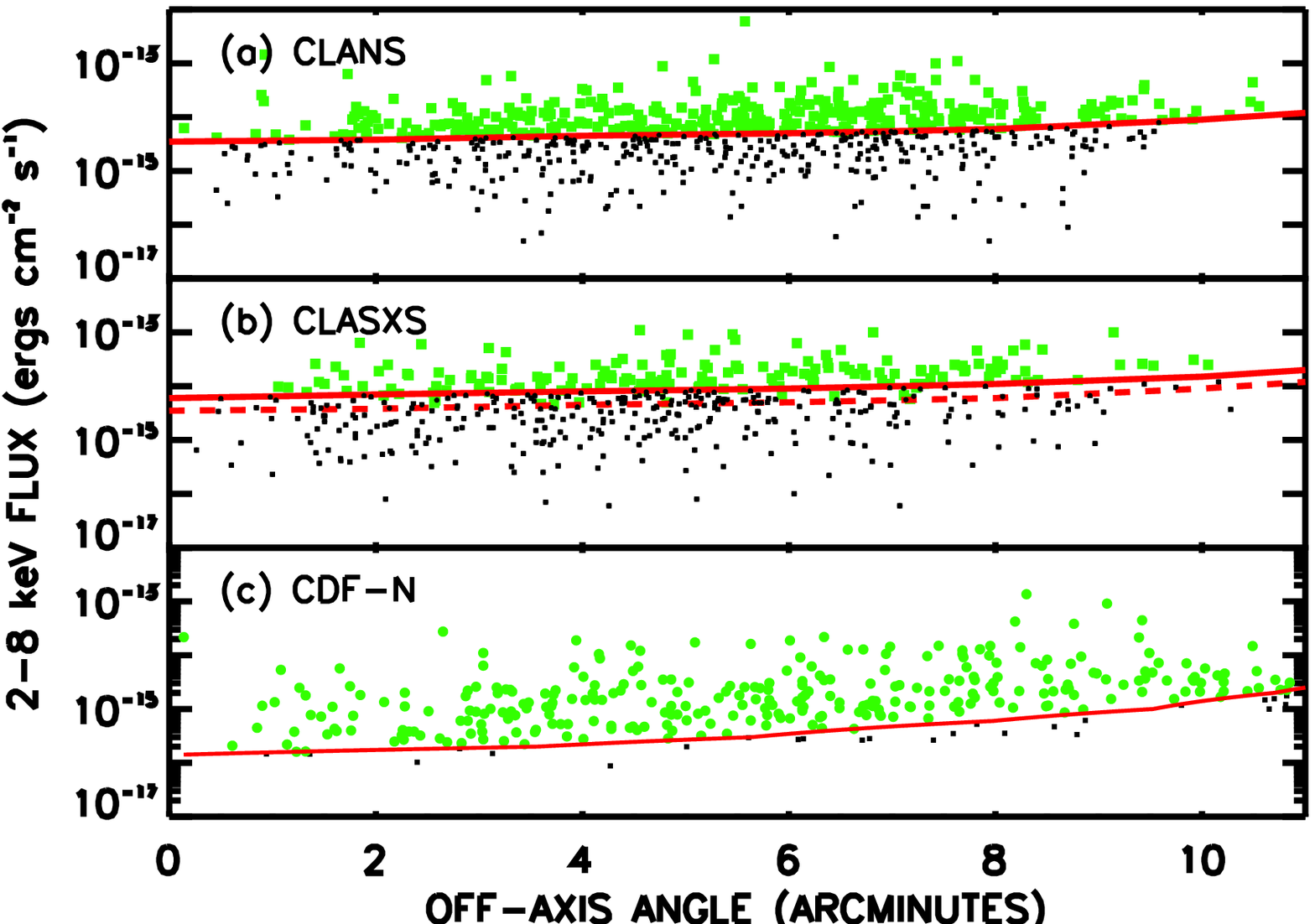}{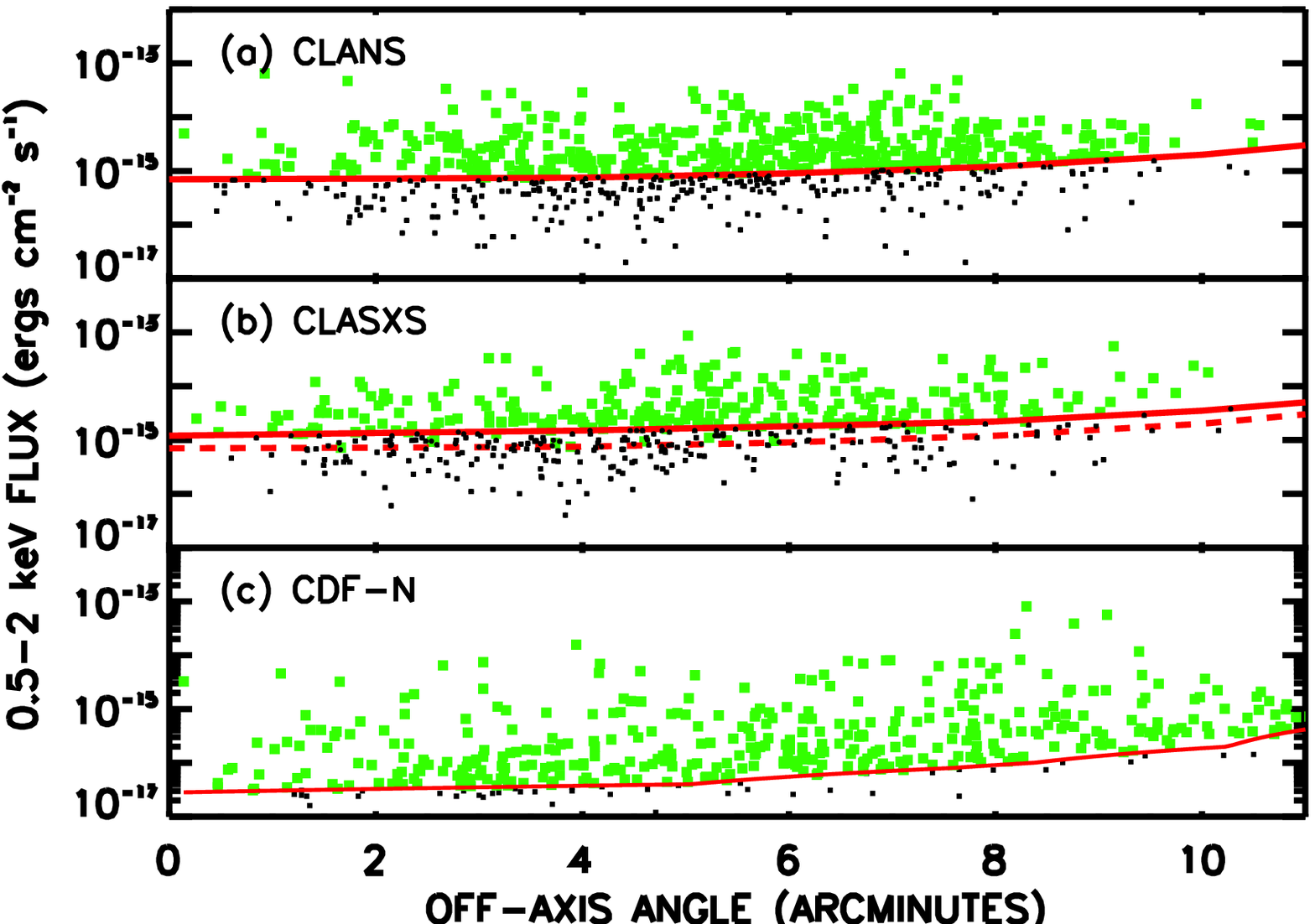}
\caption{$2-8~\rm keV$ and $0.5-2~\rm keV$ flux versus off-axis angle for the
  X-ray sources in the CLANS, CLASXS, and CDF-N fields (\emph{red
    lines}, $S/N=3$ flux limits at the off-axis angle of each source detected
  by \emph{wavdetect}; \emph{dashed red lines}, $S/N=3$ flux limits
  for the deeper $73~$ks 
  pointing in the CLASXS field; \emph{green
    circles}, sources with fluxes  
  greater than the corresponding flux limit; \emph{black squares}, sources with
  fluxes below the corresponding flux limit).}
\label{fluxlim}
\end{figure}

The red lines in Figure \ref{fluxlim} show the flux limits (for a
$S/N=3$) versus the off-axis angle. The
green circles (black squares) identify the sources whose fluxes are
greater (less) than the corresponding flux limit. Because all of the CLANS
pointings have exposure times of $\sim$70~ks, there is only a single
red line that increases slightly in flux as the off-axis
angle increases. In the CLASXS field, however, the LHNW-1 pointing has
an exposure time of 73~ks, while the other pointings are all
$\sim$40~ks. Thus, the dashed red line located at slightly lower fluxes than the
solid red line shows the flux limits corresponding to the deeper 73~ks
ACIS-I pointing. The bottom plots in Figure \ref{fluxlim} show the
Alexander et al.~(2003) flux limits (for a $S/N=3$) for the CDF-N field.  

%\clearpage
%\clearpage

\subsection{Effective Areas}
\label{fluxlimsec}
\label{effarea}

Yang et al.~(2004, 2006) used Monte Carlo simulations of the CLASXS
40~ks and 70~ks $2-8~\rm keV$ and $0.5-2~\rm keV$ band ACIS-I images to examine the
incompleteness in their number counts. The sensitivity of the
source detection drops with off-axis angle due to vignetting
effects, quantum efficiency changes across the field, and the
broadening of the point-spread functions. Using \emph{wavdetect}
on their simulated images, Yang et al.~obtained an estimate of the detection
probability function, or probability that a source would be detected
by \emph{wavdetect}, at different fluxes and off-axis angles. 

The differences at small and large off-axis angles between their 2004
and 2006 simulations are attributable to two methodological
changes. First, in their 2004 simulations they used many fewer
simulated sources, causing an undersampling at small off-axis
angles. Second, in their 2004 simulations they did
not use large wave scales, causing a quick
drop in source detection at off-axis angles greater than $6\arcmin$. 

\begin{figure}
\epsscale{1.}
\plotone{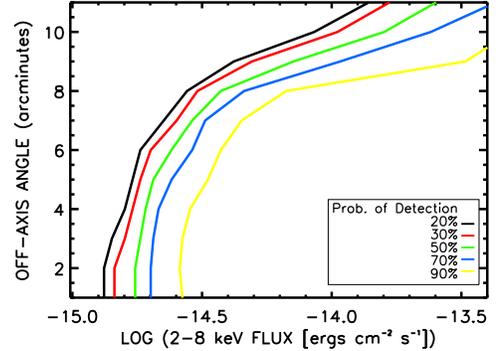}
\caption{Probability of source detection as a function of off-axis
  angle and $2-8~\rm{keV}$ flux for the 67~ks central CLANS pointing. To
  obtain these results, we interpolated from the Yang et al.~(2006)
  simulations for 40~ks and 70~ks exposure times.}   
\label{sims}
\end{figure}

We interpolated from their 40~ks and
70~ks 2006 simulations to the exposure times for the nine pointings in
both the CLANS and CLASXS fields, respectively (see Table \ref{obssum table}
for the CLANS field exposure times and Yang et al.~2004 Table 1 for the CLASXS
field exposure times) to determine the probability of detection for a
source at a given off-axis angle and flux in each pointing. Figure
\ref{sims} shows the probability of source detection as a function of
off-axis angle and $2-8~\rm{keV}$ flux for the central CLANS pointing
(\emph{Chandra}/SWIRE Lockman 5 in Table \ref{obssum table}) with an
exposure time of 67~ks. As expected, the probability of detecting a
source at a given off-axis angle decreases as the flux of the source
decreases.

We then determined the effective sky area for each pointing using the
formula 
\begin{equation}
A(f_x)=\int_0^a 2 \pi a ~da ~D(a,f_x) ~g(a),
\end{equation}
 where $a$ is the off-axis angle from the pointing center, $g(a)$ is the
 geometric fraction, and $D(a,f_x)$ is the detection probability,
 equivalent to an incompleteness factor. 

\begin{figure}
\epsscale{1.6}
\plottwo{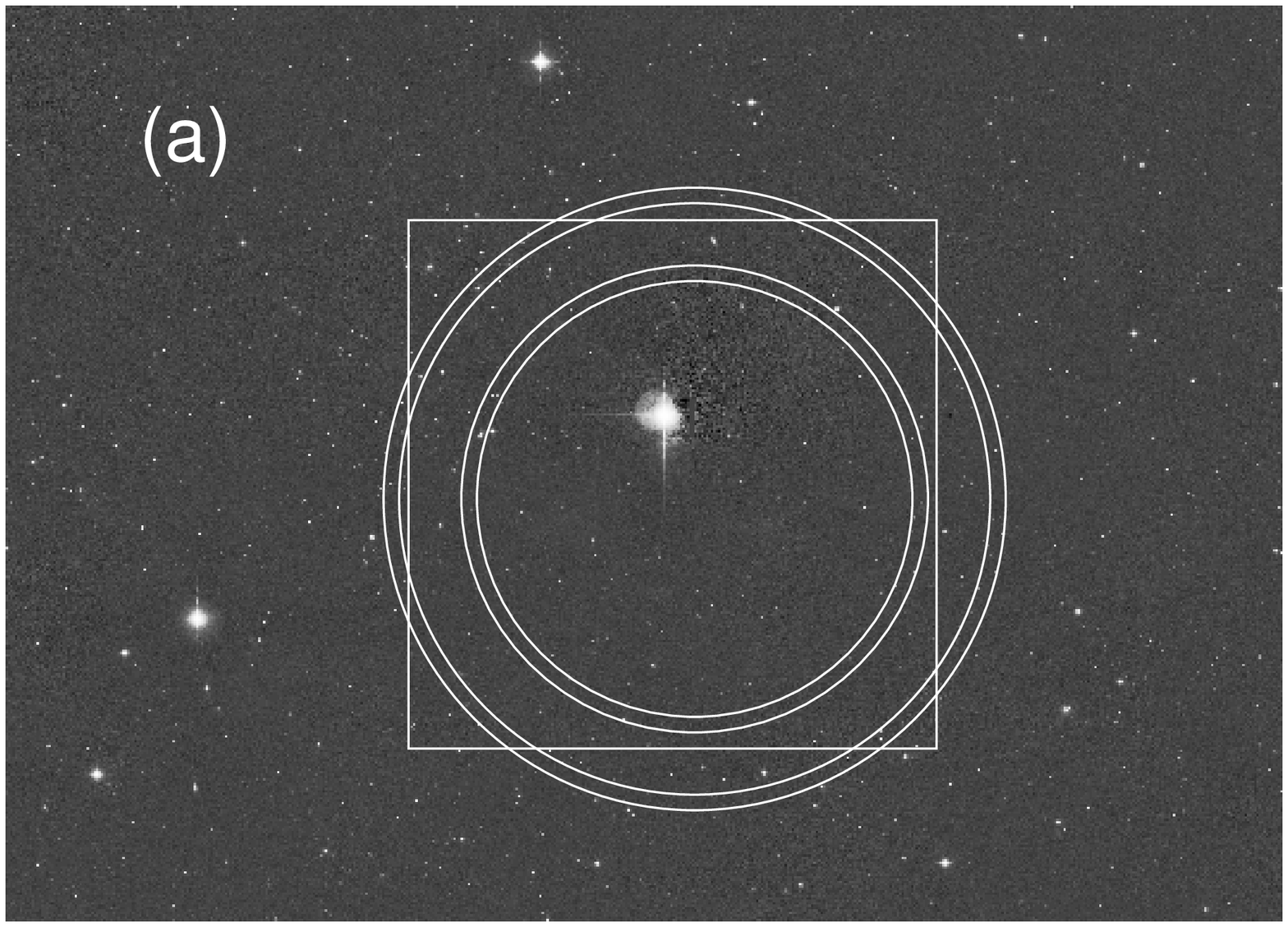}{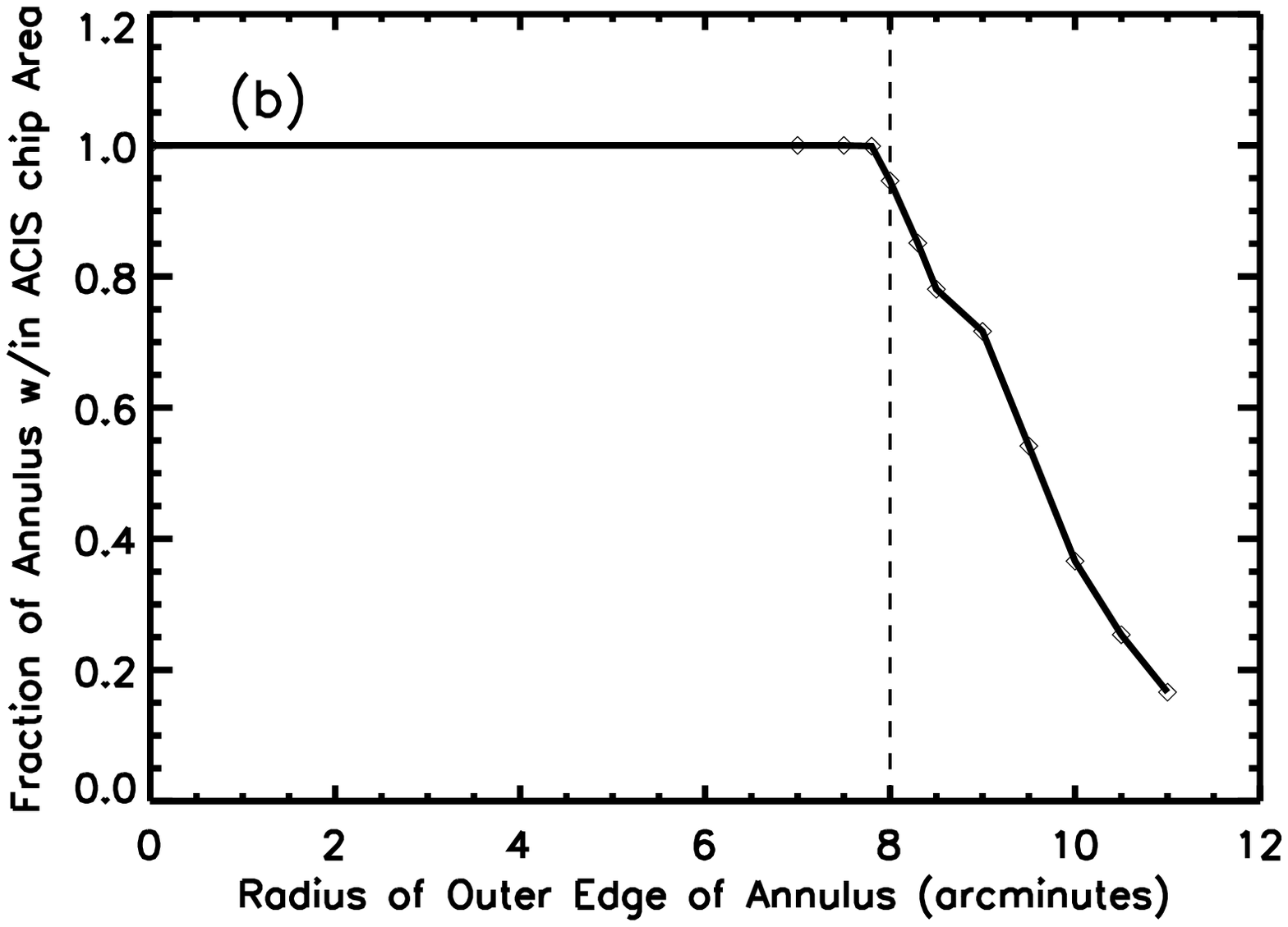}
\caption{(a) ACIS-I chip outline overlaid on the ULBCam $H$-band image of the
  CLANS field. Two annuli (off-axis angles $7\arcmin-7\farcm5$
  and $9\farcm5-10\arcmin$) are shown in
  white. Note the offset of the chip center from the pointing
  center. (b) Fraction of the total area covered by an annulus of width
  $0\farcm5$ that is contained within
  the ACIS-I chip versus the annulus' outer radius.}  
\label{annuli}
\end{figure}

Figure \ref{annuli} shows the geometric fraction, or fraction of the
total area covered by an annulus of width $0\farcm5$ and outer radius
$a$ that is contained within the ACIS-I chip. Figure \ref{annuli}a shows our
ULBCam $H$-band image (see \S \ref{ulbcam}) of the
CLANS field overlaid with
the ACIS-I chip outline (the square) and two annuli with off-axis angles
$7\arcmin-7\farcm5$ and $9\farcm5-10\arcmin$. Figure \ref{annuli}(b) shows
that the geometric fraction is 1 for all annuli with outer radii less than
$\sim8\arcmin$. At $a>8\arcmin$, $g(a)$ decreases rapidly. 

We determined $A(f_x)$ for 5 different limiting detection
probabilities, 20\%,
30\%, 50\%, 70\%, and 90\%. In the determination of $A(f_x)$, if the
simulated $D(a,f_x)$ at the off-axis angle of a source of flux $f_x$
were found to be below this limiting detection probability, then we
assigned $D(a,f_x)=0$ for that source.

Because the sensitivity of \emph{Chandra} degrades at large off-axis
angles, we exclude sources with off-axis angles greater
than $8\arcmin$ in our number counts determination (see \S \ref{numcts
  sec}). In the CLANS field at these off-axis angles there is little
overlap between pointings, so $\Omega(f_x)\approx\sum A_i(f_x)$. However, in the
CLASXS field the pointings overlap 
significantly (as shown in Figure \ref{position figure}). We used the
ds9 funtools function that determines area within region files to
subtract the overlapping area from the total $\Omega(f_x)$ value. We also
made sure that any area that had already been subtracted as part of the
geometric fraction determination was not subtracted a second
time at this stage.  

\begin{figure}[!h]
\epsscale{2.2}
\plottwo{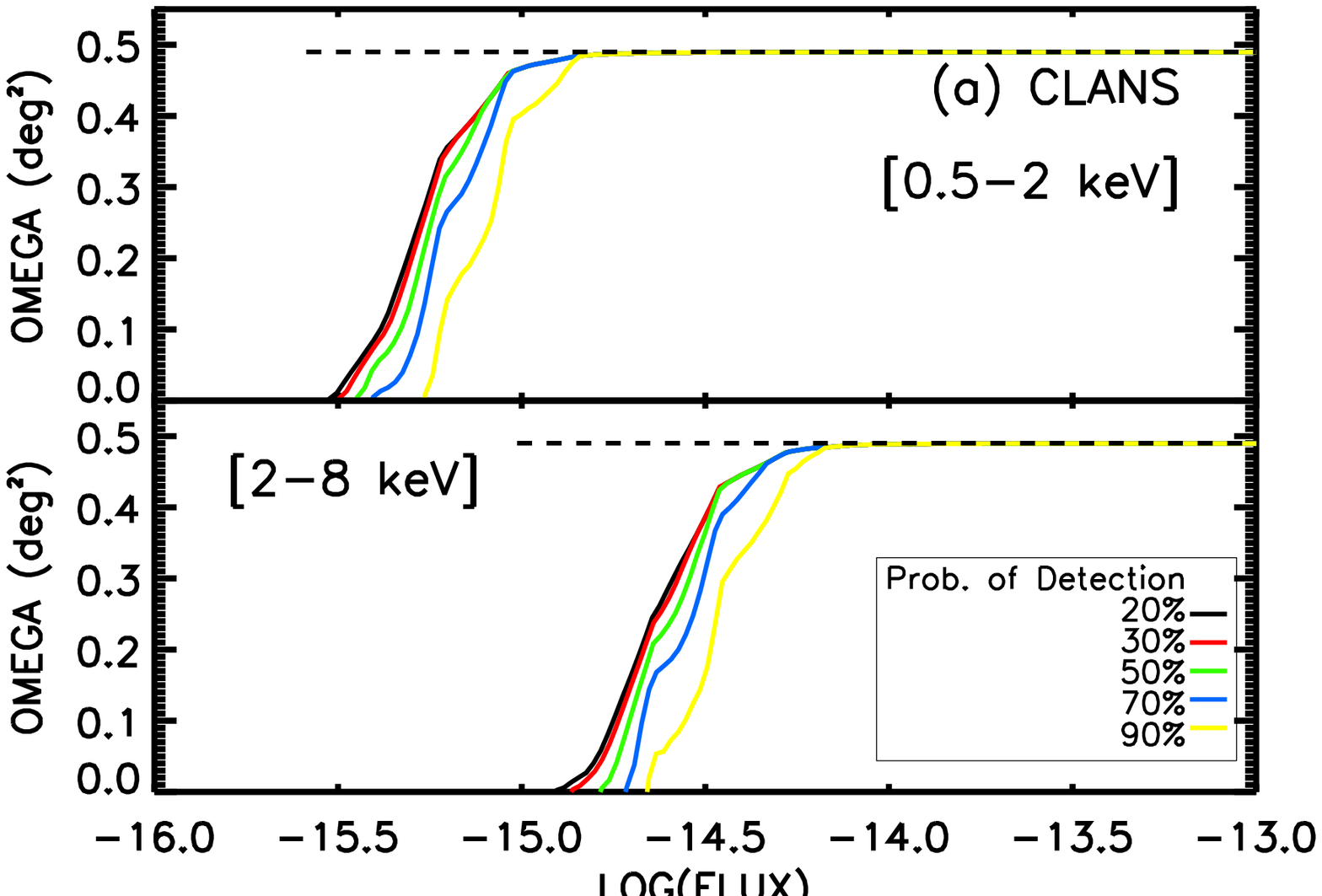}{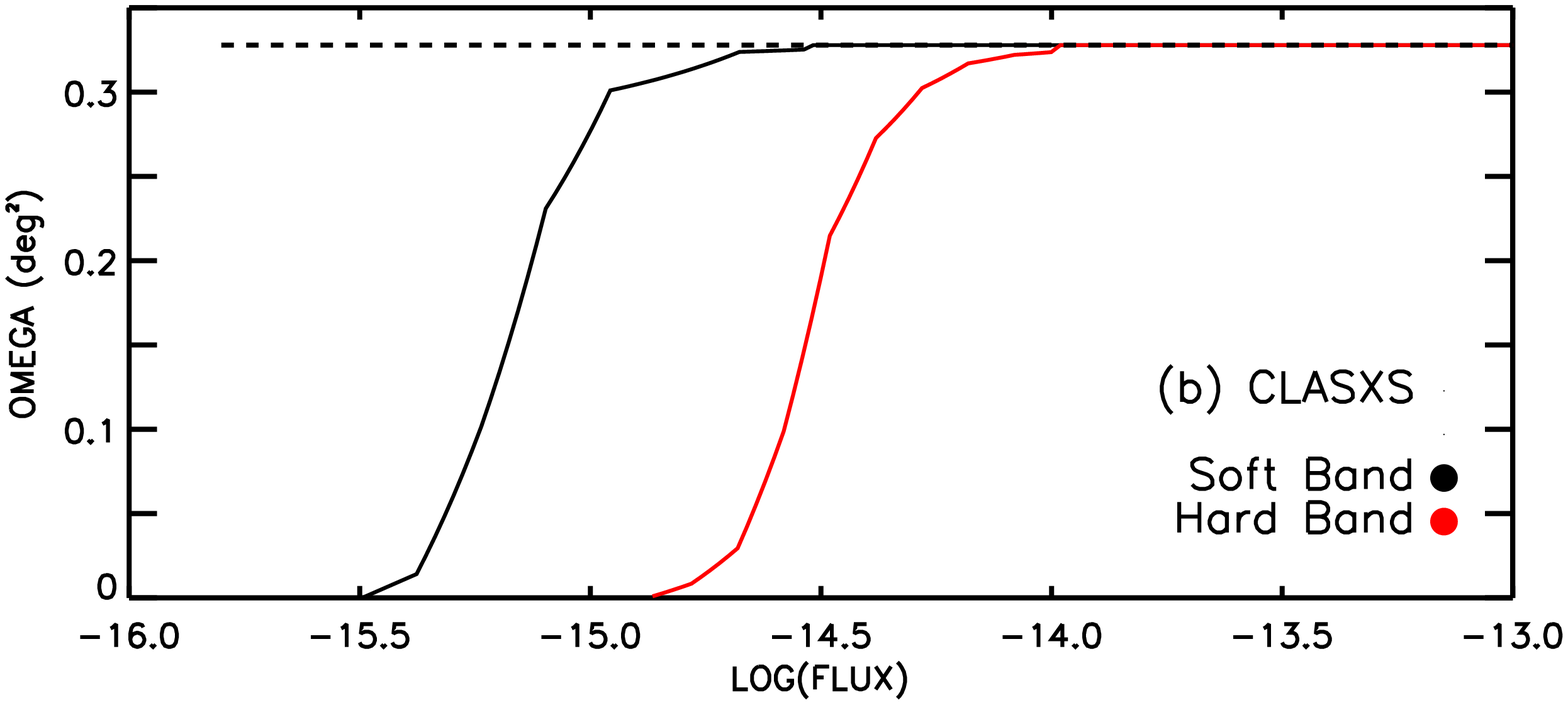}
\caption{Total true effective area, $\Omega$, versus $0.5-2~\rm keV$ and $2-8~\rm keV$
  flux for off-axis angles
  $<8\arcmin$ for (a) the CLANS field (for the 20\%, 30\%, 50\%,
  70\%, and 90\% probability of detection cases) and (b) the CLASXS field (for
  only the 30\% probability of detection case). The dashed line shows
  the effective area if $D(a,f_x)=1$ is set for all fluxes and
  off-axis angles (i.e., where $\Omega=9\times \int_0^a 2 \pi a ~da ~ g(a)$
  minus any overlapping areas between pointings).}   
\label{fluxarea}
\end{figure}

Figure \ref{fluxarea} shows the total true effective area, $\Omega$, versus
flux for (a) the 20\%, 30\%, 50\%, 70\%, and 90\% probability of
detection cases for the CLANS field and (b) the 30\% probability of detection
case for the CLASXS field. The dashed line shows
the effective area if $D(a,f_x)=1$ is set for all fluxes and off-axis
angles, in other words, where $\Omega=9\times \int_0^a 2 \pi a ~da
~ g(a)$ minus any overlapping areas between pointings.   

This figure shows that the small off-axis angles (or small $\Omega$
values) probe the faintest fluxes, since the \emph{Chandra}
sensitivity is highest close to the pointing center. Figure
\ref{fluxarea}a also shows that we can probe to fainter fluxes at the
expense of more incompleteness, since the lower we push our probability of
detection, the fainter the flux probed at a given off-axis angle. 

In our calculation of the log$N-$log$S$ distribution (see \S \ref{numcts
  sec}), we use the results from the 30\% probability of detection
case. Although this is a fairly
arbitrary choice, we chose it to maximize depth while not using too
low of a probability of detection.   

%\clearpage

\pagebreak

\section{OPTICAL AND IR IMAGING}
\label{photometric}

In this section we present new $g',r',i',z',J,H$, and $K$ photometry for
the CLANS X-ray sources (see Table \ref{clansopt table}), new
$u,g',i',J,H$, and $K$ photometry, as well as updated $B,V,R,I,$ and $z'$
photometry, for the CLASXS X-ray sources (see Table \ref{clasxsopt table} and
\S \ref{clasxsupdate}), and new $J,H,$ and $K_s$ photometry for the
CDF-N X-ray sources (see Table \ref{nir table}). We also present the
\emph{Spitzer} $3.6~\mu$m and $24~\mu$m data for all three fields
(see \S \ref{spitzer}).  

\subsection{MegaCam}
The CFHT MegaCam/Megaprime camera has a 1 deg$^2$ field of view on the
3.6 m CFHT telescope. The focal plane is covered with 36 $2$k$\times
4.5$k EEV CCD detectors with excellent response between 3200~\AA\ and
9000~\AA\ (Aune et al.~2003; Boulade et al.~2003).  

We used MegaCam to obtain deep $g',r',i'$, and $z'$-band data
for the CLANS field and deep $u,g'$, and $i'$-band data for the
CLASXS field. Observations were taken in sets of 7-point
dithered exposures forming a circular pattern. The radius of the
dither circle was $7\farcs5$, which was enough to fill the chip gaps
but minimized the chip overlap. 

The CFHT operates in queue observing mode for MegaCam observations,
which ensures uniform image quality and photometry, and a standard
reduction pipeline (Elixir) is provided by CFHT and TERAPIX. The queue
observer takes calibration frames each night, which the Elixir
pipeline uses to calibrate the data
(Magnier \& Cuillandre 2004). The pipeline corrects for bias, dark
current, flatfielding, and scattered light, with the final
photometric calibration better than 1\% across the field of view. 

The TERAPIX data processing center provided further reduction of our data
including astrometric and photometric calibration, sky subtraction,
and image combination. At TERAPIX the
`qualityFITS' data quality assessment tool was used to visually inspect the
calibrated images provided by Elixir and any defective
images were rejected. Images with a
seeing larger than $1\farcs25$ in the $g',r',i',$ and $z'$-band and
larger than $1\farcs35$ in the $u$-band were also rejected. 

\subsection{WIRCam}
We obtained our $K_s$-band photometry for the CDF-N using WIRCam on
CFHT. We used our own IDL program to stack the images into a
final mosaic, remove cosmic rays, establish a background
estimation, align the astrometry to the USNO catalog, and derive a
final zeropoint via comparison with 2MASS. More details of the
observations and reductions can be found in Barger et al.~(2008).

\subsection{ULBCam}
\label{ulbcam}
We carried out deep $J$- and $H$-band imaging of the three fields
using the Ultra-Low Background Camera (ULBCam) on the University of
Hawaii 2.2 m telescope during $2003-2006$. ULBCam consists of four
2k$\times$2k HAWAII-2RG
arrays (Loose et al.~2003) with a total $16\arcmin\times16\arcmin$ field of
view. We took the images using a 13-point dither pattern with
$\pm30\arcsec$ and $\pm60\arcsec$ dither steps in order to cover the
chip gaps. We flattened the data using median sky flats from each
dither pattern and corrected the image distortion using the
astrometry in the USNO-B1.0 catalog (Monet et al.~2003). We combined the
flattened, sky-subtracted, and warped images to form the
final mosaic. A more extensive description of the data reduction and a
detailed analysis is given in R.~Keenan et al.~(2008, in preparation).

\subsection{WFCAM}
We used WFCAM on the 3.8 m UKIRT telescope for our $K$-band
photometry of the CLANS and CLASXS fields. The WFCAM camera has an 0.8
deg$^2$ field and is composed of four Rockwell Hawaii-II
2048$\times$2048 18~$\mu$m pixel array detectors, with a pixel scale
of $0\farcs4$. Acquired data are shipped from UKIRT to the Cambridge
Astronomical Survey Unit (CASU). At CASU the WFCAM pipeline processes the
data to produce stacked imaging data. The reduction steps
are described in detail by Dye et al.~(2006).
Briefly, dark frames secured at the beginning of the night are
subtracted from each target frame to remove bad pixels and
amplifier glow and to reset anomalies from the raw frames. Twilight
flats are then used to flatfield the data. The dithered frames are
combined by weighted averaging (using a confidence map derived largely
from the flatfield frames) to produce stacked frames. 

The WFCAM Science Archive (WSA) archives all reduced WFCAM data. Each
stacked frame contains four individual images, one for each of the
four WFCAM cameras. After we retrieved the reduced data from WSA, we
used our own IDL program to apply another background subtraction
and astrometry check, to combine the four tiles into one frame, and to
align the astrometry to the USNO catalog. A more extensive description of
the data reduction is given in R.~Keenan et
al.~(2008, in preparation).   

\subsection{Spitzer}
\label{spitzer}

To improve our photometric redshift determinations in the CLANS and
CLASXS fields, we obtained
$3.6~\mu$m data from the \emph{Spitzer} Wide-Area Infrared
Extragalactic Survey (SWIRE; Lonsdale et al.~2003) Legacy Science
Program. We used the online NASA/IPAC Infrared Science
Archive via GATOR to retrieve the $3.6~\mu$m and $24~\mu$m
photometric catalogs, using a search radius of 2$\arcsec$ around our
NIR counterpart source locations. The $3.6~\mu$m ($24~\mu$m) catalog has
a $5\sigma$ limit of $5~\mu$Jy ($230~\mu$Jy) (Polletta et al.~2006).

Polletta et al.~(2006)
carried out an extensive analysis of the probability of false matches to the
\emph{Spitzer} IRAC catalogs for the CLANS field. Of the 774 X-ray sources
in their catalog, they expect $\approx19$ false assocations. We find a
similar result of 23 false matches to the IRAC catalog using a simple
randomization of our 761 CLANS X-ray sources. We find
only 2 false matches to the MIPS catalog, which has many fewer sources
than the IRAC catalog. The IRAC
and MIPS data provide complete coverage for the CLANS field
area. However, while the MIPS data provide complete coverage for the
CLASXS field area, the IRAC data only cover 305 of the 525 CLASXS
X-ray sources. 

For the CDF-N X-ray sources, we obtained the relevant $3.6~\mu$m and
$24~\mu$m data from the Great Observatories Origins
Deep Survey-North (GOODS-N) \emph{Spitzer} Legacy Science Project. We
followed the method described in detail in Wang et al.~(2006) to 
determine the $3.6~\mu$m \emph{Spitzer} magnitudes from the GOODS-N
first, interim, and second data release products (DR1, DR1+, DR2;
M.~Dickinson et al.~2008, in preparation). For the $24~\mu$m data, we
used the
DR1 + MIPS source list and the version 0.36 MIPS map provided by the
\emph{Spitzer} Legacy Program. The MIPS catalog is flux-limited and
complete at $80~\mu$Jy and the median $1\sigma$ sensitivity of the
MIPS map is $6.4~\mu$Jy. The $5\sigma$ sensitivity limit for the
$3.6~\mu$m data is $0.327~\mu$Jy (Wang et al.~2006). Of the 503 CDF-N X-ray
sources, 379 are in the area covered by the GOODS-N $3.6~\mu$m and
$24~\mu$m micron observations.  

Of the 761 CLANS, 305 CLASXS, and 379 CDF-N X-ray sources observed in
the IRAC $3.6~\mu$m band, 633 CLANS, 249 CLASXS, and 351
CDF-N sources are detected to the limits of the SWIRE and GOODS-N surveys. 

Of the 761 CLANS, 525 CLASXS, and 379 CDF-N X-ray sources observed in
the MIPS $24~\mu$m band, 222 CLANS, 119 CLASXS, and 200 CDF-N
sources are detected to the limits of the SWIRE and GOODS-N surveys.   
%\clearpage
\begin{table}
\begin{scriptsize}
\caption{CLANS Optical \& Near-IR Specifications}
\label{clansopt table}
\begin{tabular}{lcccc}
\tableline\tableline
Filter & Telescope & Average Seeing & $3\sigma$ Limit & Total Area\tablenotemark{a} \\
 & & (arcsec) & (AB mag) & (deg$^2$) \\
\tableline
$g'$ & CFHT     & 0.8    & 26.9  & 0.96\\
$r'$ & CFHT     & 1.1     &26.5 & 1.16\\
$i'$ & CFHT     & 0.9     &26.0 & 1.21\\
$z'$ & CFHT     & 0.9    &25.3 & 1.33\\
$J$ & UH 2.2 m   & 0.9 &24.0 & 0.86\\
$H$ & UH 2.2 m  & 0.9  &23.2 & 0.83\\
$K$ & UKIRT   & 1.2 &22.1 & 0.74\\
\tableline
\end{tabular}
\end{scriptsize}
\footnotesize
\tablenotetext{a}{For comparison, the CLANS X-ray survey covers $0.6$~deg$^2$.}
\end{table}

\begin{table}
\begin{scriptsize}
\caption{CLASXS Optical \& Near-IR Specifications}
\label{clasxsopt table}
\begin{tabular}{lcccc}
\tableline\tableline
Filter & Telescope & Average Seeing & $3\sigma$ Limit & Total Area\tablenotemark{a}\\
 & & (arcsec) & (AB mag) & (deg$^2$)\\
\tableline
$u$ & CFHT     & 1.3    & 25.9 & 1.16\\
$g'$ & CFHT     & 0.9   &27.1 & 0.96\\
$i'$ & CFHT     & 1.0    &25.3 & 1.12\\
$J$ & UH 2.2 m   & 0.9 &24.3 & 0.76\\
$H$ & UH 2.2 m   & 1.1 &23.1 & 0.68\\
$K$ & UKIRT   & 1.0  &22.1 & 0.74\\
\tableline
\end{tabular}
\end{scriptsize}
\footnotesize
\tablenotetext{a}{For comparison, the CLASXS X-ray survey covers $0.45$~deg$^2$.}
\end{table}
\begin{table}
\begin{scriptsize}
\caption{CDF-N Near-IR Specifications}
\label{nir table}
\begin{tabular}{lcccc}
\tableline\tableline
Filter & Telescope & Average Seeing & $3\sigma$ Limit & Total Area\tablenotemark{a}\\
 & & (arcsec) & (AB mag) & (deg$^2$) \\
\tableline
$J$ & UH 2.2 m   & 1.2 &24.7 & 0.25 \\
$H$ & UH 2.2 m   & 1.2 &24.8 & 0.29\\
$K_s$ & CFHT      & 0.8  &24.2 & 0.29\\
\tableline
\end{tabular}
\end{scriptsize}
\footnotesize
\tablenotetext{a}{For comparison, the CDF-N X-ray survey covers $0.124$~deg$^2$.}
\end{table}
%\clearpage
\subsection{Optical and NIR Source Detection and Photometry}
\label{photodet}

We performed source detection and determined optical and NIR magnitudes using
SExtractor (Bertin \& Arnouts 1996). We set the detection
threshold at eight contiguous pixels with counts at least
2$\sigma$ above the local sky background. Using the ASSOC\_PARAMS function,
we searched
a $2\farcs5$ radius around the X-ray point source location in the
$z'$-band image to determine the nearest optical counterpart location. We then
used this $z'$-band optical counterpart location to determine the
magnitudes in the other bands. If there was no $z'$-band optical
counterpart but there was a $g'$-band counterpart within the
$2\farcs5$ radius, we used this location instead. We then checked
all sources by eye to ensure that the same source was being used
to determine the optical and NIR magnitudes. In the CDF-N, for
which we only present new NIR photometry, we used the Barger et
al.~(2003) X-ray source optical counterpart locations.

For sources with $z'<21$, we used the MAG\_AUTO magnitudes. For $z'>21$,
we used the $3\arcsec$ aperture-corrected magnitudes. We determined these
by taking the ${3\arcsec}$ MAG\_APER magnitude for each
source and then applying an offset. We calculated the offset as the
median difference between the $3\arcsec$ MAG\_APER and $6\arcsec$
MAG\_APER magnitudes for sources with $18<z'<20$. The
typical offset is 0.2 magnitudes with a $1\sigma$ error of 0.06. As
expected, there is a slight trend towards smaller offsets with increasing
magnitude ($\delta_{offset}<0.05$ magnitudes). Between $18<z'<20$, the
MAG\_AUTO and $6\arcsec$ MAG\_APER magnitudes generally agree well, with the
MAG\_AUTO magnitudes 0.01-0.03 magnitudes fainter than the $6\arcsec$
MAG\_APER magnitudes. See R.~Keenan et al.~(2008, in preparation) for a
more detailed discussion.

As stated above, in the catalogs we
include magnitudes for sources with counts at least
$2\sigma$ above the local sky background (as determined by
SExtractor). We provide the aperture-corrected $3\sigma$ limits of the
images in Tables
\ref{clansopt table}, \ref{clasxsopt table}, and \ref{nir table},
which show the depth of our optical and NIR data. We determined these
limits by laying down random $3\arcsec$ apertures away from objects on
the images. We masked out the areas around bright stars to prevent
erroneous detections and measurements from scattered light. We derived
the aperture-corrected $3\sigma$ limit for each image from
%\begin{equation}
\begin{eqnarray}
3\sigma &=&-2.5\log(3~RMS\sqrt{\pi\times(1\farcs5/platescale)^2}) \nonumber \\
& & + ZP~+OFFSET,
\end{eqnarray}
%\end{equation}
where RMS is the average standard deviation of the blank
apertures, ZP is the zeropoint for the image in question, and OFFSET
is the aperture-correction as discussed in the previous paragraph. A more
detailed description of this method is
given in R.~Keenan et al.~(2008, in preparation). 

Typical $1\sigma$ errors on the CLANS $g',r',i',z',J,H,K,3.6~\mu$m and $24~\mu$m
fluxes are 0.2, 0.3, 0.4, 0.9, 3.3, 7.2, 4.8, 10.2, and $629.9\times10^{-30}$
ergs~cm$^{-2}$~s$^{-1}$~Hz$^{-1}$, respectively. For the 
CLASXS field, typical $1\sigma$ errors for the
$u,g',i',J,H,K,3.6~\mu$m and $24~\mu$m fluxes are 
0.2, 0.2, 0.9, 2.4, 5.4, 3.6, 10.2, and $629.9\times10^{-30}$
ergs~cm$^{-2}$~s$^{-1}$~Hz$^{-1}$, respectively. And typical $1\sigma$
errors for the CDF-N $J,H,K,3.6~\mu$m and $24~\mu$m fluxes are 1.5,
1.3, 0.8, 0.5, and
$64.0\times10^{-30}$ ergs~cm$^{-2}$~s$^{-1}$~Hz$^{-1}$, respectively. 

\subsection{CLASXS: Updated Optical Photometry}
\label{clasxsupdate}

The original CLASXS catalog of Steffen et al.~(2004) presented $B, V,
R, I$, and $z'$ photometry from CFHT and Subaru observations for 521
of the 525 X-ray point sources. When comparing
the spectral energy distributions (SEDs) using these data with our
more recent $u, g'$, and $i'$-band CFHT observations, we found that the
original zeropoints for all but the $B$-band data were incorrect. The
zeropointing errors reflect the difficulty of calibrating the Subaru
Suprime-Cam images. Due to Subaru's large collecting area, calibration stars
saturate even with the shortest possible exposure times. Also, the Steffen et
al.~(2004) CFHT data were taken using CFH12K, which does not operate
in queue mode, so no automatic calibrations were done (in contrast
with the standard MegaCam queue mode procedure).  

To determine the zeropoint offsets, we compared SEDs
created from the Steffen et al.~(2004) CFH12K data with SEDs made from
the more recently obtained $u,g'$, and $i'$-band MegaCam data. We restricted our
sample to sources with $21<g'<23$. The zeropoint offsets listed in
Table \ref{zpcorr table} are the mean of the values needed to shift
the SEDs made from the original CFHT data onto the SEDs made from the
new $u,g'$, and $i'$-band data. 

We did this separately for the SEDs with a more obvious bluer slope
(ascending SED as frequency increases) and those with a more obvious redder
slope (descending SED as frequency increases), and found no major
differences in the offsets between these two types of sources. We
found that the CFHT $B$-band data had a negligible offset while the
$R$- and $z'$-band data were best re-zeropointed using an offset of
-0.6. 

We followed the same method to re-zeropoint the original Subaru
$V, R, I$, and $z'$-band data (see Table \ref{zpcorr table}).

%\clearpage
\begin{table}[!h]
\begin{small}
\caption{CLASXS Zeropoint Offsets}
\label{zpcorr table}
\begin{tabular}{rrrrrr}
\tableline\tableline
 & $B$ & $V$ & $R$ & $I$ & $z'$ \\
\tableline
Subaru & \nodata & -0.5 & -0.5 & -0.4 & -0.3 \\
CFHT & 0. & \nodata & -0.6 & \nodata & -0.6 \\ 
\tableline
\end{tabular}
\end{small}
\end{table}
%\clearpage

\section{REDSHIFT INFORMATION}
\label{spectroscopictop}

\subsection{Spectroscopic Observations}
\label{spectroscopic}

We have carried out extensive spectroscopic observations of the X-ray
sources in the CLANS, CLASXS, and CDF-N fields. All of the CLANS
spectroscopic observations are presented for the first time in this
article. In addition, we have obtained some new spectroscopic
observations for the CLASXS and CDF-N fields. 

For the CLANS field we
obtained all the optical spectra using the Deep
Extragalactic Imaging Multi-Object Spectrograph (DEIMOS; Faber et
al.~2003) on the 10 m Keck II telescope. We used the 600 line
mm$^{-1}$ grating, which yielded a resolution of 3.5~\AA~and a
wavelength coverage of 5300~\AA. The exact central wavelength depends
on the slit position, but the average was 7200~\AA. Each $\sim$1 hr
exposure was broken into three subsets, with the objects stepped along
the slits by $1\farcs5$ in each direction.  

The DEIMOS spectroscopic reductions follow the same
procedures used by Cowie et al.~(1996) for the Low-Resolution Imaging
Spectrograph (LRIS; Oke et al. 1995) reductions. In brief, we removed
the sky contribution by subtracting the median of the dithered
images. We then removed cosmic rays by registering the images and using a
cosmic ray rejection filter as we combined the images. We also removed geometric
distortions and applied a profile-weighted extraction to obtain the
spectrum. We did the wavelength calibration using
a polynomial fit to known sky lines rather than using calibration
lamps. We inspected each spectrum individually and measured a redshift
only for sources where a robust identification was possible. The
high-resolution DEIMOS spectra can resolve the doublet
structure of the [O II] $\lambda\lambda$3727 and 3729 lines, allowing
spectra to be identified by this doublet alone. For sources not
identified by this doublet, only redshift identifications based on multiple
emission and/or absorption lines were included in the sample. We find
that the spectroscopic redshifts for the non-broad-line AGNs
(broad-line AGNs) are accurate to $\sim 0.001$ ($\sim 0.005$). 

Steffen et al.~(2004) presented a spectroscopic redshift catalog
for the CLASXS X-ray sources. Most of the spectroscopically
observed sources were observed using DEIMOS, following the same
procedures as for the CLANS field, but some of the brighter sources
($I<19$) were observed using HYDRA (Barden et al.~1994) on the WIYN
3.5~m telescope. However, for any source observed with HYDRA for
which Steffen et al.~(2004) were unable to determine a redshift and
classification, they re-observed the source with DEIMOS. Thus, 
only a small number of the CLASXS sources have redshifts and
classifications based on their HYDRA spectra. Subsequent to Steffen et
al.~(2004), we obtained 11 additional spectra in the CLASXS field
using DEIMOS. We have redshift identifications and classifications for
all 11. In Table \ref{CLASXSopt table}, \emph{s} indicates the spectroscopic
redshifts from Steffen et al.~(2004). 

Barger et al.~(2003) presented a spectroscopic redshift catalog for
the 2 Ms CDF-N X-ray sources. The spectra were obtained with either 
DEIMOS or LRIS. Fainter objects were
observed a number of times with DEIMOS such that the total exposure
times for these
sources are greater than the $\sim 1$ hr used for the CLANS and CLASXS
sources. For this reason, the redshift identification rate for the
CDF-N, as compared with the CLANS and CLASXS fields, is slightly
higher. We supplement these CDF-N spectroscopic observations
with eight redshifts obtained by Chapman et al.~(2005) and
Swinbank et al.~(2004), adopting the Swinbank et al.~(2004) NIR redshifts over 
the Chapman et al.~(2005) optical redshifts, where available, because
redshift measurements are generally more reliable when they are made
from emission-line features. Where there are redshifts from both
sources, the redshifts are within $<0.008$ of each other. We acquired
one more spectroscopic redshift from Reddy et al.~(2006). The source
classifications, however, are not available for the Chapman et 
al.~(2005), Swinbank et al.~(2004), and Reddy et al.~(2006)
redshifts. Subsequent
to Barger et al.~(2003), we obtained 49 additional spectra in the
CDF-N field using DEIMOS. We have redshift
identifications and classifications for 39 of these. In Table
\ref{cdfnspect table}, \emph{a}, \emph{b}, \emph{c}, and \emph{d}
indicate the spectroscopic redshifts from Barger et al.~(2003),
Swinbank et al.~(2004), Chapman et al.~(2005), and Reddy et
al.~(2006), respectively.

\subsection{Spectroscopic Completeness}

\begin{figure}
\epsscale{2}
\plottwo{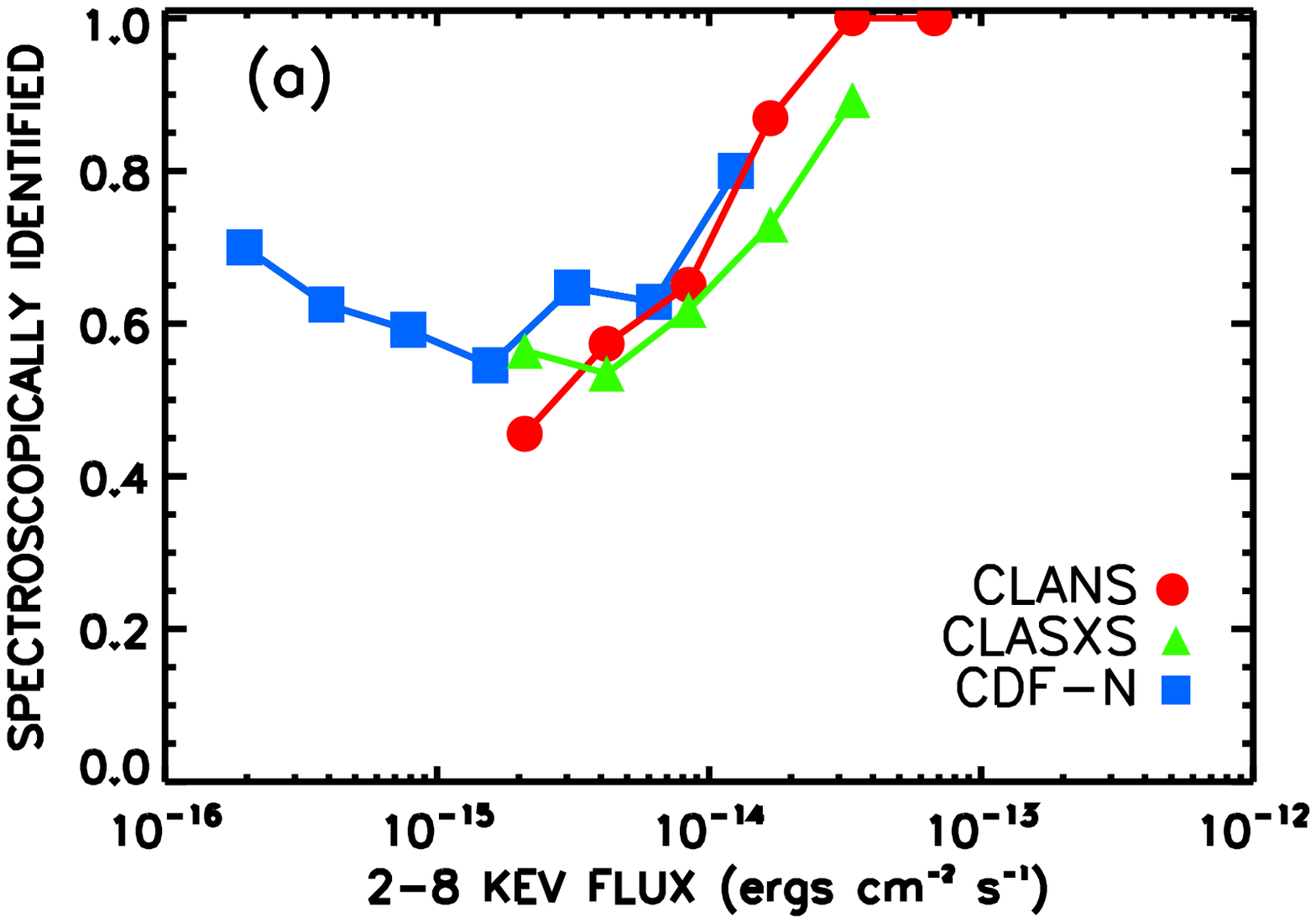}{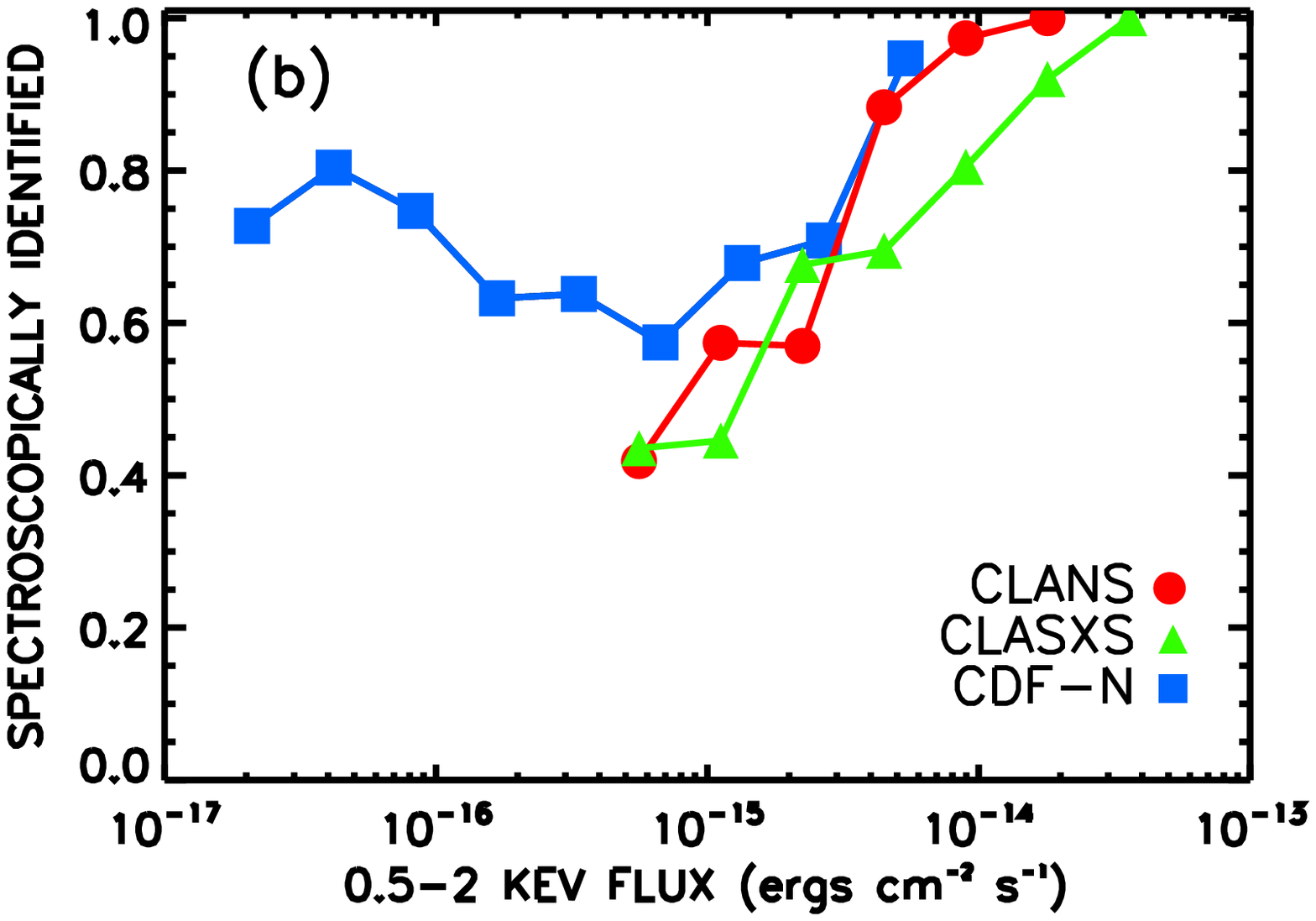}
\plottwo{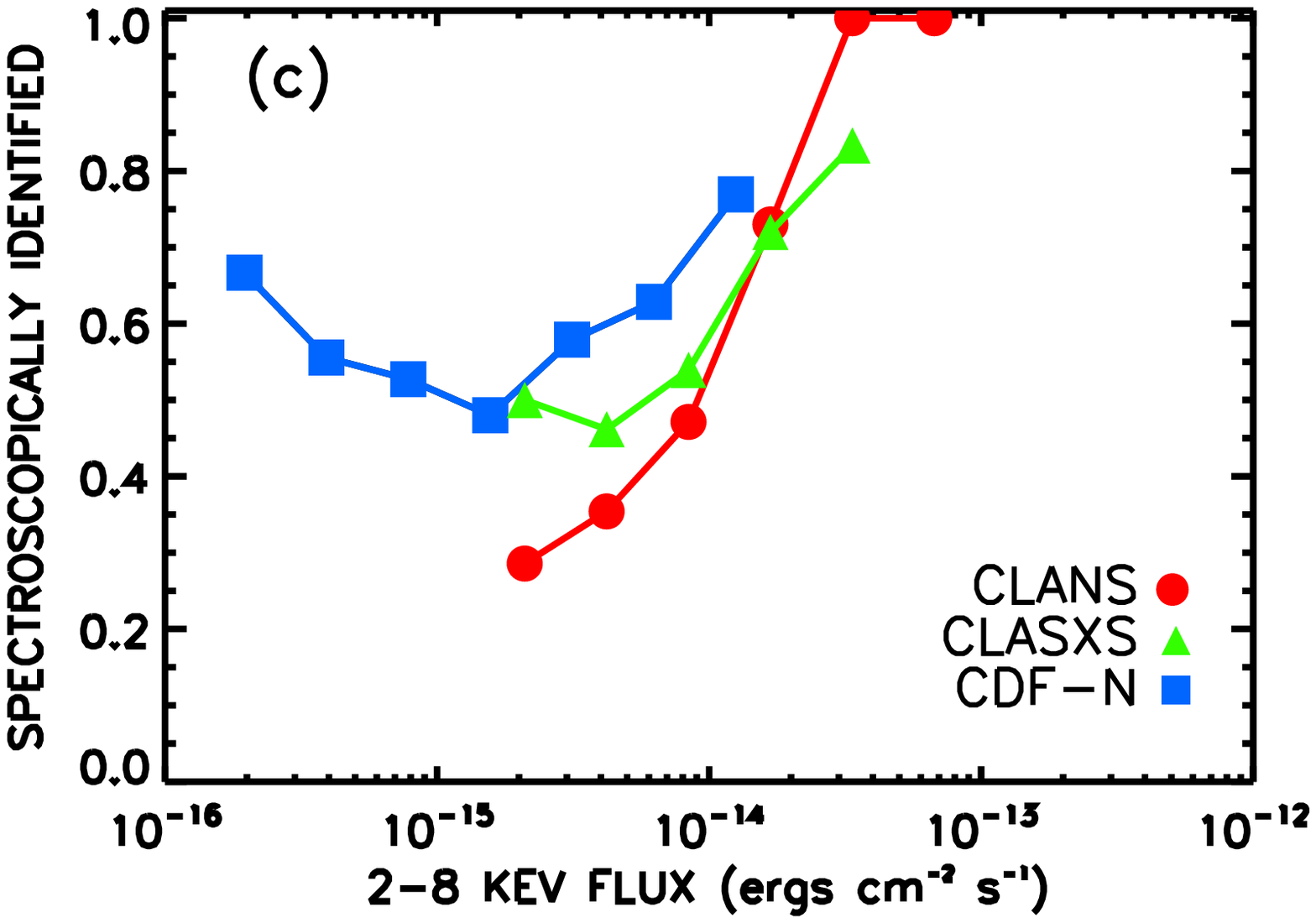}{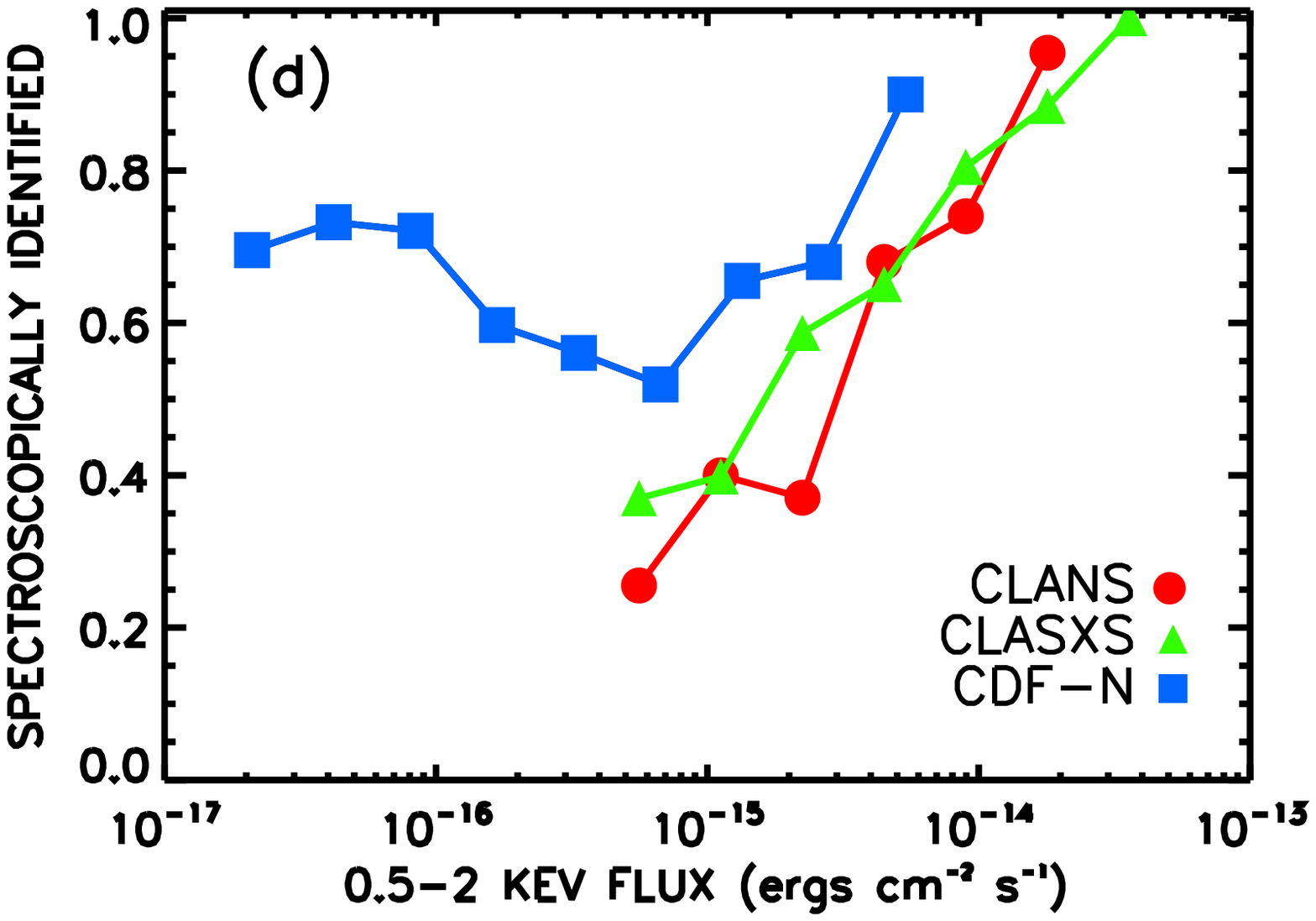}
\caption{(a) Fraction of \emph{spectroscopically observed} sources
  that are spectroscopically
  identified in the $2-8~\rm keV$ band (\emph{red circles}, CLANS;
  \emph{green triangles}, CLASXS; \emph{blue squares}, CDF-N). Only
  flux bins containing more than 10 sources are plotted. (b) Same as
  (a) but in the $0.5-2~\rm keV$ band. (c) Fraction of \emph{all} sources
  that are spectroscopically identified in the $2-8~\rm keV$ band. (d)
Same as (c) but in the $0.5-2~\rm keV$ band.}
\label{spect figure}
\end{figure}

Figure \ref{spect figure} shows the useful flux ranges of the three
\emph{Chandra} surveys. Specifically, Figures
\ref{spect figure}a and \ref{spect figure}b show the fraction of
spectroscopically observed sources in each flux bin that are spectroscopically
identified and Figures \ref{spect figure}c and \ref{spect figure}d show
the fraction of all sources in each flux bin that are
spectroscopically identified (see Table \ref{class table} for the
actual numbers for each field). Above $10^{-14}$
ergs~cm$^{-2}$~s$^{-1}$ ($2-8~\rm keV$), 79\%, 71\%, and 82\% of all
the CLANS, CLASXS, and CDF-N X-ray sources, respectively, have
spectroscopic redshifts.

The higher spectroscopic completeness at high X-ray fluxes (i.e.,
$f_{2-8~\rm{keV}}>10^{-14}$ ergs~cm$^{-2}$~s$^{-1}$) is partly due to
the fact that at these fluxes the sample is dominated by broad-line
AGNs, and broad-line AGNs are straightforward to identify. In
addition, at fainter X-ray fluxes the sources tend to be optically
fainter, making the redshift identifications at these fluxes more
difficult. In particular, the intermediate-flux, optically normal
galaxies at $z\sim2$ are the most difficult to identify. At the
faintest X-ray fluxes \emph{Chandra} begins to detect star forming galaxies
at lower redshifts, and it is the appearance of this population that
again improves our spectroscopic completeness.

%\clearpage
%\clearpage

\subsection{Photometric Redshifts}
\label{photoz}

It is possible to extend the redshift information to fainter
magnitudes using photometric redshifts. To determine photometric redshifts
for the X-ray sources in the CLANS, CLASXS, and CDF-N fields, we used
the template-fitting
method described in Wang et al.~(2006; see also P\'{e}rez-Gonz\'{a}lez
et al.~2005). In this method one builds comparison templates using
sources within the sample. 

Following the prescription in Wang et al.~(2006), we created SEDs for
our sources using the optical through IR data to match with template
SEDs. The CLANS field has 8 bands of coverage ($g', r', i', z', J,
H, K, 3.6~\mu$m), the CLASXS field has 11 bands of coverage ($u, B, g',
V, R, i', z', J, H, K, 3.6~\mu$m), and the CDF-N field has 10 bands of
coverage ($U, B, V, R, I, z', J, H, K_s, 3.6~\mu$m). We used the $J$-band
fluxes to normalize the bolometric fluxes for each source. 

\begin{figure}[!h]
\epsscale{1}
\plotone{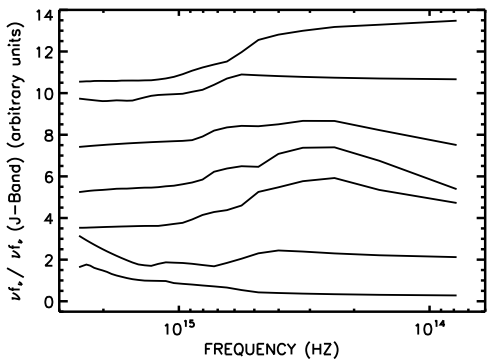}
\caption{CDF-N template SEDs created by taking the median $\nu
  f_{\nu}/\nu f_{\nu~(J-band)}$ values of the rest-frame SEDs with
  similar shapes.}  
\label{templ}
\end{figure}

We then used the
spectroscopically identified sources in each field to construct seven distinct
templates over the frequency range from $6\times10^{13}$ to
$3\times10^{15}$ Hz. Each template was created by taking the median
$\nu f_{\nu}/\nu f_{\nu~(J-band)}$ values of the rest-frame SEDs with
similar shapes. Figure \ref{templ} shows the templates for the CDF-N
field. In this work, we have not created templates specific to each
spectroscopic class, although the majority of the SEDs used to create
the templates with large FUV to J-band ratios are broad-line
AGNs. Trouille et al.~(2008, in preparation) discuss in detail the
differences between the SEDs for broad-line AGNs and for non-broad-line
AGNs for our sample.
 
Next we determined the photometric redshifts by finding the best-fit,
via least-squares minimization, between our individual source SEDs and
these templates. The best-fit in least-squares minimization is the one
in which the sum of the squared residuals has its least value; a
residual being the difference between the observed value and the
template value. Sources whose best-fit corresponded to a value
greater than 4.5 were not assigned a photometric redshift. This optimized the
number of sources for which we could determine photometric redshifts
in each field while
minimizing the number of seriously discrepant sources. Throughout, we
only used sources with reliable magnitude determinations in at least 4,
5, and 5 of the wavebands for the CLANS, CLASXS, and CDF-N fields,
respectively.

\begin{figure}
\epsscale{1.}
\plotone{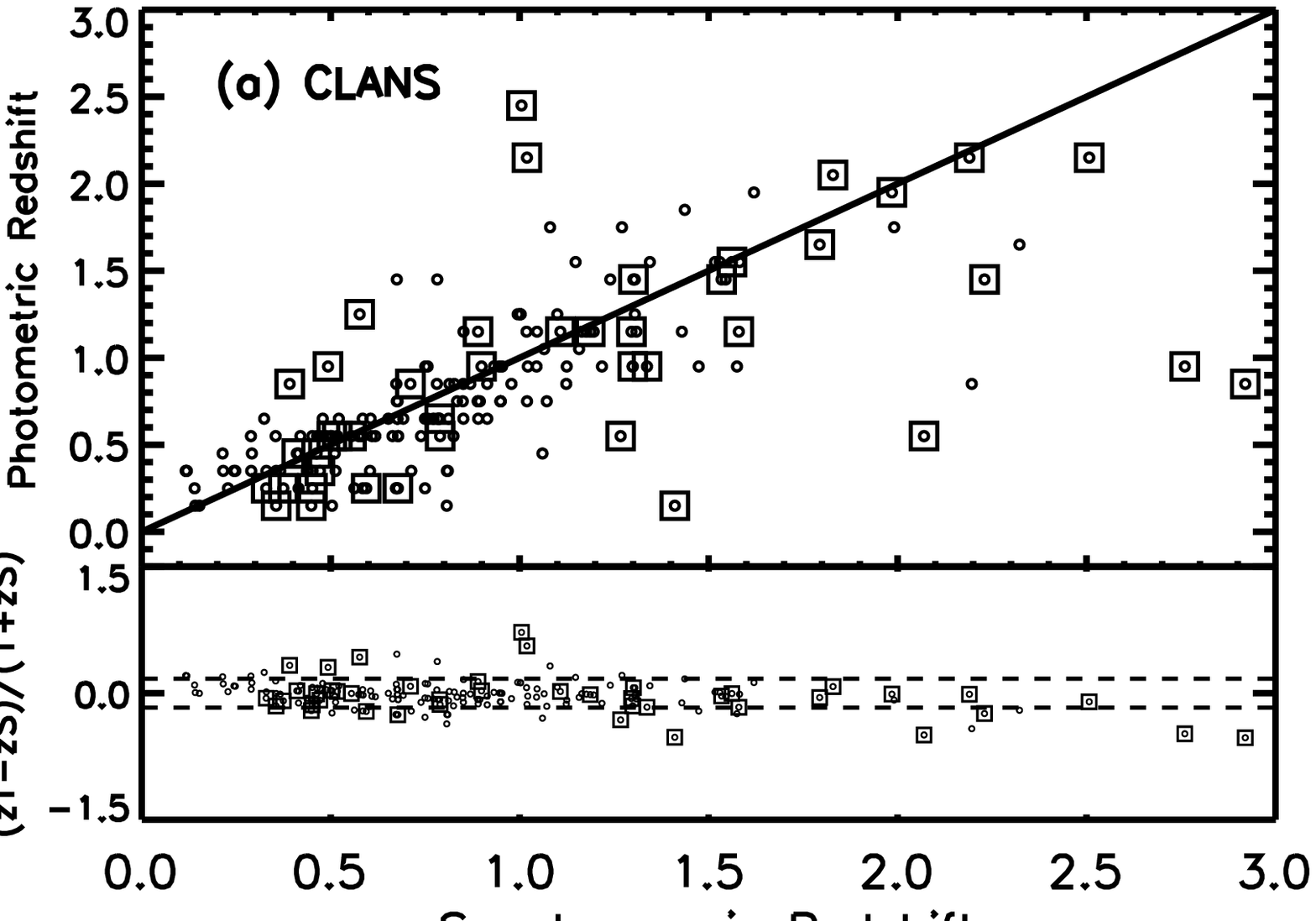}
\plotone{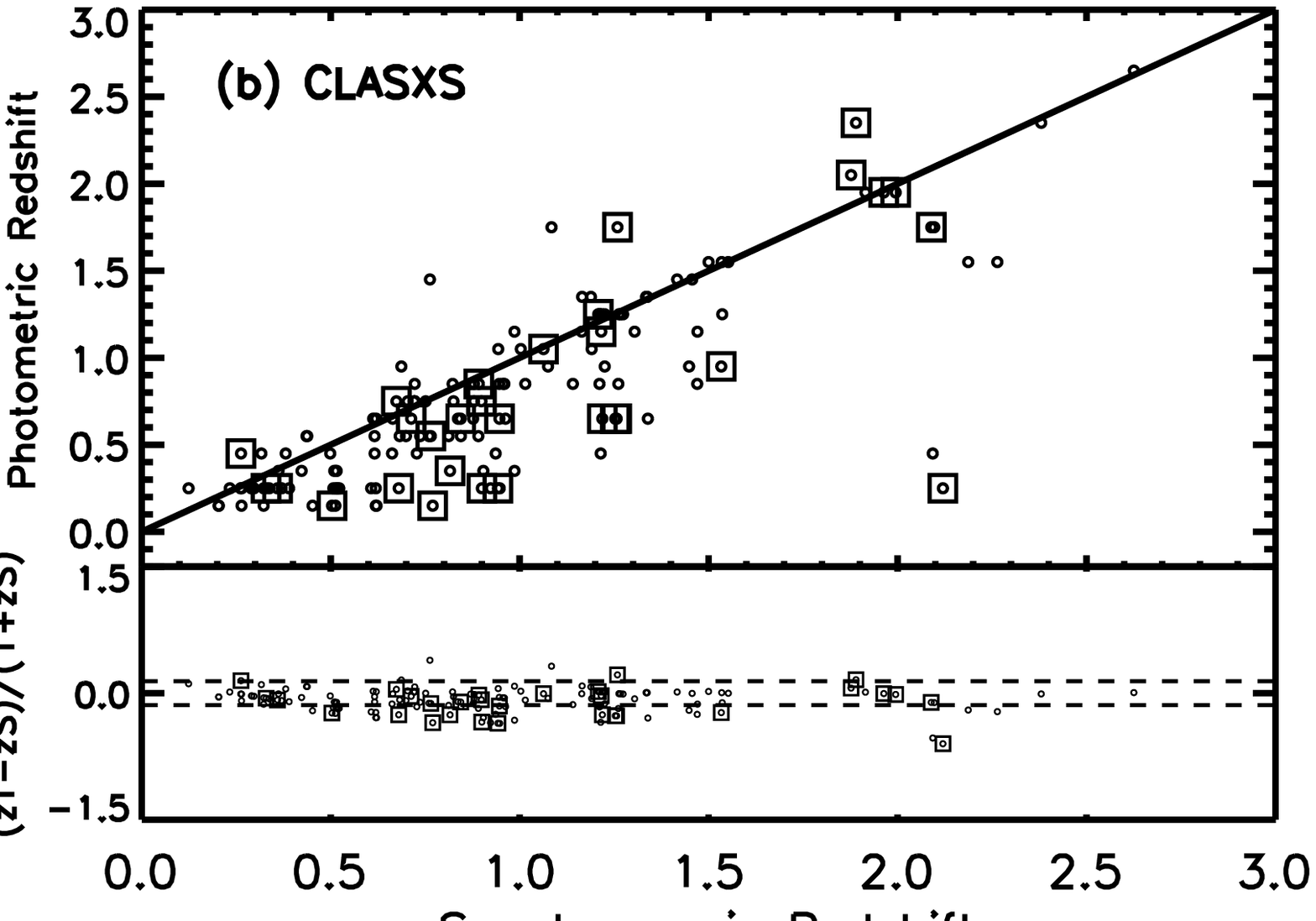}
\plotone{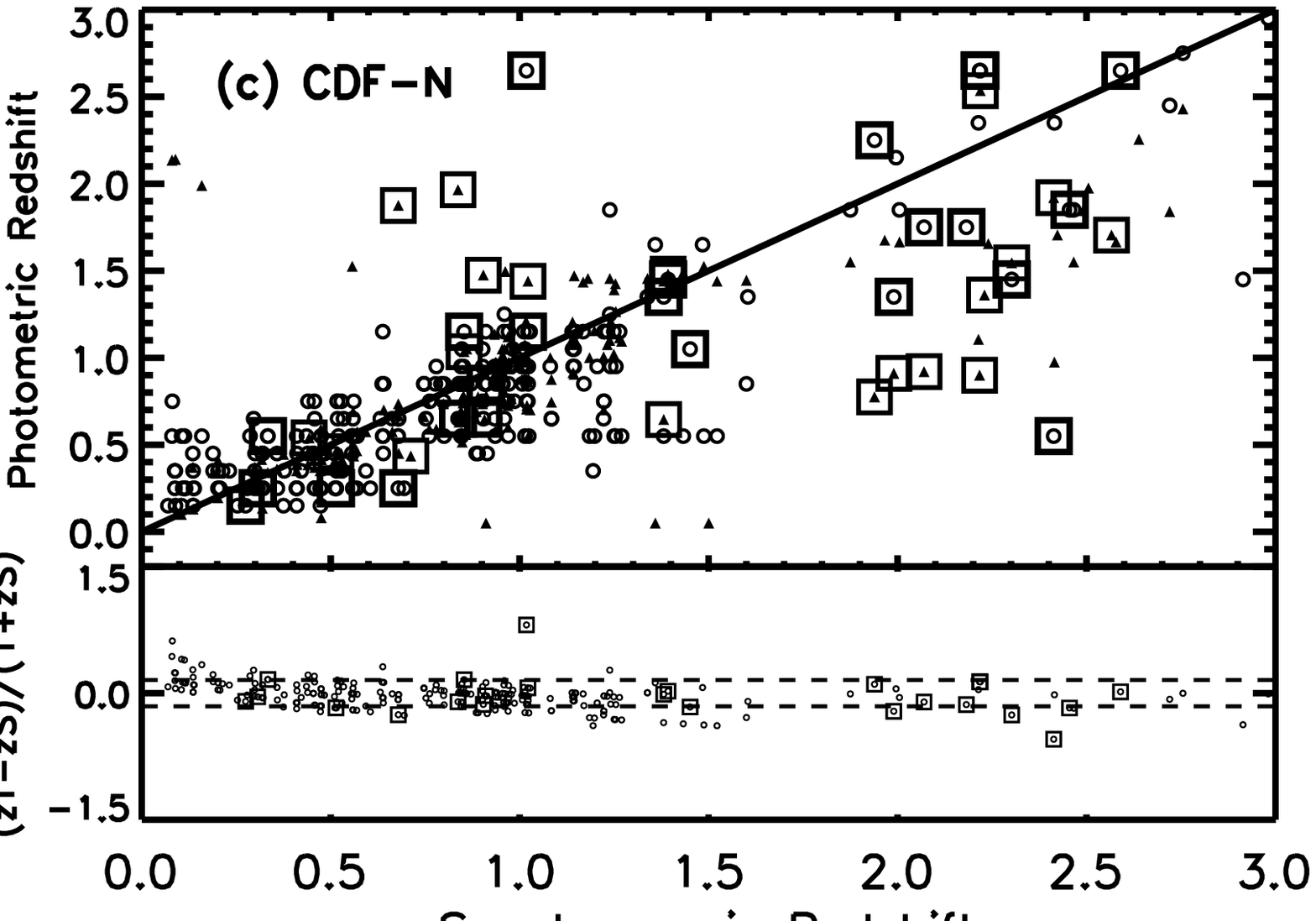}
\caption{Comparison of the photometric redshifts with the spectroscopic
redshifts for the spectroscopically identified sample in the (a)
CLANS, (b) CLASXS, and (c) CDF-N fields (\emph{open circles},
our template-fitting technique results; \emph{large open squares}, broad-line
AGNs; \emph{filled triangles in c only}, \emph{HyperZ}
results). The panels at the base of each figure show the photometric
redshift residuals for our template-fitting results for
each field. Dashed lines enclose the 1$\sigma$ error interval for
$(z_T-z_S)/(1+z_S)$.} 
\label{spectrovsphoto figure}
\end{figure}

In Figure \ref{spectrovsphoto figure}
we compare our photometric redshifts ($z_T$) with the spectroscopic
redshifts ($z_S$)
for the spectroscopically identified sources. Of the 327, 260, and 307
spectroscopically identified sources in the
CLANS, CLASXS, and CDF-N fields that are not stars, we detected 270,
223, and 296, respectively, in at least the
minimum number of bands. We were able to determine photometric
redshifts for $\sim80$\% of these. 

Figure \ref{spectrovsphoto figure}
also shows the photometric redshift residuals for each field
(\emph{base panels}). The dashed lines enclose the 1$\sigma$ error
interval for $(z_T-z_S)/(1+z_S)$. The redshift dispersion for the
combined sample is $\sigma_z\sim 0.16(1+z_S)$. The redshift dispersion
improves slightly to $\sigma_z\sim 0.11(1+z_S)$ when only considering
the absorbers in our sample and degrades to $\sigma_z\sim
0.2(1+z_S)$ for the broad-line
AGNs (\emph{these are marked in the figure with large open
  squares}). The effect of this is
mitigated by the fact that in our spectroscopically observed sample,
the identification of the broad-line AGNs is highly complete (see \S
\ref{optclass}) and so these sources generally do not require photometric
redshifts. Therefore, this is only of importance to the
relatively small percentage ($10-30$\%) of spectroscopically
unobserved sources in
our three fields. With $<6$\% of the sources for which we are able to
determine photometric redshifts being seriously discrepant sources
(i.e., having
$[z_T-z_S]/[1+z_S]>2~\sigma$), the method robustly
places sources in the correct redshift range.  

Of the sources for which we could not determine photometric redshifts
(sources whose least-squares minimization exceeded our threshold of 4.5),
$\sim$50\% have SEDs that lack the smoothness needed to be fit by any of our
individual templates. The other $\sim$50\% are smooth but have SEDs
that differ too much from any of the templates to be fit by them. 

Of the 425, 245, and 182 spectroscopically unidentified sources in the
CLANS, CLASXS, and CDF-N fields, we detected 286, 166, and 134,
respectively, in at least the minimum number of bands. Of these, we
were able to determine photometric redshifts for 234, 134, and 107,
respectively (i.e., again, for about 80\% of them).   

In order to test the reliability of our template-fitting
program, we compared our CDF-N photometric redshift determinations with
those we determined using the publicly available photometric redshift code
\emph{HyperZ} (Bolzonella et al.~2000). We used the CDF-N
field because it has the highest level of
spectroscopic completeness of the three fields and excellent
multiwavelength coverage. \emph{HyperZ} measures
photometric redshifts by finding the best fit, defined by the $\chi^2$
statistic, to the observed SED from a library of galaxy templates. We
used the eight Bruzual \& Charlot (1993) templates provided with the program
covering a range of galaxy types (burst, elliptical, S0, Sa, Sb, Sc,
Sd, Im). 

Imposing a $\chi^2<15$ and $P(\chi)<1$ limit appears to optimize the
completeness of the \emph{HyperZ} photometric redshifts
while minimizing the number of seriously discrepant sources. We
determined this by comparing our \emph{HyperZ} photometric redshifts
with our spectroscopic redshifts for the spectroscopically identified
sources. The triangles in Figure
\ref{spectrovsphoto figure}c show the 190 \emph{HyperZ} photometric
redshifts for the spectroscopically identified sources
in the CDF-N field. The redshift dispersion for the \emph{HyperZ}
results is $\sigma_z\sim 0.29(1+z_S)$, as compared to $\sigma_z\sim
0.15(1+z_S)$ using our template-fitting approach on the
CDF-N. Restricting the sample to only broad-line AGNs, both
methods exhibit a degraded goodness-of-fit. The redshift
dispersion for the broad-line AGNs using \emph{HyperZ} is
$\sigma_z\sim 0.30(1+z_S)$, as compared to $\sigma_z\sim
0.24(1+z_S)$ using our template-fitting approach on the CDF-N
broad-line AGNs. 

Of the spectroscopically unidentified sources in the CDF-N field that
were detected in at least the minimum number of bands (134),
we were able to determine photometric redshifts for 75 of
them using \emph{HyperZ}. For comparison, we were able to determine
photometric redshifts for 107 of them using our template-fitting technique. 

\begin{figure}
\epsscale{1.}
\plotone{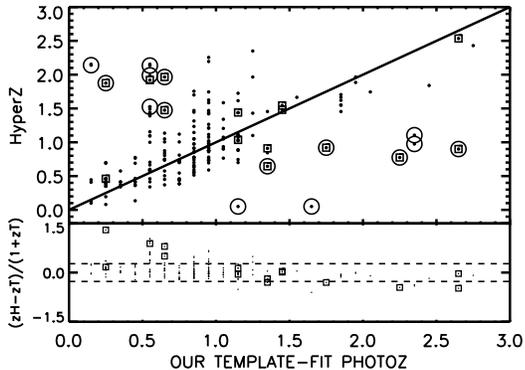}
\caption{Our template-fit photometric redshifts versus \emph{HyperZ}
  photometric redshifts for all the sources (spectroscopically
  identified or unidentified) in the CDF-N field for which we were
  able to determine photometric redshifts (\emph{solid circles}). Open
  squares indicate the broad-line AGNs. Open circles designate the
  15 $(z_H-z_T)/(1+z_T)>\sigma$ outlier sources with
  known spectroscopic redshifts. For all 15 of these sources, our
  template-fit photometric redshifts are within $\Delta z<0.5$ of the
  spectroscopic redshifts.}
\label{hyperz figure}
\end{figure}

Figure \ref{hyperz figure} shows the \emph{HyperZ} photometric
redshifts versus our template-fit photometric redshifts for all the
sources (spectroscopically identified or unidentified) in the
CDF-N for which we were able to determine photometric redshifts. The
panel at the base of this figure shows the photometric redshift
residuals. The dashed lines enclose the 1$\sigma$ error interval,
which is $\sigma_z\sim 0.25(1+z_S)$. We have overlaid open circles on the 15
$(z_H-z_T)/(1+z_T)>\sigma$ sources with known
spectroscopic redshifts. For
all 15 of these sources, our template-fit photometric redshifts are within
  $\Delta z <0.5$ of the spectroscopic redshifts. Of the remaining
  sources, less than 5\% of the \emph{HyperZ} CDF-N photometric
  redshifts exhibit $(z_H-z_T)/(1+z_T)>2\sigma$. 

Given that our template-fitting method provides a more reliable,
self-consistent, and complete photometric redshift determination
method than \emph{HyperZ}, we only use our template-fit
photometric redshifts hereafter.

%\clearpage

\subsection{Redshift and Flux Distributions}

%\clearpage
\begin{figure}
\epsscale{1}
\plotone{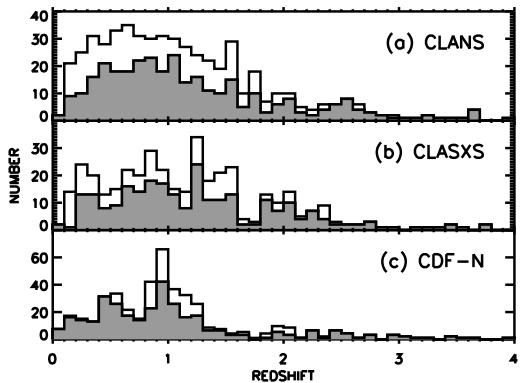}
\caption{Number of sources with spectroscopic (\emph{shading}) 
  and photometric (\emph{open}) redshifts versus
  redshift for the CLANS, CLASXS, and CDF-N samples. Bin size $\Delta z=0.1$.} 
\label{redshift hist figure}
\end{figure}

\begin{figure}
\epsscale{1.1}
\plotone{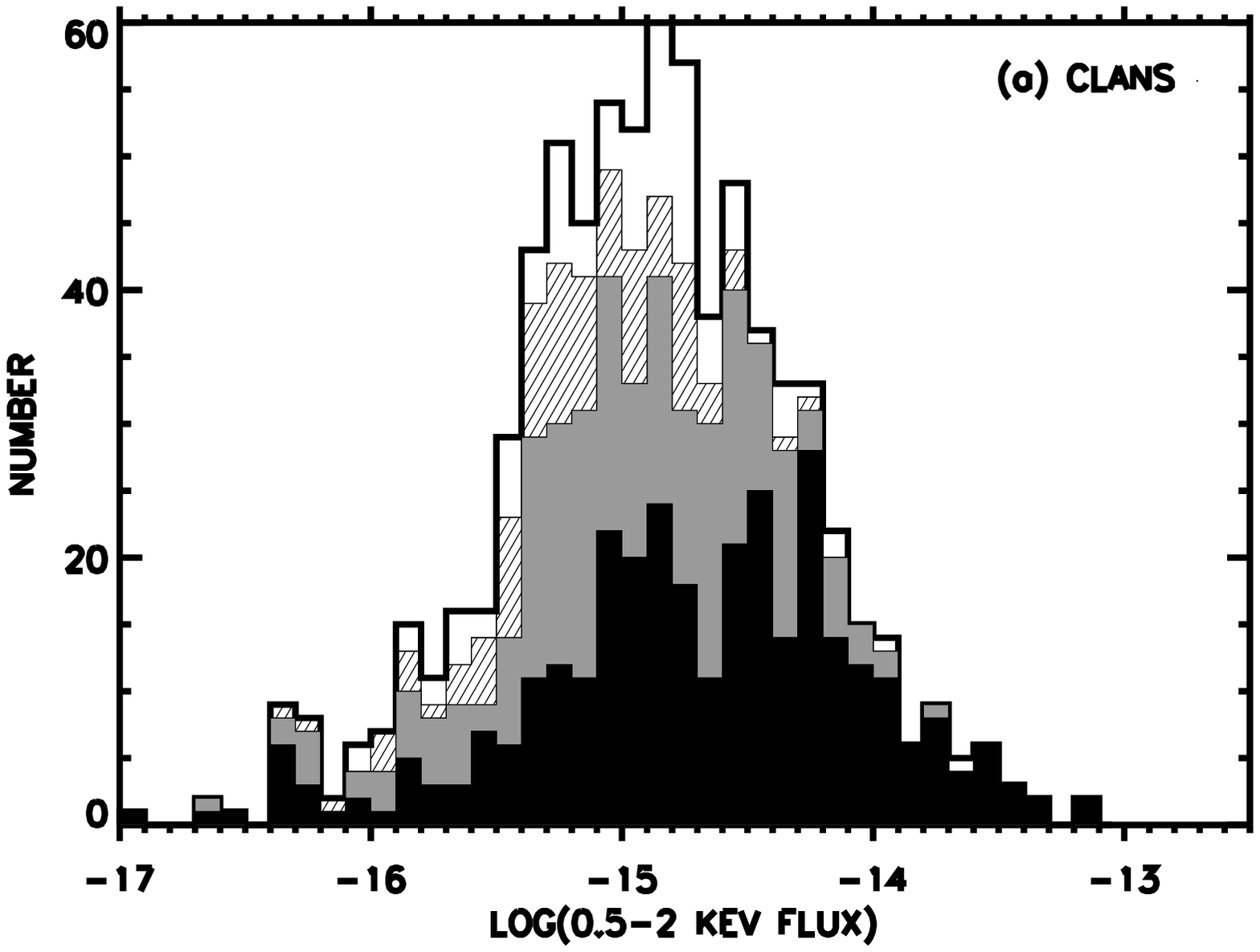}
\plotone{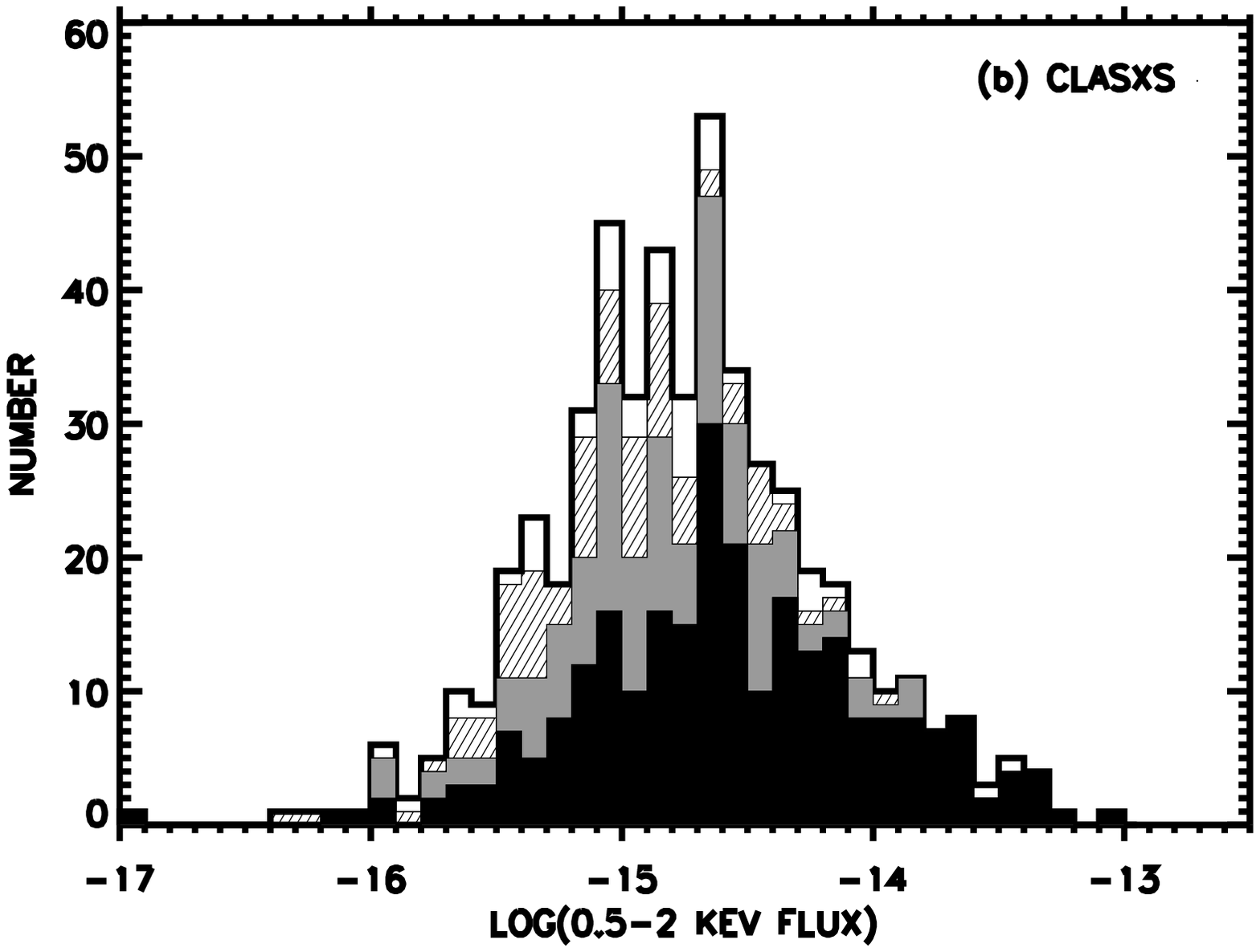}
\plotone{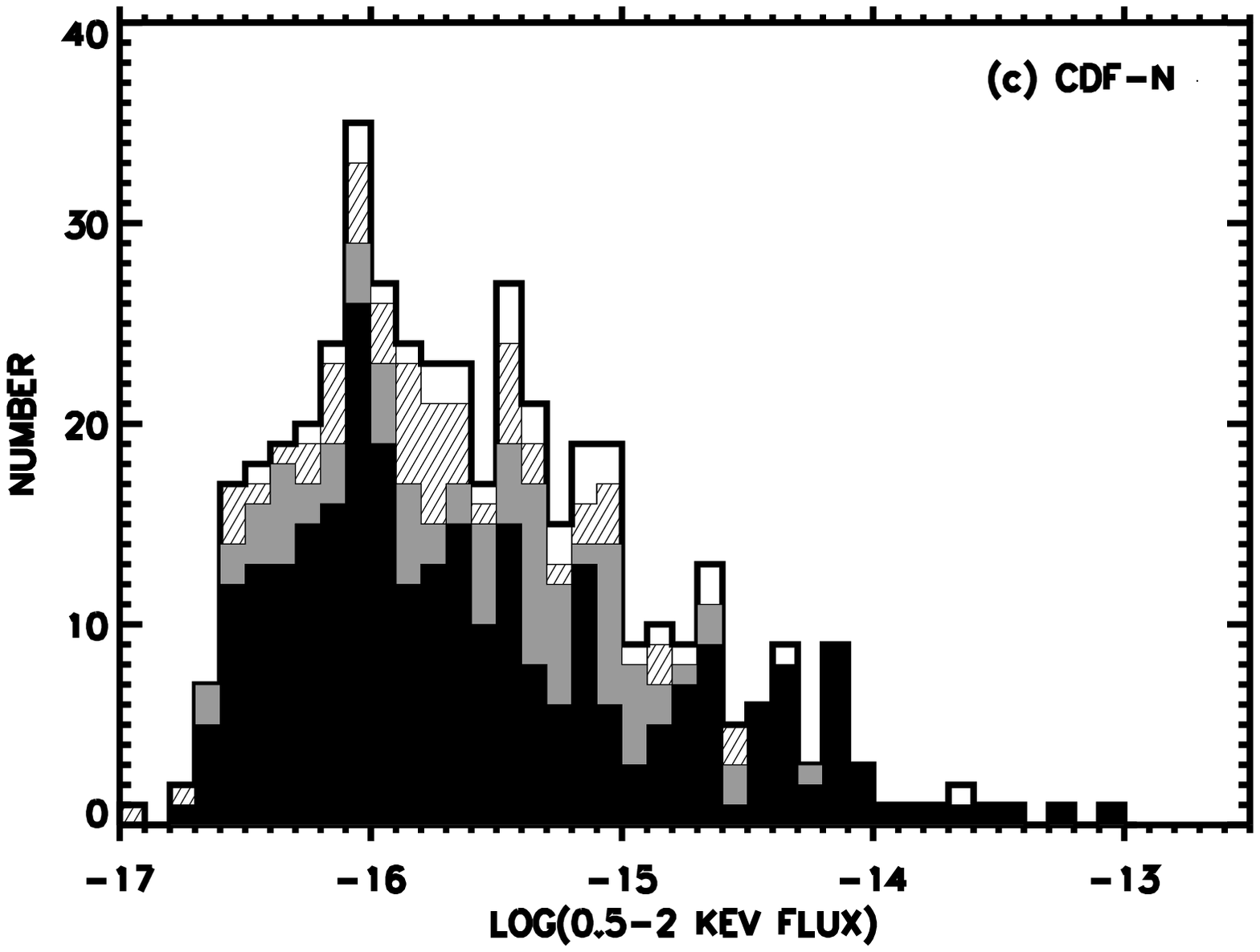}
\caption{$0.5-2~\rm{keV}$ flux
  distributions for the (a) CLANS, (b) CLASXS, and (c) CDF-N X-ray
  sources (\emph{filled}, spectroscopic redshifts; \emph{shading}, photometric redshifts; \emph{hatched}, spectroscopically
  observed but unidentified and no photometric redshift determined;
  \emph{open}, spectroscopically unobserved and no photometric
  redshift determined).} 
\label{softspect figure}
\end{figure}

\begin{figure}
\epsscale{1.1}
\plotone{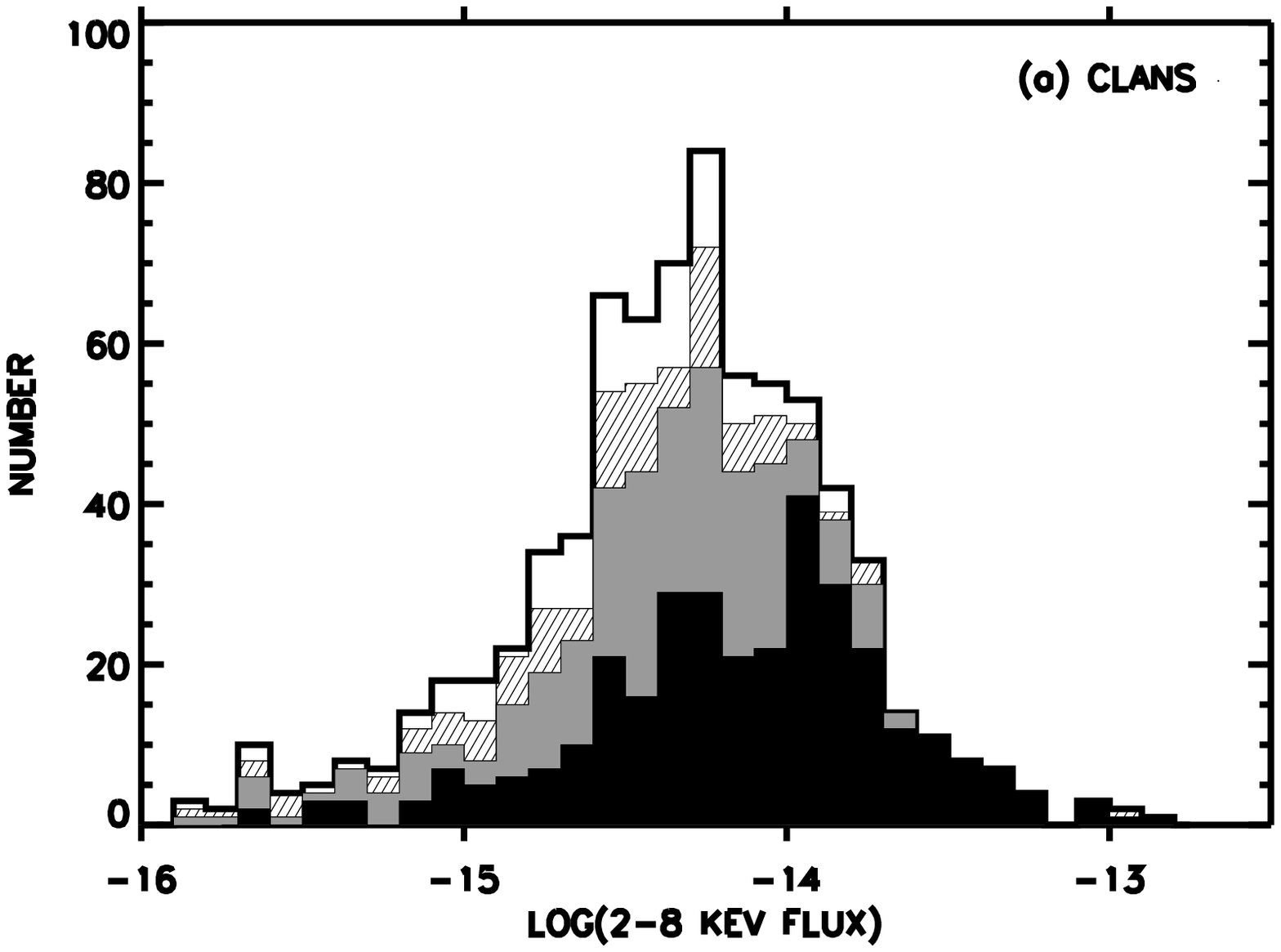}
\plotone{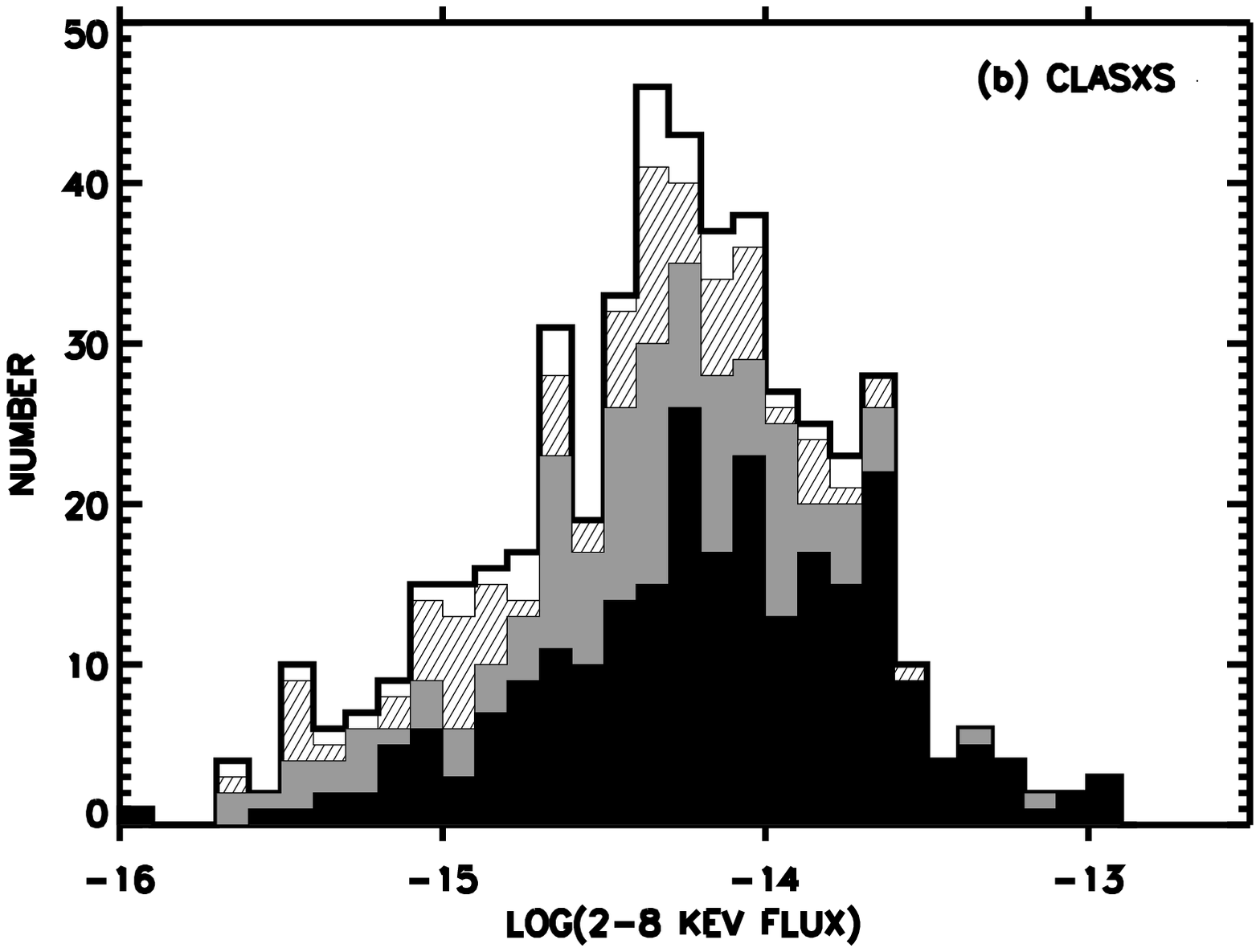}
\plotone{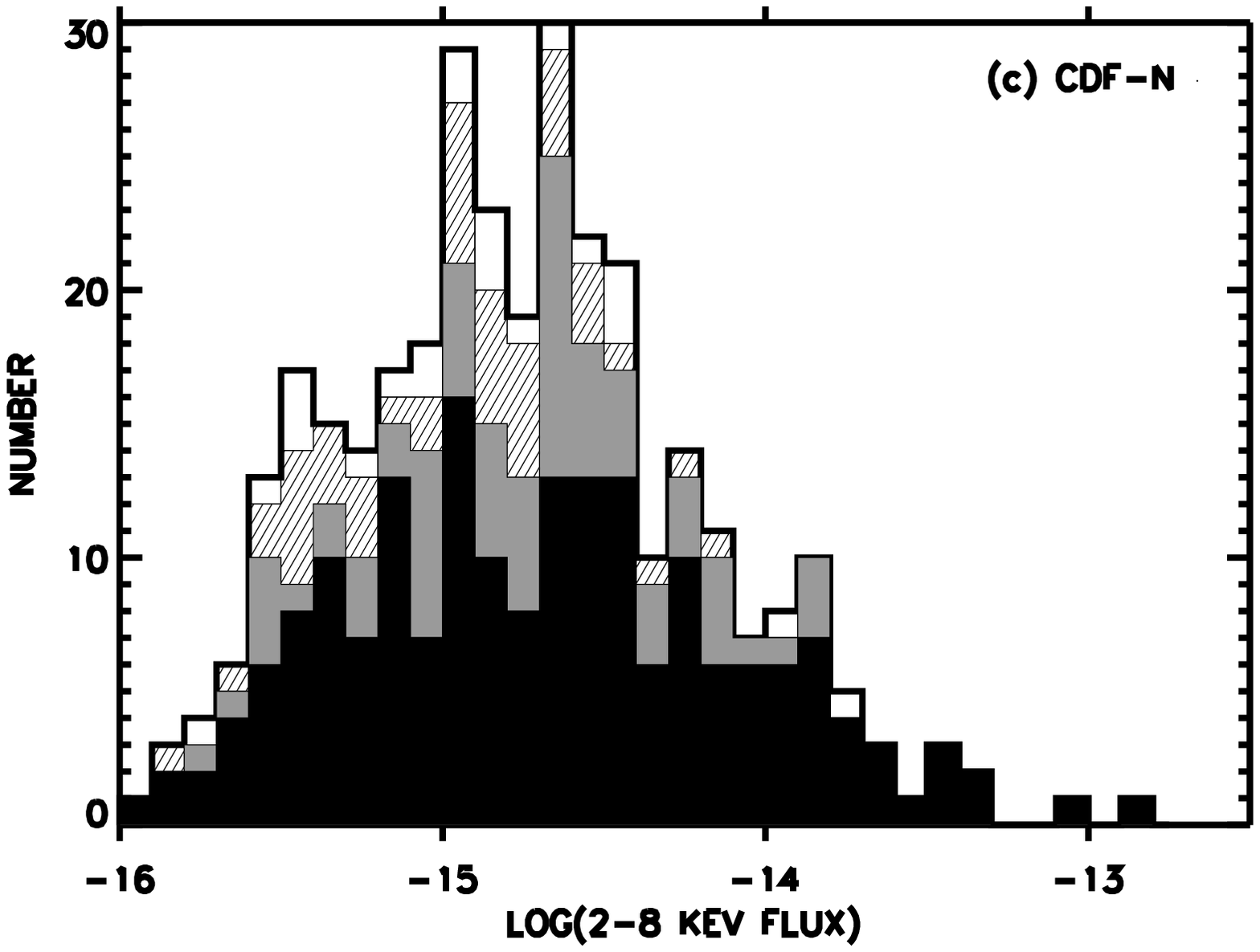}
\caption{$2-8~\rm{keV}$ flux distributions for the (a) CLANS, (b)
  CLASXS, and (c) CDF-N X-ray sources (\emph{filled}, spectroscopic
  redshifts; \emph{shading}, photometric redshifts; \emph{hatched},
spectroscopically observed but unidentified and no photometric
redshift determined; \emph{open}, spectroscopically
unobserved and no photometric redshift determined).}
\label{hardspect figure}
\end{figure}
%\clearpage

Figure \ref{redshift hist figure} shows the redshift distribution
of sources in the three fields with spectroscopic (\emph{shading}) and
photometric (\emph{open}) redshifts using a low-resolution ($\Delta z=0.1)$
binning. There are no obvious structures at $z<1$ in the CLANS and
CLASXS spectroscopic redshift distributions, unlike the apparent
excesses found at $z=0.48$ and $0.94$ in the CDF-N sample by Barger et
al.~(2003) and at $z=0.674$ and $0.734$ in the CDF-S sample by Gilli
et al.~(2003) and Szokoly et al.~(2004). There may be a grouping of
CLASXS sources at $z\sim1.2$. While there is some structure in the
redshift distributions for both the CLANS and CLASXS fields, the
differences between the redshift
distributions for those fields versus the CDFs show that
we have sampled a large enough area in the CLANS and CLASXS
fields to suffer less cosmic variance compared to the CDFs.

Figures \ref{softspect figure} and \ref{hardspect figure} show the
$0.5-2~{\rm keV}$ and $2-8~{\rm keV}$ flux distributions for the CLANS,
CLASXS, and CDF-N sources. The spectroscopically identified sources
(\emph{filled}), photometrically identified sources (\emph{shading}),
spectroscopically observed but neither spectroscopically 
nor photometrically identified sources (\emph{hatched}), and
spectroscopically unobserved and photometrically unidentified sources
(\emph{open}) are indicated. With photometric redshifts for more
than $50$\% of the spectroscopically
unidentified sources below $10^{-14}$ ergs~cm$^{-2}$~s$^{-1}$
($2-8~\rm keV$), we are able to determine $2-8~\rm keV$ X-ray luminosities
for $\sim75$\% of all the sources in the full sample (see \S \ref{Lx}).  

%\clearpage
\section{OPTICAL SPECTRAL CLASSIFICATION}
\label{optclass}

We have classified the spectroscopically identified \emph{Chandra}
sources into four optical spectral classes according to the
prescription in Barger et al.~(2005). We call
sources without any strong emission lines [EW([OII])$< 3$~\AA~or
EW(H$\alpha$ + NII) $< 10$~\AA] \emph{absorbers}; sources with strong
Balmer lines and no broad or high-ionization lines \emph{star
  formers}; sources with [NeV] or CIV lines or strong
[OIII] [EW([OIII] $\lambda5007) > 3$EW(H$\beta$)] \emph{high-excitation
sources}; and, finally, sources with optical lines having FWHM line
widths greater than 2000 km s$^{-1}$ \emph{broad-line AGNs}. See
Barger et al.~(2005) for a more detailed discussion of the use of 2000
km s$^{-1}$ as the dividing line for broad-line AGNs. 

In an effort to have our classifications for the sources in the three
fields be as uniformly determined as possible, several of the authors
repeated the classifications for all of the sources in all three
fields and came to a consensus. We note that there may be some
redshift bias in the classifications. Above $z\sim 0.65$, at which
point MgII enters the bandpass, we
can expect to have a uniform classification of broad-line AGNs, though
there may be a few anomolous cases. At lower redshifts, however, there
may be a number of sources classified as broad-line Seyferts which
would not be identified as broad-line AGNs at high redshifts. 

Table \ref{class table} gives the number of sources by optical
spectral type for the CLANS, CLASXS, and CDF-N samples.    

%\clearpage
\begin{table}
\begin{small}
\caption{Breakdown of the X-ray Samples by Field and Spectral Class}
\label{class table}
\begin{tabular}{l c c c c}
\tableline\tableline
Category & CLANS & CLASXS & CDF-N \\
\tableline
Total           & 761 & 525 & 503 \\
Observed        & 533 & 468  & 459 \\
Identified      & 336 & 280 & 312\tablenotemark{a} \\
Broad-line      & 126 & 103 & 39 \\
High-excitation & 92 & 57 & 42 \\ 
Star formers    & 87 & 77 & 146 \\
Absorbers       & 22 & 23 & 71 \\
Stars           & 9 & 20 & 14 \\
\tableline
\end{tabular}
\end{small}
\footnotesize
\tablenotetext{a}{Nine additional sources in the CDF-N field have spectroscopic redshifts from Swinbank et al.~(2004), Chapman et al.~(2005), and Reddy et al.~(2006) but no optical spectral classifications. We spectroscopically observed eight of the nine sources but were unable to determine identifications.}
\end{table}
%\clearpage

%\clearpage
%\pagebreak
\section{OPTICAL/IR COUNTERPARTS CATALOGS}
\label{catalogs}

\subsection{CLANS}

In Table \ref{CLANSopt table} we present the all new optical, NIR, and MIR
magnitudes and spectroscopic information for the CLANS X-ray point
source catalog. We ordered the 761 sources by increasing right
ascension and labeled each with the Table \ref{xray table} source
number (col [1]). Columns (2)
and (3) give the right ascension and declination coordinates of
the optical/NIR counterparts in decimal degrees. Columns (4)-(10)
provide the aperture-corrected, broadband $g', r', i', z', J, H,$ and
$K$ magnitudes. Columns (11) and (12) give the $3.6~\mu$m
and $24~\mu$m \emph{Spitzer} magnitudes. Column (13) lists the
spectroscopic redshifts, column (14) lists the photometric
redshifts determined using our template-fitting technique, and column
(15) lists the optical spectral classifications.  

\subsection{CLASXS}

In Table \ref{CLASXSopt table} we present the new and updated or
existing optical, NIR,
and MIR magnitudes and spectroscopic information for the CLASXS X-ray point
source catalog. For any
source with both CFHT and Subaru data in the $R$- and $z'$-bands, we used
the CFHT magnitude. We
ordered the 525 X-ray 
sources by increasing right ascension and labeled each with the Yang
et al.~(2004)
source number (col [1]). Columns (2) and (3) give the right ascension
and declination coordinates of the optical/NIR counterparts in decimal
degrees. Columns (4)-(13) provide the new and updated
aperture-corrected broadband $u, g', B, V, R, i', z',
J, H,$ and $K$ magnitudes. Columns (14) and (15) give the $3.6~\mu$m
and $24~\mu$m \emph{Spitzer} magnitudes. Column (16) lists the updated
spectroscopic redshifts, column (17) lists the
photometric redshifts determined using our template-fitting technique,
and column (18) lists the optical spectral classifications.   

\subsection{CDF-N}

In Table \ref{cdfnspect table} we give the optical, NIR, MIR, and
spectroscopic information for the CDF-N X-ray point source catalog. We
ordered the 503 sources by increasing right
ascension and labeled each with the Alexander et al.~(2003) Table 3a source
number (col [1]). Columns (2) and (3) provide the right ascension and
declination coordinates of the optical/NIR counterparts in decimal
degrees. Columns (4)-(9) provide the Barger et al.~(2003) $U,B,V,R,I,$
and $z'$ magnitudes (from the images of Capak et al.~2004). Columns
(10)-(12) give the new $J,H,$ and $K_s$ magnitudes. Columns (13) and
(14) give the $3.6~\mu$m and $24~\mu$m \emph{Spitzer} magnitudes. Column
(15) lists the updated spectroscopic redshifts, 
column (16) lists the photometric redshifts determined using our
template-fitting technique, and column (17) lists the optical spectral
classifications. 

%\pagebreak
\begin{table*}
\begin{scriptsize}
\caption{CLANS Optical/IR Counterparts Catalog}
\label{CLANSopt table}
\begin{tabular}{ccccccccccccccc}
\tableline\tableline
    \#&       RA&      DEC&   $g'$&   $r'$&   $i'$&   $z'$&   $J$&   $H$&   $K$&$3.6~\mu$m& $24~\mu$m&z$_{spec}$&z$_{phot}$& class\\
(1) & (2) & (3) & (4) & (5) & (6) & (7) & (8) & (9)& (10)&(11) & (12) & (13) & (14) & (15) \\
\tableline
     1&  -99.000&  -99.000&-99.00&-99.00&-99.00&-99.00&-99.00&-99.00&-99.00&-99.00&-99.00&    -1.00&-99.00&   -1\\
     2&  160.642&   59.176& 23.04& 22.40& 21.91& 21.78& 21.02& 19.80&-99.00& 19.50&-99.00&     1.43& -3.00&    4\\
     3&  160.642&   59.169& 25.43& 25.04& 24.40& 23.71& 22.53& 21.93&-99.00&-99.00&-99.00&    -1.00&  1.35&   -1\\
     4&  -99.000&  -99.000&-99.00&-99.00&-99.00&-99.00&-99.00&-99.00&-99.00&-99.00&-99.00&    -1.00&-99.00&   -1\\
     5&  -99.000&  -99.000&-99.00&-99.00&-99.00&-99.00&-99.00&-99.00&-99.00& 20.87&-99.00&    -1.00&-99.00&   -1\\
    ..&       ..&       ..&    ..&    ..&    ..&    ..&    ..&    ..&    ..&    ..&    ..&       ..&    ..&    ..\\
\tableline
\end{tabular}
\end{scriptsize}
\footnotesize
\tablecomments{Table 11 is available in its entirety in the electronic edition of the \emph{Astrophysical Journal Supplement}. A portion is shown here for guidance regarding its form and content. All magnitudes are in AB magnitudes.\\
Typical photometric uncertainties are given in \S \ref{photodet}.\\
Magnitude $=-99$, source not detected ($2\sigma$ significance). \\
$z_{spec} = 0$ and corresponding class $= -99$, source spectroscopically observed but neither the redshift nor the class could be identified. \\
$z_{spec} = -1$ and corresponding class $= -1$, source not yet spectroscopically observed. \\
$z_{spec} = -2$ and corresponding class $= -2$, source is a star. \\
$z_{phot} = -3$, source has a spectroscopic redshift.\\
$z_{phot} = -99$, source has neither a spectroscopic nor a photometric redshift. \\
class $=0$, absorbers; class $=1$, star formers; class $=3$, high-excitation sources; class $=4$, broad-line AGNs.}
\end{table*}

%{\rotate
\begin{table*}
\begin{tiny}
\caption{CLASXS Optical/IR Counterparts Catalog}
\label{CLASXSopt table}
\begin{tabular}{cccccccccccccccccc}
\tableline\tableline
    \#&       RA&      DEC&   $u$&   $B$&   $g'$&   $V$&   $R$&   $i'$&   $z'$&   $J$&   $H$&   $K$&$3.6~\mu$m& $24~\mu$m&z$_{spec}$&z$_{phot}$& class\\
(1) & (2) & (3) & (4) & (5) & (6) & (7) & (8) & (9)& (10)&(11) & (12) & (13) & (14) & (15) & (16) & (17) & (18)\\
\tableline
     1&  157.731&   57.556& 24.33&-99.00& 23.69&-99.00& 22.44& 21.91&-99.00& 20.17& 20.30&-99.00& -1.00&-99.00&    -1.00$^s $&  0.55&   -1\\
     2&  157.748&   57.646&-99.00&-99.00& 23.74&-99.00& 21.97& 21.44&-99.00& 20.44& 20.17&-99.00& -1.00& 17.84&    -1.00$^s $&  0.45&   -1\\
     3&  157.764&   57.614& 23.96&-99.00&-99.00&-99.00& 22.80& 22.65&-99.00& 22.06& 21.88&-99.00& -1.00&-99.00&     0.00$^s $&  0.25&  -99\\
     4&  157.774&   57.630& 23.80&-99.00& 23.78&-99.00& 23.04& 22.96& 23.05& 22.56&-99.00&-99.00& -1.00&-99.00&     0.00$^s $&-99.00&  -99\\
     5&  157.800&   57.590& 25.29& 24.51& 24.27&-99.00& 23.24& 22.71& 22.39& 21.94& 21.76&-99.00& -1.00&-99.00&     0.00$^s $&  0.35&  -99\\
    ..&       ..&       ..&    ..&    ..&    ..&    ..&    ..&    ..&    ..&    ..&    ..&    ..&    ..&    ..&       ..&    ..&    ..\\
\tableline
\end{tabular}
\end{tiny}
\footnotesize
\tablecomments{Table 12 is available in its entirety in the electronic edition of the \emph{Astrophysical Journal Supplement}. A portion is shown here for guidance regarding its form and content. All magnitudes are in AB magnitudes.\\
Typical photometric uncertainties are given in \S \ref{photodet}.\\
Magnitude $=-99$, source not detected ($2\sigma$ significance). \\
 z$_{spec} = 0$ and corresponding class $= -99$, source spectroscopically observed but neither the redshift nor the class could be identified. \\
z$_{spec} = -1$ and corresponding class $= -1$, source not yet spectroscopically observed. \\
z$_{spec} = -2$ and corresponding class $= -2$, source is a star. \\
$z_{phot} = -3$, source has a spectroscopic redshift.\\
$z_{phot} = -99$, source has neither a spectroscopic nor a photometric redshift. \\
class $=0$, absorbers; class $=1$, star formers; class $=3$, high-excitation sources; class $=4$, broad-line AGNs.\\
Reference for z$_{spec}$: s=Steffen et al.~(2004). All other spectroscopic redshifts presented here for the first time.}
\end{table*}
%}
%\clearpage
\begin{table*}
\begin{scriptsize}
\caption{CDF-N Optical/IR Counterparts Catalog}
\label{cdfnspect table}
\begin{tabular}{ccccccccccccccccc}
\tableline\tableline
    \#& RA&DEC&       $U$&       $B$&       $V$&       $R$&       $I$&$z^{\prime}$&       $J$&       $H$&     $K_s$&3.6$~\mu$m& 24$~\mu$m&z$_{spec}$&z$_{phot}$& class\\
(1) & (2) & (3) & (4) & (5) & (6) & (7) & (8) & (9)& (10)&(11) & (12) & (13) & (14) & (15) & (16) & (17)\\
\tableline
     1&   188.813&    62.235&      26.4&      25.7&      25.3&      25.1&      25.2&      24.8&     -99.0&     -99.0&     23.12&    -99.00&    -99.00&     -1.00  &    -99.00&   -1\\
     2&   188.820&    62.261&      25.4&      24.9&      24.9&      24.5&      24.7&      24.5&     -99.0&      22.1&     21.41&    -99.00&    -99.00&      0.00$^a $&    -99.00&  -99\\
     3&   188.828&    62.264&      21.3&      20.2&      19.1&      18.6&      18.2&      18.0&      17.0&      16.7&     16.62&    -99.00&    -99.00&      0.14$^a $&     -3.00&    0\\
     4&   188.831&    62.228&      27.2&      26.7&      25.6&      24.4&      23.6&      23.2&     -99.0&      22.3&     21.70&    -99.00&    -99.00&     -1.00  &      0.65&   -1\\
     5&   188.839&    62.274&      23.2&      22.8&      22.4&      21.8&      21.4&      21.2&      20.6&      20.2&     20.06&    -99.00&    -99.00&      0.56$^a $&     -3.00&    1\\
    ..& ..& ..&        ..&        ..&        ..&        ..&        ..&        ..&        ..&        ..&        ..&        ..&        ..&        ..&        ..&   ..\\
\tableline
\end{tabular}
\end{scriptsize}
\footnotesize
\tablecomments{Table 13 is available in its entirety in the electronic edition of the \emph{Astrophysical Journal Supplement}. A portion is shown here for guidance regarding its form and content. All magnitudes are in AB magnitudes.\\
Typical photometric uncertainties are given in \S \ref{photodet}.\\
Magnitude $=-99$, source not detected ($2\sigma$ significance). \\
z$_{spec} = 0$ and corresponding class $= -99$, source spectroscopically observed but neither the redshift nor the class could be identified. \\
z$_{spec} = -1$ and corresponding class $= -1$, source not yet spectroscopically observed. \\
z$_{spec}>0$ and class $= 10$, spectroscopic redshift from the literature without a corresponding spectrum.\\
z$_{spec} = -2$ and corresponding class $= -2$, source is a star. \\
$z_{phot} = -3$, source has a spectroscopic redshift.\\
$z_{phot} = -99$, source has neither a spectroscopic nor a photometric redshift. \\
class $=0$, absorbers; class $=1$, star formers; class $=3$, high-excitation sources; class $=4$, broad-line AGNs.\\
Reference for z$_{spec}$: a=Barger et al.~(2003), b=Swinbank et al.~(2004), c=Chapman et al.~(2005), d=Reddy et al.~(2006). All other spectroscopic redshifts presented here for the first time.}
\end{table*}

%\clearpage
%see tab11.tex, tab12.tex, and tab13.tex for the full version of these tables (to be included in the electronic edition). 

%having trouble rotating these tables in this preprint aastex format. i do realize tab11 and tab12, in particular, are too long to be placed horizontally on the page as they are now.

\pagebreak
\clearpage
\section{X-ray Luminosities}
\label{Lx}

%\clearpage
\begin{figure}
\epsscale{2.2}
\plottwo{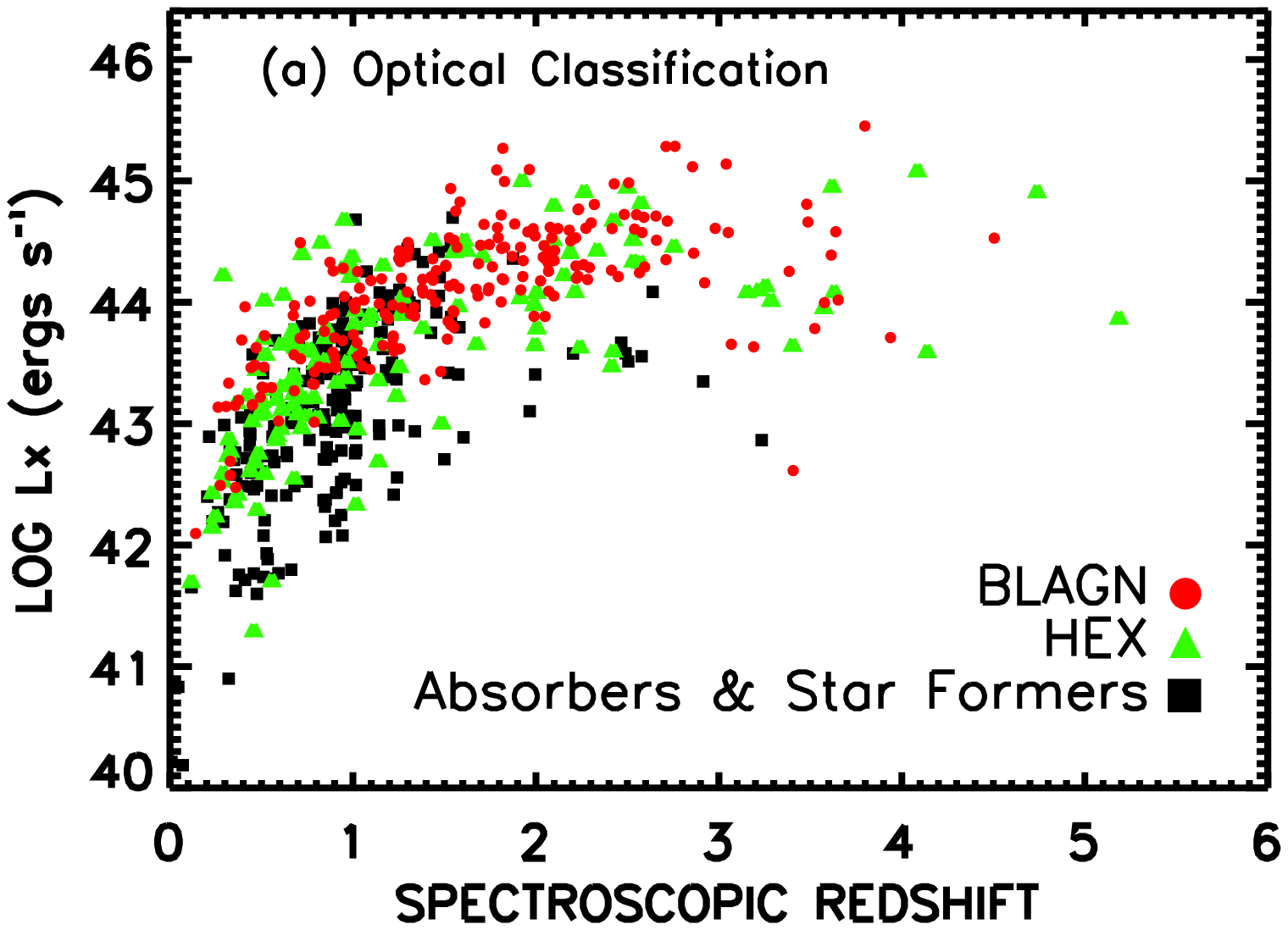}{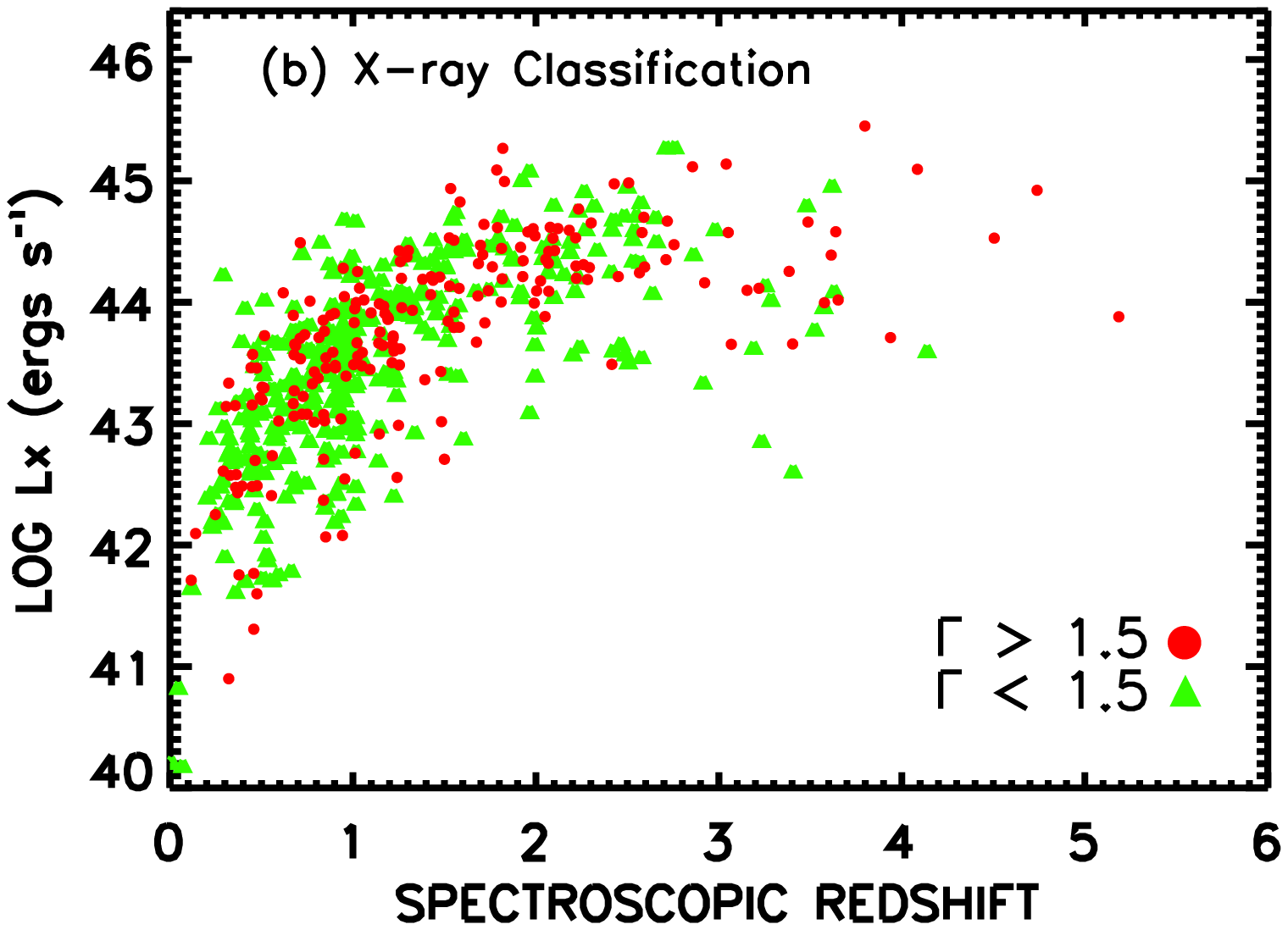}
\caption{(a) Rest-frame $2-8~\rm keV$ X-ray luminosity versus redshift for the
  CLANS, CLASXS, and CDF-N sources according to their optical
  classification (\emph{red circles}, broad-line AGNs; \emph{green
    triangles}, high-excitation narrow-line sources [HEX]; \emph{black
    squares}, absorbers and star formers). We only include sources with
  spectroscopic redshifts, $L_X>10^{40}$ ergs~s$^{-1}$, and an X-ray detection
  significance greater than $3\sigma$. (b) Same as (a) but now
  classifying the sources according to their
  X-ray spectral properties (\emph{red circles}, $\Gamma>1.5$; \emph{green
    triangles}, $\Gamma<1.5$).} 
\label{lx_z opt}
\end{figure}
%\clearpage

We used the X-ray fluxes and redshifts of our sources to
calculate their rest-frame $2-8~\rm keV$ X-ray luminosities. We limited our
sample to sources with an X-ray detection significance greater than
$3\sigma$. For the CLANS sources we determined the $3\sigma$
significance level using the $1\sigma$ error bars on the $2-8~\rm keV$
and $0.5-2~\rm keV$ flux values listed in
Table \ref{xcat table1}. For the CLASXS and
CDF-N sources we used the $1\sigma$ error bars from Yang et
al.~(2004) and Alexander et al.~(2003), respectively. 

At $z<3$, we calculated the
luminosities from the observed-frame $2-8~\rm keV$ fluxes,
and at $z>3$, we used the observed-frame
$0.5-2$~keV fluxes. One advantage of using the observed-frame $0.5-2~\rm keV$
X-ray fluxes at high redshifts is the increased sensitivity, since the
$0.5-2~\rm keV$ \emph{Chandra} images are deeper than the $2-8~\rm keV$
images. In addition, at $z=3$, observed-frame $0.5-2~\rm keV$ corresponds
to rest-frame $2-8~\rm keV$, providing the best possible match to the
lower redshift data. 

In calculating the $K-$corrections for obtaining the rest-frame
$2-8~\rm keV$ luminosities, we
assumed an intrinsic $\Gamma=1.8$. Since the $K-$correction factor for
emission modeled by a power law follows the form $[1+z]^{\Gamma - 2}$,
these corrections are very small. If we instead assumed the individual
photon indices used in the flux calculations (rather than the
universal power-law index of $\Gamma=1.8$ adopted here) to calculate
the $K-$corrections, we would find only a small difference in the
rest-frame luminosities (Barger et al.~2002). 

%
%06-09-08: the following split equation will not work without amsmath and/or amsthm which are not allowed. So I have tried to recreate it the best I could without needing these packages. If the editor can make the equation resemble even more closely the original intention, I would be very grateful (although the current form is acceptable). Thank you!
%
%In short, the rest-frame $2-8~\rm keV$ luminosity
%\begin{equation}
%L_X = f \times 4\pi d_L^2 \times K-\text{correction using} 
%\begin{cases}
%K-\text{correction}=(1+z)^{-0.2} & \text{$z<3$} \\
%f=f_{2-8~\text{keV}}   
%\\
%K-\text{correction}=\frac{1}{4}(1+z)]^{-0.2} & \text{$z\ge3$} \\
%f=f_{0.5-2~\text{keV}}
%\end{cases}
%\end{equation}
%
In short, the rest-frame $2-8~\rm keV$ luminosity, $L_X$, equals $f \times 4\pi d_L^2 \times K-$correction.
\begin{equation}
\hspace{1.5 mm} z<3: \hspace{2 mm} K-{\rm corr.}=(1+z)^{-0.2} \hspace{2 mm} {\rm \&} \hspace{2 mm} f=f_{2-8~{\rm keV}} 
\end{equation}
\begin{equation}
z\ge3: \hspace{2 mm}  K-{\rm corr.}=\frac{1}{4}(1+z)^{-0.2} \hspace{2 mm} {\rm \&} \hspace{2 mm} f=f_{0.5-2~{\rm keV}} 
\end{equation}

The $\frac{1}{4}$ factor in the $z \ge 3$
$K-$correction is a result of normalizing so that there is no
$K-$correction if $z=3$, when the observed-frame $0.5-2$~keV corresponds
exactly to the rest-frame $2-8$~keV.  

Figure \ref{lx_z opt}a shows rest-frame $2-8~\rm keV$ X-ray luminosity versus
spectroscopic redshift for the CLANS, CLASXS, and CDF-N X-ray
sources split by their optical spectral classification. This figure illustrates
the Steffen effect, in which the broad-line AGNs dominate the number densities
at higher X-ray luminosities, while the non-broad-line AGNs dominate at
lower X-ray luminosities (Steffen et al.~2003).   

Figure \ref{lx_z opt}b shows the same relation, but now the sources
in the three fields are split by their X-ray classifications. For the
CLANS and CLASXS X-ray sources, we converted the individual hardness
ratios, HR, into $\Gamma$ values, as described in \S \ref{hr}. For the
CDF-N X-ray sources, we used the Alexander et al.~(2003) values for
$\Gamma$. We note that $\sim 1/4$ of the sources
in Figure \ref{lx_z opt}b are within $1\sigma$ of $\Gamma=1.5$. 

Figures \ref{lx_z opt}a and \ref{lx_z opt}b demonstrate that while the
optical and
X-ray classifications show generally the same broad differentiation
between the classes, there is clearly a different scatter or mixing of
the objects, which reflects the properties of each individual
source. In Yencho et al.~(2008) we present the X-ray luminosity functions by spectral
type, and in L.~Trouille et al.~(2008, in preparation) we present a
detailed analysis of the optical and X-ray spectral characteristics of
the individual objects in our sample.  

\section{NUMBER COUNTS}
\label{numcts sec}

In order to determine the amount of field to field variation within
our sample and to compare our survey with those by others, we have
determined the differential number counts for each individual field
and for the full sample.
 
In the following, we have excluded any CLANS or
CLASXS sources with off-axis angles greater than $8\arcmin$ because
the sensitivity of
\emph{Chandra} drops significantly at large off-axis angles and
because of the field overlap in the CLASXS region. Using
the Yang et al.~(2006) simulations (see \S \ref{fluxlimsec}), we
have also excluded any
CLANS and CLASXS sources with fluxes less than the flux
threshold for the source's particular off-axis angle location and
pointing exposure time (see Figure \ref{sims}). In this section, we
use only the flux thresholds and $\Omega(S_i)$ values for a 30\%
probability of detection. See \S
\ref{effarea} and Figure \ref{fluxarea} for a discussion of the
$\Omega(S_i)$ values used for the CLANS and CLASXS fields. 

The Poisson fluctuations
in the source fluxes could result in an overestimation of the number
counts close to the detection limits. This is known as the Eddington
bias. For the CLANS and CLASXS fields, the detection threshold is
below the `knee' of the log$N-$log$S$ relation, and the Eddington bias
is relatively small. We have not included the Eddington bias
correction in our determination of the number counts for these fields
(including the Eddington bias correction from the Yang et al.~2004
simulations does not change the results significantly). 

We used the Alexander et al.~(2003) $\Omega(S_i)$ values (see
their Figure 19) and flux threshold values for the CDF-N field. We
also applied the Bauer et al.~(2004) recovered flux 
corrections (see their Figure 2) to account for the few percent
increase at all fluxes due to an additional aperture correction not
originally accounted for in Alexander et al.~(2003), an increase
for faint sources due to photometry errors (see their Figure 3), and a
decrease for faint sources due to the Eddington bias. In the following,
we have excluded any CDF-N sources with off-axis angles greater than
$10\arcmin$ as well as any CDF-N sources with fluxes less than the
flux threshold for the source's particular off-axis angle location.   

We computed the differential log$N-$log$S$ relations for the full and
individual samples using the formula  
\begin{equation}
\frac{dN}{dS}=\sum_{S_1< S_i <S_2}
\frac{1}{\Omega_{tot}(S_i) \times  \Delta S}
\end{equation} 
in units of deg$^{-2}$ per 10$^{-15}$ ergs~cm$^{-2}$~s$^{-1}$. $S_1$
and $S_2$ are the minimum and maximum fluxes in each bin. $\Delta
S = S_2 - S_1$. We estimated the error bars in the number counts by the error
propagation rule using Gehrels (1986) statistics. 

\begin{figure}
\epsscale{2.2}
\plottwo{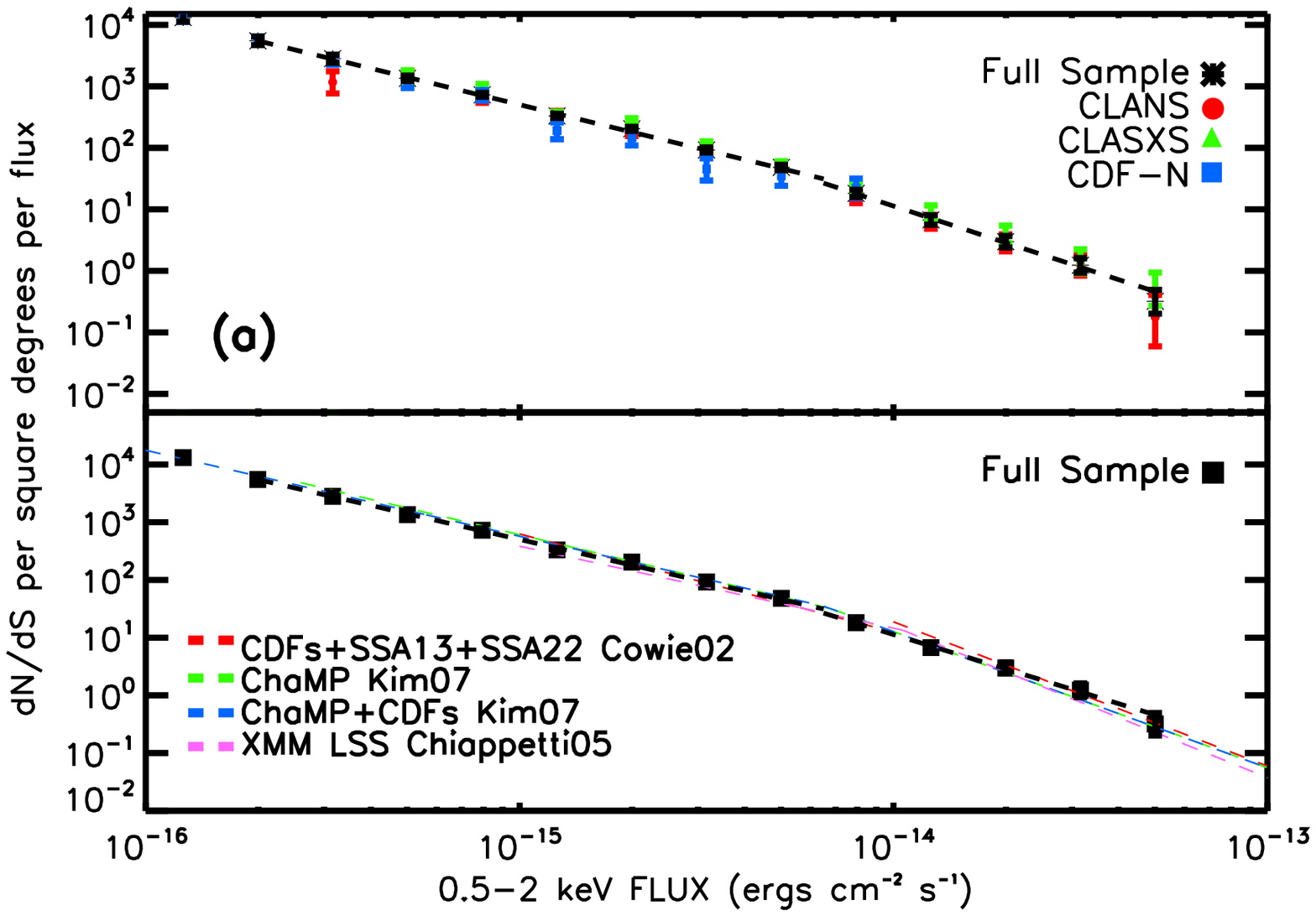}{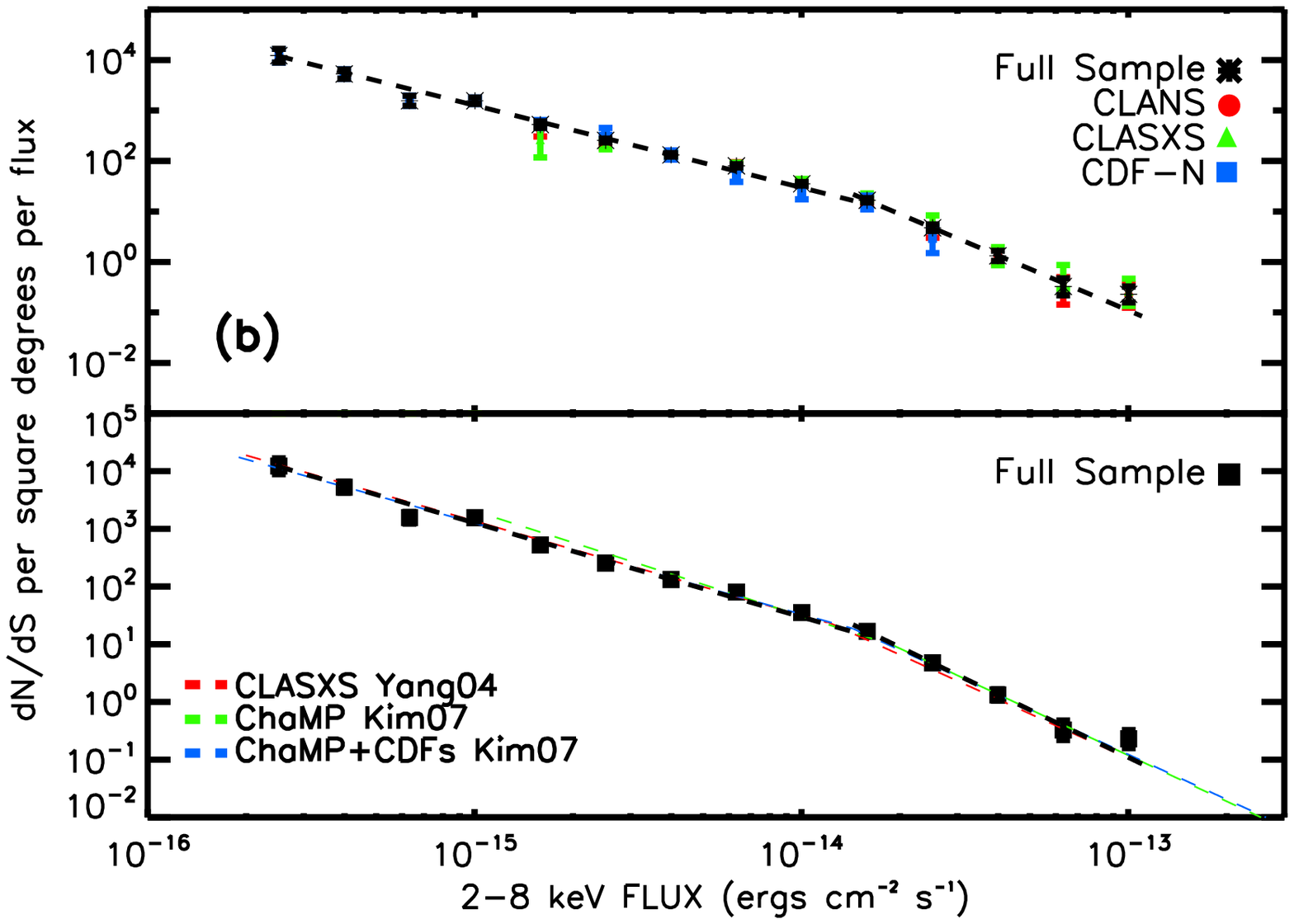}
\caption{(a) $0.5-2~\rm keV$ band differential number counts
  (\emph{black asterix}, full sample; \emph{red circles}, CLANS;
  \emph{green triangles}, CLASXS; \emph{blue squares}, CDF-N; \emph{black
  dashed line}, best-fit to the full sample). Only results for bins
with at least four sources are shown. Bottom plot in (a) shows
the full sample \emph{(black squares)} overplotted with results
from other surveys, as indicated. (b) Same as (a) but for the $2-8~\rm keV$
band differential number counts.} 
\label{diffcts1}
\end{figure}

Figure \ref{diffcts1} shows the (a) $0.5-2~\rm keV$ and (b) $2-8~\rm keV$
differential number counts for the
full sample, as well as for each field individually. The bottom plots
compare the differential number counts for our full sample with those
from other surveys. We parametrized our best-fits to the full
sample, shown as black dashed lines in the plots, by means of a broken power
law of the form
%
%06-09-08: the following split equation will not work without amsmath and/or amsthm which are not allowed. So I have tried to recreate it the best I could without needing these packages. If the editor can make the equation resemble even more closely the original intention, I would be very grateful (although the current form is acceptable). Thank you!
%
%\begin{equation}
%dN/dS=
%     \begin{cases}
%     n_{0,faint}(S/10^{-14})^{-\alpha_{faint}} & \text{(S $< S_{break}$)}
%\\
%     n_{0,bright}(S/10^{-14})^{-\alpha_{bright}} &  \text{(S $> S_{break}$)}
%     \end{cases}
%\end{equation}
%
\begin{equation}
dN/dS= n_{0,faint}(S/10^{-14})^{-\alpha_{faint}}  \hspace{5 mm}  (S < S_{break})\\
\end{equation}
\begin{equation}
\hspace{12 mm}= n_{0,bright}(S/10^{-14})^{-\alpha_{bright}} \hspace{4 mm}  (S > S_{break})
\end{equation}

We determined the best-fits using error-weighted least-squares
fits. Table \ref{powerlaw} shows the best-fit parameters. The quoted
errors are 1$\sigma$ formal errors on the fits. 

\begin{figure*}
\epsscale{1}
\plottwo{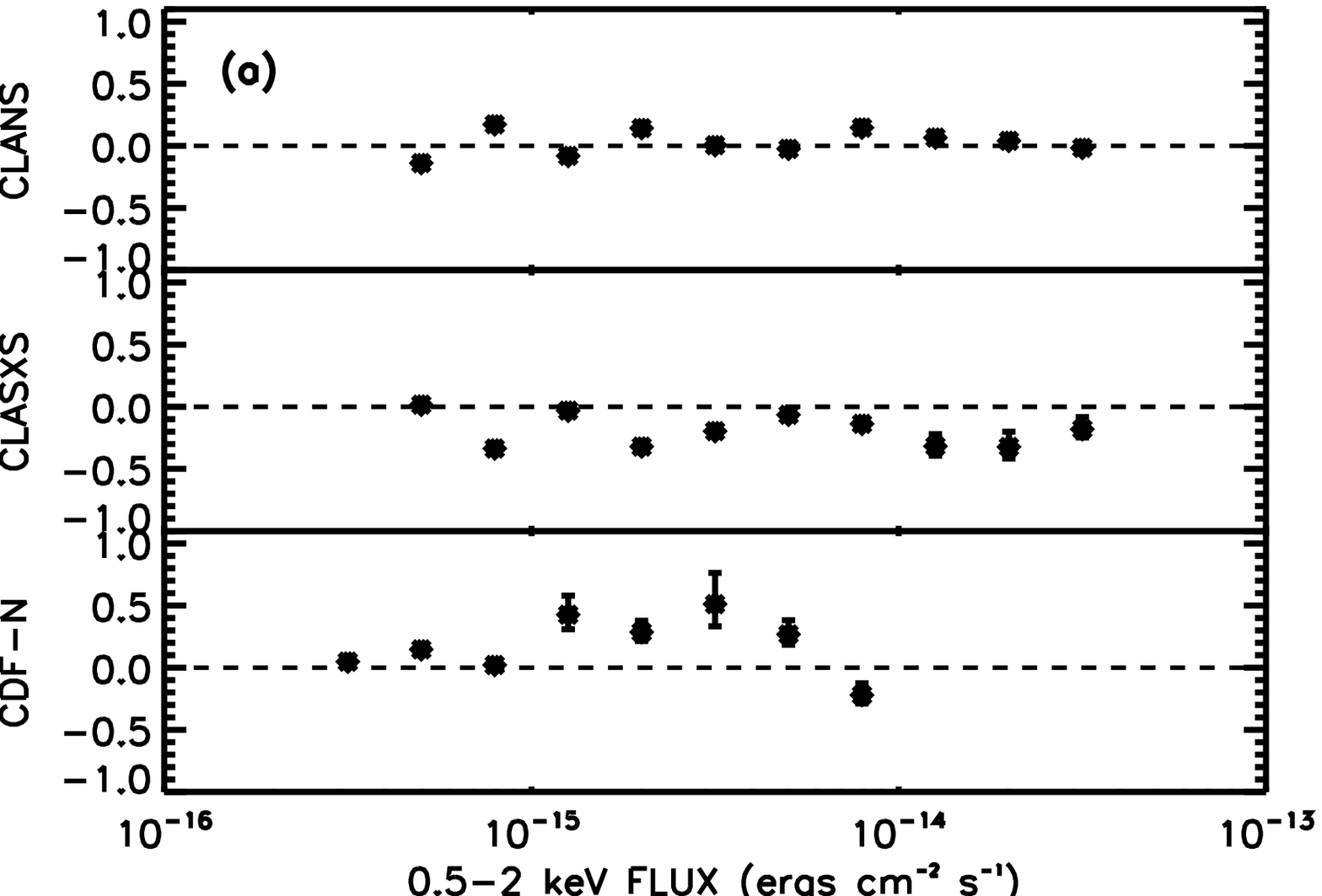}{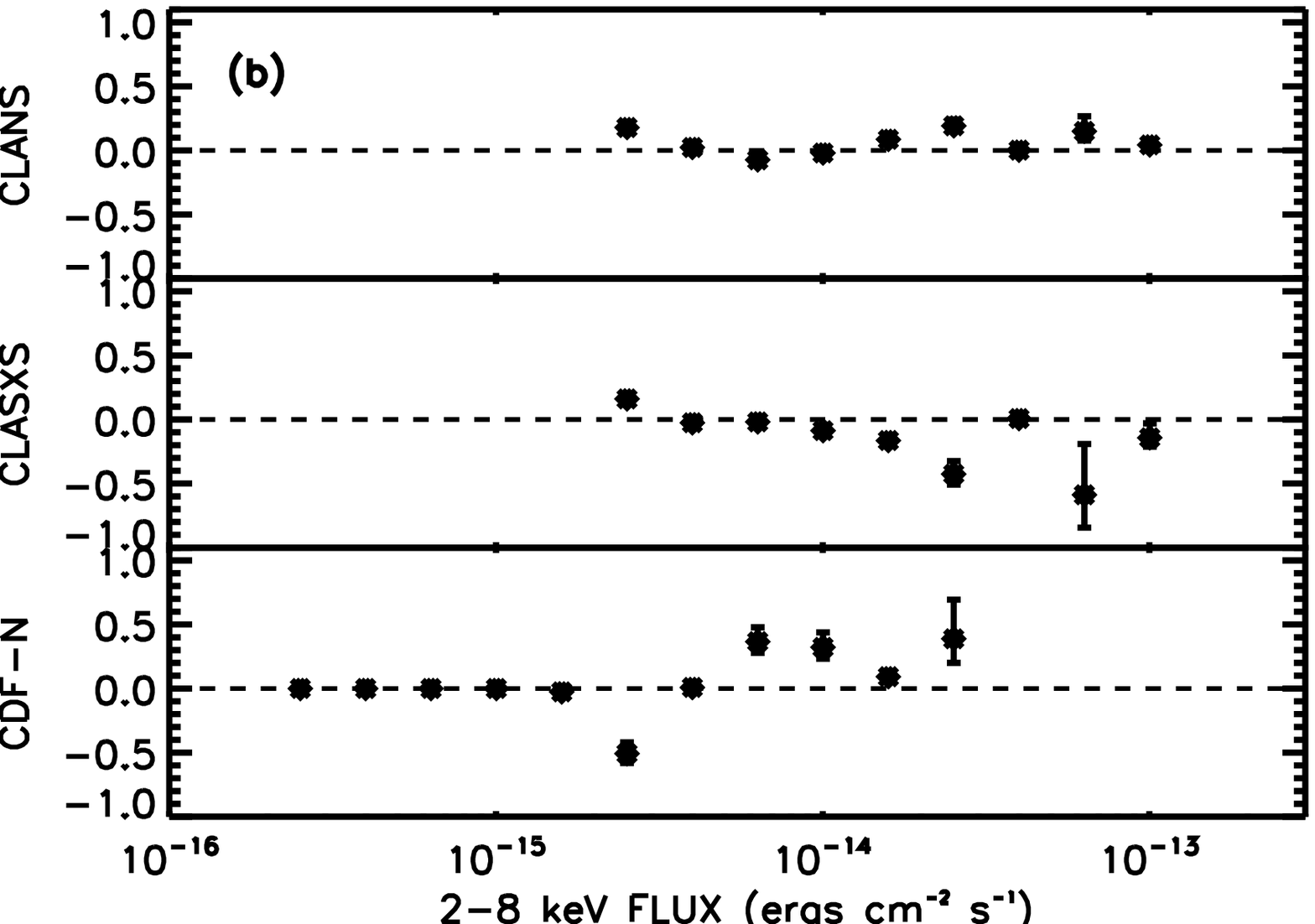}
\caption{$0.5-2~\rm keV$ and $2-8~\rm keV$ band differential number count residuals,
  defined as 1 minus the ratio of $dN/dS$ for the individual
sample and $dN/dS$ for the full sample, for the CLANS, CLASXS, and
CDF-N fields. Only results for bins with at least four sources are shown.} 
\label{diffcts div}
\end{figure*}

Figure \ref{diffcts div} shows the (a) $0.5-2~\rm keV$ and (b)
$2-8~\rm keV$ differential number counts fractional residuals for each field
separately, defined as 1 minus the ratio of $dN/dS$ for the individual
sample and $dN/dS$ for the full sample. The fields agree reasonably
well, exhibiting a variance of $<10\%$ in almost all of the $0.5-2~\rm
keV$ and $2-8~\rm keV$ flux bins.   

\begin{figure*}
\epsscale{1}
\plottwo{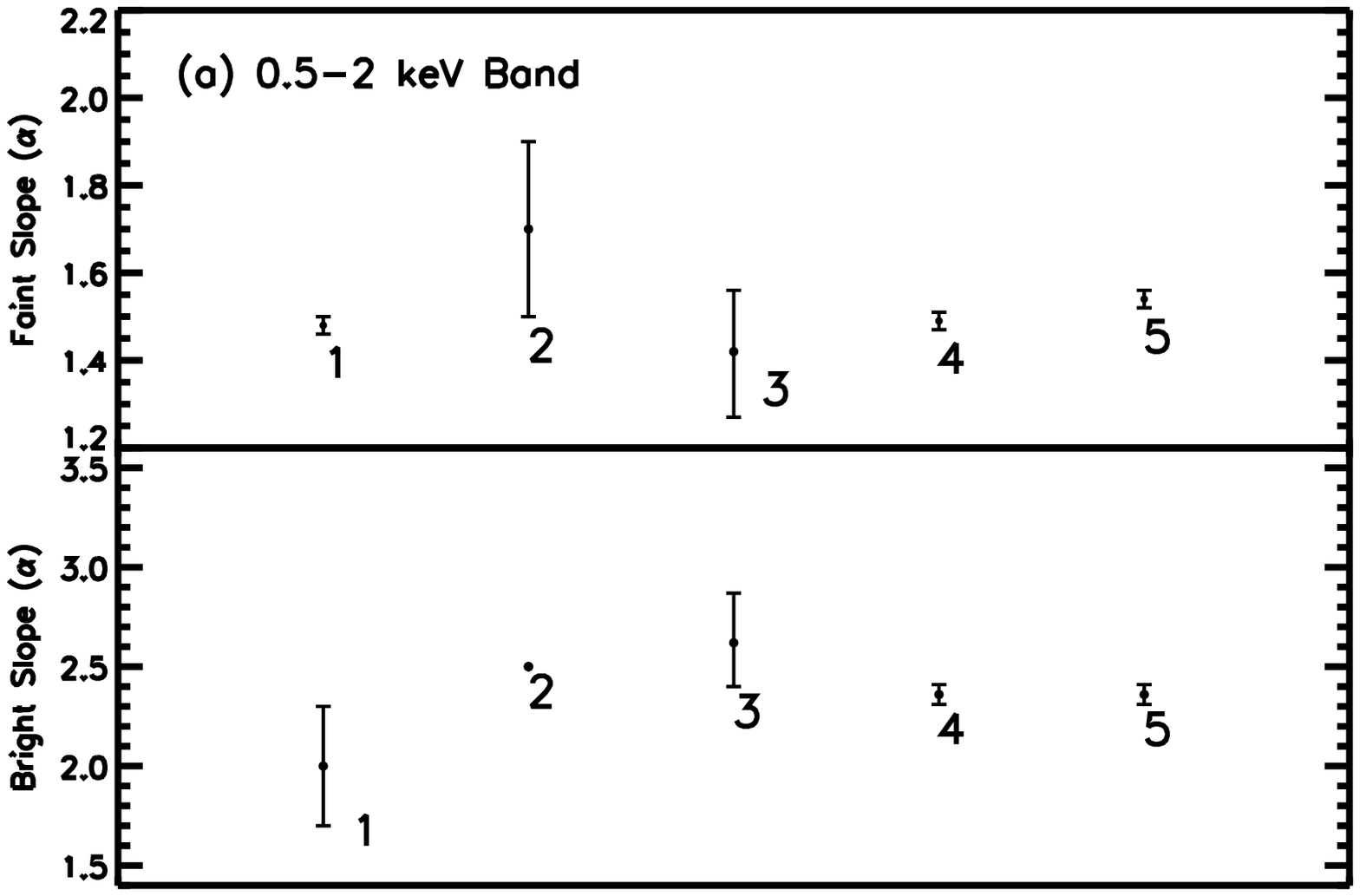}{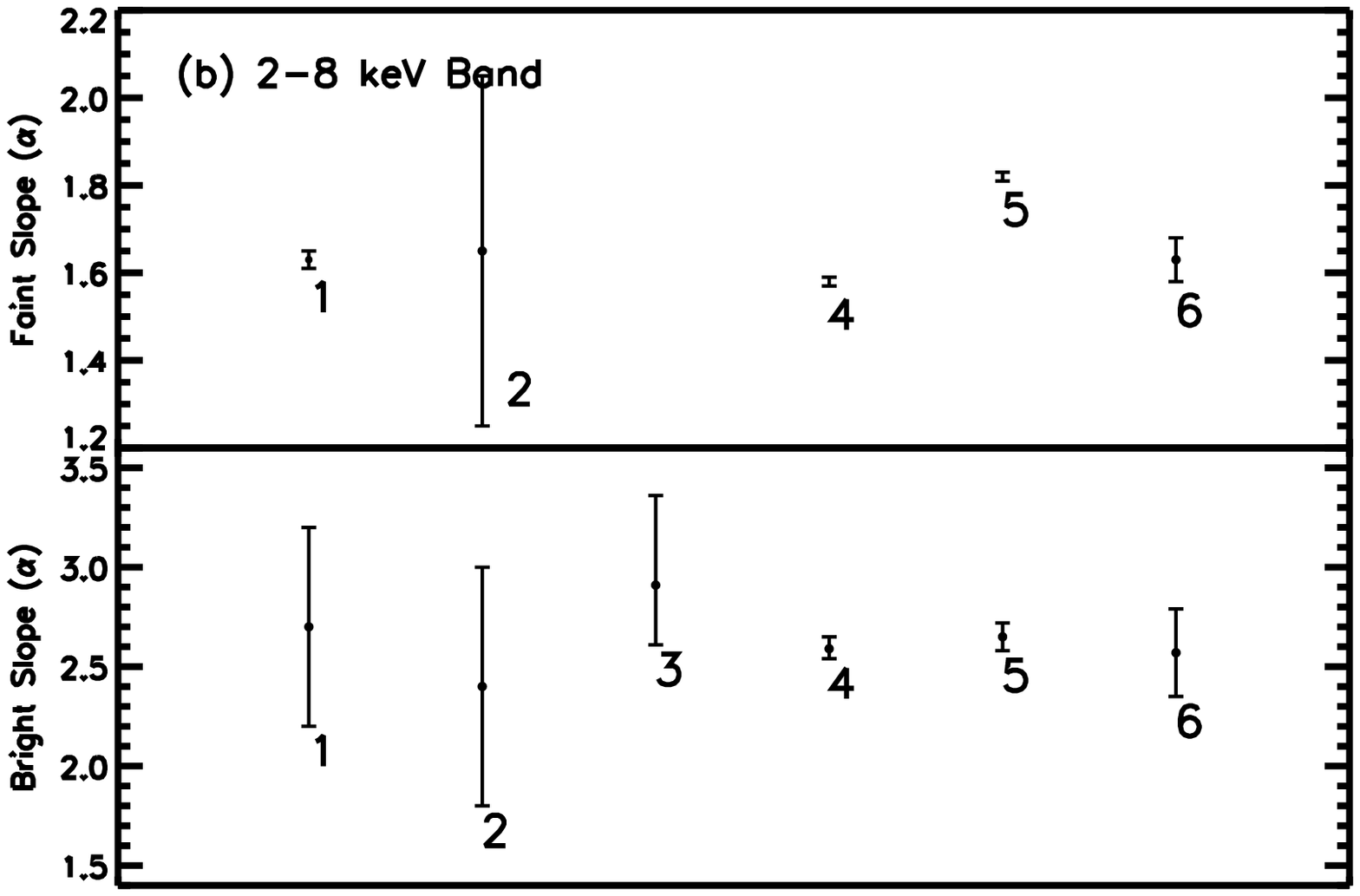}
\caption{ (\emph{Top}) Faint and (\emph{bottom}) bright power indices
  of the differential number counts for this and previous studies in
  (a) the $0.5-2~\rm keV$ band and (b) the $2-8~\rm keV$ band. The
  references are as follows: (1) this study; (2) Yang et al.~(2004) CLASXS; (3)
  Chiappetti et al.~(2005) XMM LSS; (4) Kim et al.~(2007b) ChaMP+CDFs; (5)
  Kim et al.~(2007b) ChaMP; (6) Cowie et al.~(2002)
  CDFs+SSA13+SSA22. Note that Yang et al.~(2004) fixed the bright
  slope as 2.5 in the $0.5-2~\rm keV$ band, so there is no error.}  
\label{compdiffcts}
\end{figure*}

In Figure \ref{compdiffcts} we directly compare the best-fit slopes
for the (a) $0.5-2~\rm keV$ and (b) $2-8~\rm keV$
faint and bright ends of our differential number counts with those of
other surveys (see Kim et al.~2007b for an exhaustive
comparison of differential number counts determinations
from surveys to date). The best-fit slopes of $1.48\pm0.02$ and $1.64\pm0.02$,
respectively, for the faint end of our full sample $0.5-2~\rm keV$ and
$2-8~\rm keV$ band
differential number counts agree well with other surveys. However, we note that
our results at the faint end are not independent from the Kim et
al.~(2007b) and Cowie et al.~(2002) results (points 4 and
6 in Figure \ref{compdiffcts}), since these also include the CDF-N. At the
bright end, our uncertainties are larger. However, the $0.5-2~\rm keV$ band
differential number counts best-fit slope of $2.0\pm0.3$ is within
the errors of the
other surveys (note that the Yang et al.~2004 slope was fixed
at 2.5 with no error). The actual value of the slope is
slightly shallower than that of the other surveys. In the $2-8~\rm
keV$ band, the best-fit slope of
$2.7\pm0.5$ agrees reasonably well with the results from other
surveys.

%\clearpage

%\clearpage
%\pagebreak
\begin{table*}
\begin{small}
\centering
\caption{Broken Power Law Best-fit Parameters for the Full Sample}
\label{powerlaw}
\begin{tabular}{c|cc|c|cc}
\tableline\tableline 
Diff. Num. Counts        &    Faint End       &    &  Break Flux   &  Bright End& \\
& $n_{0}$ & $\alpha$ & & $n_{0}$ & $\alpha$\\
\tableline
$0.5-2~\rm keV$ & $16.5\pm1.2$ &$1.48\pm0.02$ &$6.5\times10^{-15}$ & $11.3\pm0.4$ & $2.0\pm0.3$\\
$2-8~\rm keV$  & $29.7\pm1.2$ &$1.64\pm0.02$ &$1.44\times10^{-14}$ & $59.0\pm27.0$ &$2.7\pm0.5$\\
\tableline
\end{tabular}
\end{small}
\end{table*}
%\clearpage

\clearpage
\section{SUMMARY}

In this paper, we compiled the database for the OPTX project, which
combines data from the CLANS, CLASXS, and CDF-N fields to create one
of the most spectroscopically complete samples of \emph{Chandra} X-ray
sources to date. With this database, we can analyze the effect of
spectral type on the shape and evolution of the X-ray luminosity
functions and compare the optical and X-ray spectral properties of the
X-ray sources in our sample.  

In detail, we presented the first
X-ray, infrared, and optical photometric and
spectroscopic catalogs for the CLANS field. The CLANS X-ray survey
covers 0.6 deg$^2$ and 
reaches fluxes of $7\times 10^{-16}$ ergs~cm$^{-2}$~s$^{-1}$ in the
$0.5-2~\rm keV$ band and $3.5\times 10^{-15}$ ergs~cm$^{-2}$~s$^{-1}$ in
the $2-8~\rm keV$ band. We presented $g',r',i',z',J,H,K,3.6~\mu$m, and
$24~\mu$m photometry for the 761 X-ray sources in the sample. We
spectroscopically observed 533 of the CLANS sources, obtaining
redshift identifications for 336. We extended the redshift information
to fainter magnitudes using photometric redshifts, which resulted in an
additional 234 redshifts.  

We also presented new and updated CLASXS photometry and some new
spectroscopy, along with existing data. Using new $u, g',$ and $i'$
magnitudes, we corrected the zeropoints for the original Steffen et
al.~(2004) $V,
R, I,$ and $z'$ photometry (the $B$-band zeropoint was fine). We also
obtained $J, H, K, 3.6~\mu$m, and $24~\mu$m
photometry for the 525 X-ray sources in the sample. Since the
publication of the Steffen et al.~(2004) redshift catalog for this
field, we have obtained an additional 11 spectra and identified
redshifts for all 11. As a result, of the now 468
spectroscopically observed CLASXS sources, we have presented redshift
identifications for 280. We extended the redshift information for the
CLASXS field to include an additional 134 photometric redshifts. 

Furthermore, we presented new CDF-N $J,H,$ and $K_s$ photometry and
some new optical spectroscopy, along with existing data. We also
included the GOODS-N \emph{Spitzer} $3.6~\mu$m and $24~\mu$m
detections. Since the publication of the Barger et al.~(2003) redshift
catalog for this field, we have obtained an additional 49 spectra and
identified redshifts for 39. As a result, of the 503 X-ray
sources in the CDF-N field, we have spectroscopically observed 459,
obtaining redshift identifications for 312. We
extended the redshift information for the CDF-N field to include an
additional 107 photometric redshifts.  

Finally, we determined the differential X-ray number counts for each
survey individually and for the full sample. Our differential number counts
for the combined sample agree well with the results from other X-ray surveys. 

\acknowledgements
 We thank John Silverman for alerting us to the zeropoint problems
 with the optical data for the CLASXS X-ray sources. We thank the
 TERAPIX team for the work they do to produce the CFHT data. L.~T.~was
 supported by a National Science Foundation Graduate Research
 Fellowship and a Wisconsin Space Grant Consortium Graduate Fellowship
 Award during portions of 
 this work. We also gratefully acknowledge support from NSF grants AST
 0239425 and AST 0708793 (A.~J.~B.~) and AST 0407374 and AST 0709356
 (L.~L.~C.~), the University of
 Wisconsin Research Committee with funds granted by the Wisconsin
 Alumni Research Foundation (A.~J.~B.~), and the David and Lucile
 Packard Foundation (A.~J.~B.~). This article is part of L.~T.'s
 Ph.D.~thesis work at the University of Wisconsin-Madison.

\end{document}